\documentclass{aa}
\usepackage{url}
\usepackage[colorlinks,citecolor=blue]{hyperref}
\usepackage{graphicx}
\usepackage[dvipsnames]{xcolor}
\usepackage{amsmath}
\usepackage{natbib}
\usepackage{txfonts}
\usepackage{float}
\usepackage{adjustbox}
\usepackage{MnSymbol} 
\usepackage{mhchem} 
\usepackage{physics}
\usepackage{placeins} 
\usepackage{threeparttable} 
\usepackage[percent]{overpic}
\usepackage{caption}
\usepackage{float}
\usepackage{dblfloatfix} 
\usepackage{cuted}
\usepackage{pict2e} 
\usepackage{booktabs}

\begin{document}

\newcommand{\edit}[1]{\textbf{#1}}
\renewcommand{\edit}[1]{#1}

\newcommand{\rr}[1]{\textbf{#1}}
\renewcommand{\rr}[1]{#1}

\title{Extending dynamical mass measurements:\\ Probing gravitational instability as a possible origin of millimeter-dust spirals} 
\author{V. Pezzotta \inst{1}
  \and S. Facchini \inst{1} 
  \and A. F. Izquierdo \inst{2,3}
  \and G. Lodato \inst{1} 
   \and C. Longarini \inst{4}
   \and J. Bae \inst{3}
   \and M. Galloway-Sprietsma \inst{3}
   \and C. Pinte \inst{5,6}
   \and C. J. Law \inst{2,7}
   \and T. Paneque-Carreño \inst{8}
    }
    
\institute{Dipartimento di Fisica, Università degli Studi di Milano, Via Celoria 16, Milano, 20133, Italy \\ 
              \email{viviana.pezzotta@unimi.it}
 \and {NASA Hubble Fellowship Program Sagan Fellow} 
 \and {Department of Astronomy, University of Florida, Gainesville, FL 32611, USA} 
 \and {Institute of Astronomy, University of Cambridge, Madingley Road, Cambridge, CB3 0HA, United Kingdom} 
 \and {School of Physics and Astronomy, Monash University, Clayton VIC 3800, Australia} 
 \and {Univ. Grenoble Alpes, CNRS, IPAG, 38000 Grenoble, France} 
 \and {Department of Astronomy, University of Virginia, Charlottesville, VA 22904, USA} 
 \and {Department of Astronomy, University of Michigan, Ann Arbor, MI 48109, USA} 
 }
\date{Received x xxxxxx xxxx/ Accepted x xxxxxxx xxxx}
\abstract {Constraining the total mass of protoplanetary disks is crucial for determining the amount of material available for planet formation. Yet providing accurate measurements of the disk mass is challenging. Investigating the gas dynamics provides a powerful tracer-independent method to precisely characterize disk masses. 
By fitting the velocity rotation curves of different molecular tracers with an accurate model including the thermal stratification and self-gravity of the disk, we constrained the stellar masses, disk masses, and scale radii for the disks around HD 97048 and WaOph 6. 
We obtained \edit{$M_\star=2.226 ^{+0.054}_{-0.049} \ M_\odot$, $M_\mathrm{d}=0.3 ^{+0.055}_{-0.061} \ M_\odot$, and $R_\mathrm{c}=172 ^{+24}_{-14}$ au for HD 97048 and $M_\star=0.956\ ^{+0.006}_{-0.006} \ M_\odot$, $M_\mathrm{d}=0.21 ^{+0.045}_{-0.038}\ M_\odot$, and $R_\mathrm{c}=647 ^{+193}_{-155}$ au} for WaOph 6. We also measured the corresponding gas-to-dust and disk-to-star mass ratios.
We efficiently extended the dynamical method to characterize embedded sources exhibiting features of absorption, for which a careful analysis is required to avoid biases in the retrieved velocity profiles. We prove the importance of including a beam smearing correction to the curves. If not included, this observational effect can systematically bias the velocity profiles, thus altering the disk mass estimates up to $\approx 45\%$. We provide comprehensive estimates of the systematic uncertainties on the best-fit parameters by bootstrapping over both the retrieved geometry and 2D thermal structure of the two disks. The overall uncertainty on the disk masses is $\approx 20\%$. Finally, we investigated the connection between disk stability and the appearance of spiral morphologies in the millimeter continuum emission by comparing the Toomre parameter of all dynamically weighed disks to date, demonstrating that disks with millimeter-dust spirals have systematically lower values of $Q$.
}
\titlerunning{Extending dynamical mass measurements: Probing GI as a possible origin of millimeter-dust spirals}
\maketitle
\nolinenumbers 
\newcommand{\eddyfont}{\ttfamily eddy\rmfamily}
\newcommand{\gofishfont}{\ttfamily GoFish\rmfamily}
\newcommand{\radialprofilefont}{\ttfamily radial{\_}profile\rmfamily}
\newcommand{\radproffont}{\ttfamily radprof \rmfamily}

\newcommand{\SF}[1]{\textcolor{olive}{#1}}
\newcommand{\VP}[1]{\textcolor{orange}{#1}}

\section{Introduction}\label{sec:intro}
Measuring the masses of protoplanetary disks is an essential yet challenging task. The amount of dust grains and gaseous material in planet-forming disks determines their evolution and sets the conditions for planet formation. The most abundant species in disks is \ce{H_2}, a molecule with no permanent electric dipole and only high energy lines, which are insensitive to the cold disk reservoir that dominates the mass. Therefore, other tracers are typically used to infer disk masses (e.g., see \citealt{miotello2023} for a review), which can be extrapolated assuming conversion factors to the \ce{H_2} mass or abundance. 

The most commonly used tracer is the millimeter continuum flux density \citep{beckwith1990}. If the emission is optically thin, the observed flux density is a direct probe of the dust mass. However, the continuum emission at millimeter wavelengths has been shown to be optically thick in many disks \citep{liu2022,kaeufer2023,xin2023}.
In addition, other significant assumptions regarding the opacity of dust grains, the temperature, and the gas-to-dust ratio are needed to convert the dust millimeter flux into a total disk mass \citep{manara2023}. Bright, optically thin emission lines of CO isotopologs can also be used to weigh disks, as CO is the second most abundant molecular species in disks and has a relatively simple and well-known chemistry \citep{vandishoeck1988}. However, large uncertainties are associated with the assumed molecular properties, such as the excitation temperature and the abundances relative to \ce{H_2} \citep{miotello2014,miotello2016,oberg2023,miotello2023}. 
A chemical tracer of CO-poor gas, \ce{N_2H^+}, has recently been used to empirically correct CO-based gas masses \citep{anderson2022,trapman2022,trapman2025}. In this case, precise constraints on the cosmic-ray ionization rate in protoplanetary disks are needed. Hence, the main traditional methods to estimate disk masses are highly dependent on the chosen tracers and their physico-chemical properties as well as the assumed constraints on the conversion factors and optical depths. 

In this context, dynamical measurements that leverage the observed gas motion represent an important step forward in obtaining an accurate, tracer-independent methodology to obtain disk masses. The azimuthal velocity of the gas represents the dominant motion in disks \citep{pinte2023}. For a single emission line, it is possible to directly derive the radial profile of the azimuthally averaged rotation velocity of the gas - i.e., the gas rotation curve. The azimuthal motion of the gas is primarily shaped by the gravitational influence of the central star. In addition to this dominant Keplerian term, the presence of pressure gradients and of the disk self-gravity also leave an imprint on the gas kinematical pattern \rr{\citep{rosenfeld2013}}, which can be directly traced through the rotation curves. These contributions depend on both the disk mass and its scale radius \edit{(i.e., the characteristic radius where the surface \rr{density} profile is exponentially tapered)}, two fundamental properties of disks that are challenging to measure. By fitting the gas rotation curve with an appropriate model including the different contributions to the gas azimuthal velocity, we can derive dynamical estimates for the stellar mass, the disk mass, and the disk scale radius simultaneously, without any assumptions on the dust or gas properties. The modeling of rotation curves, including the contribution of the disk mass, has become increasingly popular, starting from the analytical work of \citet{bertin} and the first dynamical measurements of vertically isothermal disks \citep{veronesi2021,lodato}. Recently, \citet{martire} showed that the vertical thermal stratification of the disk has to be taken into account when fitting rotation curves, as the vertical variation in the temperature modifies the pressure gradients, providing a further nonnegligible contribution in the rotation curves. \citet{veronesi2024} and \citet{andrews2024} characterized the systematic uncertainties that are intrinsic to this dynamical methodology, setting $5\%$ as the minimum disk-to-star mass ratio value for which the disk mass is measurable dynamically and estimating a typical systematic uncertainty of $25-30\%$ on the disk mass. \citet{pezzotta2025} extended the fitting procedure to a multi-molecule ($\geq2$ lines) scenario, simultaneously fitting seven tracers emitting from different heights in the disk. More recently, \citet{exoalma_longarini} extracted dynamical disk masses for ten sources included in the exoALMA Large Program sample \citep{exoALMA_1}.

\edit{In this work, we dynamically extract the stellar masses, disk masses, and disk scale radii of HD 97048 and WaOph 6, two well-known sources hosting dynamically perturbed disks with no dynamical mass measurements to date. 
These disks respectively show a kinematically detected kink in the gas emission (hinting at the presence of an embedded protoplanet, \edit{\citealt{pinte2019}}) and a clear spiral in the millimeter dust emission \edit{\citep{andrews2018,huang2018_3}}.
We first extend the dynamical methodology to analyze sources with signs of absorption in the gas emission.
For both sources, we derive the 2D thermal structures of the disks and obtain the gas rotation curves; we fitted the curves while taking into account the disk self-gravity and vertical thermal stratification, and we give a comprehensive view of the systematic uncertainties on the derived disk masses. Finally, we investigate the connection between disk stability and the appearance of spiral morphologies in the millimeter-dust emission.}

This paper is organized as follows. 
In Sect. \ref{sec:3_sources} we introduce our sample. In Sect. \ref{sec:4_methods_results} we describe the adopted methodology, and we illustrate our results. In Sect. \ref{sec:5_discussion} we compare our estimates with the literature, and we discuss our findings in the context of disk stability. Finally, in Sect. \ref{sec:6_conclusions}, we draw our conclusions.


\section{Targeted sources} \label{sec:3_sources}

\edit{An accurate estimate of the disk 2D $(R,z)$ thermal structure is a fundamental requirement to obtain precise dynamical mass measurements. The detection of multiple ($\geq2$) gas lines for each source 
is needed. For our analysis, mid-inclination disks are preferred to geometrically retrieve the molecular emitting surfaces 
and therefore recover the disk temperature vertical distribution.} Instead of simply assuming a vertically isothermal scenario, taking into account the vertical thermal stratification of disks is crucial, as neglecting the vertical gradient in the disk temperature can lead to dynamical estimates for the stellar mass $M_\star$, disk mass $M_\mathrm{d}$, and disk scale radius $R_\mathrm{c}$ \edit{differing by a factor of a few} \citep{martire,exoalma_longarini}. 
In addition to the 2D disk temperature, a precise sampling of the molecular rotation curves is needed: \rr{in Appendix \ref{app:requirements} we investigate how the number of fitted rotation curves impacts the extracted disk mass estimate, provided the 2D thermal structure of the disk. With a test on the exoALMA sample \citep{exoALMA_1}, we show that the knowledge of the 2D thermal structure is in most cases the dominant contribution to reliable disk mass measurements, with respect to the number of fitted lines.}

Considering the requirements to achieve reliable mass measurements through the dynamical method, explained above and in Appendix \ref{app:requirements}, \edit{we searched the ALMA archive for sources with observations of at least two bright CO lines, obtained with sufficient line sensitivity, spatial and spectral resolution to enable surface reconstruction}. Two lines emitting from distinct locations in the disk are required to fit the 2D disk thermal structure; the availability of multiple lines of different CO isotopologs, covering distinct regions in the vertical and radial ranges, is optimal for our goal. We only considered sources with favorable inclination angles for the reliable retrieval of the emitting surfaces (35\textdegree-75\textdegree). 
We excluded from our sample those disks that already have dynamical mass measurements from previous studies (including the sources targeted in the MAPS and exoALMA Large Programs).

From our archival search, we selected four sources: HD 97048, WaOph 6, CI Tau, Sz 91. We performed the analysis routine on all of them: due to the \edit{low signal-to-noise ratio, we could not reliably extract the molecular emitting surfaces needed to fit the 2D thermal structures for CI Tau and Sz 91. Hence, we} discarded these two sources and focused our analysis on HD 97048 and WaOph 6: we list the properties of the sources and the details of the corresponding ALMA data we used in Table \ref{tab:sources}. For each line, we used the line+continuum cube to extract the gas brightness temperature along the emitting layers with the Planck function \edit{\citep[continuum subtraction could lead to underestimating the gas temperature, ][]{weaver2018}}, and the continuum-subtracted cube for the kinematical analysis. For HD 97048, we used \ce{^12CO} and \ce{C^18O} 2 -- 1 cubes from \citet{law2022}, while \ce{^13CO} 3 -- 2 cubes were obtained from \citet{pinte2019}.
For WaOph 6, we used the publicly available \ce{^12CO} 2 -- 1 continuum-subtracted cube from the DSHARP ALMA Large Program \citep{andrews2018}, and the corresponding line+continuum cube from \citet{law2022}. \ce{^12CO} and \ce{^13CO} 3 -- 2 cubes were obtained from \citet{paneque2023}. \edit{We refer readers to the above-mentioned papers for a thorough description of data calibration and imaging processes.}

\begin{table}[h!]
\centering
\begin{minipage}{\columnwidth} 
\caption{Line transitions considered for this work for the analyzed sources.}
\label{tab:sources}
\renewcommand\arraystretch{1.2}

\begin{adjustbox}{max width=\linewidth}
\begin{tabular}{|c|c|c|c|c|c|c|c|}
\hline
            \textbf{source}  & \textbf{d} & \textbf{transition} & \textbf{$\nu$} & \textbf{beam}       & \textbf{beam PA}  & \textbf{spec. res.} & \textbf{project code}\\
                             &  [pc]      &                     & [GHz]          & [$\prime \prime$]   & [\textdegree]     & [km/s]  &             \\ 
            \hline 
               HD 97048 &  184  & \ce{^12CO} 2 -- 1 & 230.538           & 0.41 $\times$ 0.17 &  29  & 0.2 & 2015.1.00192.S \\
                        &       & \ce{^13CO} 3 -- 2 & 330.588           & 0.15 $\times$ 0.11 &  -38 & 0.12 & 2016.1.00826.S \\
                        &       & \ce{C^18O} 2 -- 1\tablefootmark{a} & 219.560 & 0.42 $\times$ 0.18 &  24 & 0.2 & 2015.1.00192.S \\
                WaOph 6 &   122 & \ce{^12CO} 2 -- 1 & 230.538           & 0.14 $\times$ 0.12 & -79 & 0.35 & 2016.1.00484.L \\    
                        &       & \ce{^12CO} 3 -- 2 & 345.796           & 0.33 $\times$ 0.25 &  87 & 0.223 & 2015.1.00168.S \\ 
                        &       & \ce{^13CO} 3 -- 2 & 330.588           & 0.41 $\times$ 0.26 &  -87 & 0.223 & 2015.1.00168.S \\ 
            \hline
\end{tabular}
\end{adjustbox}

\vspace{2pt}
\tablefoot{For each line we list the distance of the source, the emission rest frequency of the transition, the spatial and spectral resolutions of the non-continuum-subtracted data cubes, and the ALMA project codes.\\
\tablefoottext{a}{In this case, beam parameters and spectral resolution refer to the continuum-subtracted cube.
This transition was excluded from the thermal structure analysis, as only the continuum-subtracted cube was available, but we included in the rotation curve fitting, which requires continuum-subtracted data cubes.}
}
\end{minipage}

\end{table}

Both sources show remarkable features in the emission, \edit{which are discussed in detail in the following sections.}, hinting at dynamically perturbed disks. Hence, determining their disk masses is of primary value to further characterize these unique systems. 

\begin{figure}[h]
    \includegraphics[width=\columnwidth]{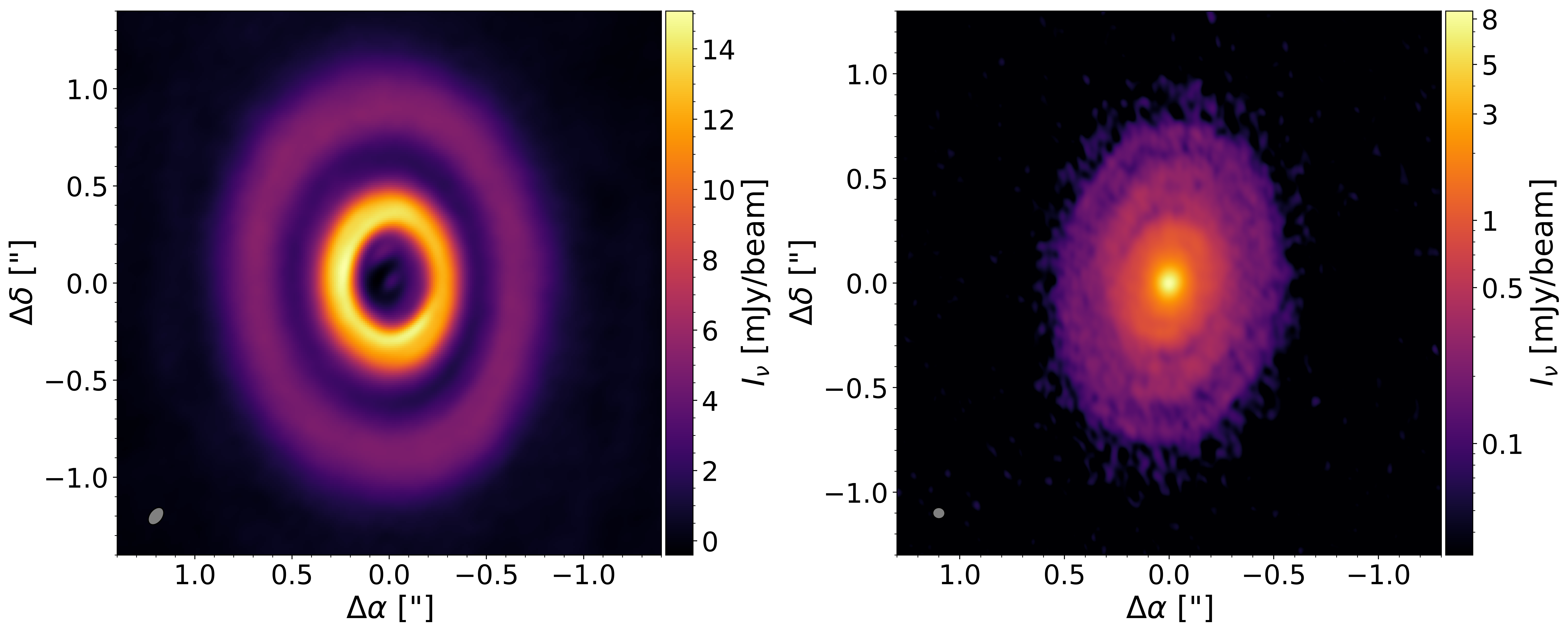}
    \caption{\rr{Dust millimeter-continuum emission for HD 97048 (0.89 mm, left panel) and WaOph 6 (1.25 mm, right panel). Data from 2016.1.00826.S \citep[PI: G. van der Plas, ][]{pinte2019} and 2016.1.00484.L \citep[PI: S. Andrews, ][]{andrews2018}. We applied a \textit{log} stretch for WaOph 6 for a better visualization of the dust structures.}}
    \label{fig:continuum}
\end{figure}

\subsection{HD 97048}
HD 97048 is a $\approx 3$ Myr old Herbig AeBe star (spectral type B9-A0, $T_\star \approx 10000$ K), located in the Chameleon I star-forming region \citep{lagage2006,vanleeuwen2007}. The protoplanetary disk around the central star was initially observed in the mid-infrared polycyclic aromatic hydrocarbon (PAH) emission by \citet{lagage2006}, and was found to be quite extended and strongly flared. Disk structures \edit{such as spiral-like features and clumps were} detected in scattered light observations with HST, at $\approx 600$ au \citep{doering2007}. The disk was observed in the near-infrared with VLT/SPHERE and Gemini/GPI in the $J-K$ bands, unveiling multiple substructures between 40 and 340 au.
The disk extends out to $\approx 350$ au in the \edit{millimeter} dust continuum emission, \edit{shown in Fig. \ref{fig:continuum} (left panel)}
, with a large $\approx 40-46$ au inner cavity, and azimuthally symmetric rings and gaps \citep{walsh2016,vanderplas2017,asensio-torres2021}. The disk molecular emission, as probed by the CO 3 -- 2 line, was observed out to $\approx 750$ au from the star, twice the radial extent of the (sub)millimeter dust disk. High spatial and spectral resolution data for the \ce{^13CO} 3 -- 2 line allowed the kinematic detection of a localized velocity kink at $130$ au, co-located with both a dust and a gas gap \citep{pinte2019}. The presence of an embedded planet of a few Jupiter masses has been suggested to explain the direct correspondence between the localized perturbation in the gas motion detected in CO and the co-located dust gap. 
An accurate and precise measurement of the disk mass is necessary to characterize the disk stability and allow further investigation of the nature of the perturbed gas flow in the disk.

\subsection{WaOph 6}
WaOph 6 is a $\approx 0.3$ Myr old T Tauri star (spectral type K6, $T_\star \approx 4200$ K) located in the $\rho$ Ophiuchus star-forming region \citep{eisner2005,ricci2010}. 
This source was observed at high spatial resolution as part of the DSHARP ALMA Large Program \citep{andrews2018}. Its protoplanetary disk hosts symmetric $m=2$ compact spiral patterns in the millimeter-dust emission, extended from $\approx 25$ au out to $\approx 75$ au, in conjunction with annular substructures \citep{huang2018_3}. A gap and a ring were detected at 79 au and 88 au, respectively, but additional substructures could be present and tightly intertwined with the complex spiral morphology \citep[\edit{see right panel of Fig. \ref{fig:continuum}, and}][]
{huang2018_2}. 

The presence of an embedded planetary companion and gravitational instability (GI) are two possible mechanisms that can cause the appearance of such millimeter-dust spiral features. In principle, GI is expected to create symmetric, logarithmic spiral arms with constant pitch angle, while a planet would induce spirals with variable pitch angles \citep{meru2017}. WaOph 6 shows a structure that can be traced back to a pair of logarithmic spirals; still, the retrieval of the pitch angle for this disk was complicated by the presence of overlapping annular substructures and large uncertainties in the radial locations of the spirals \citep{huang2018_3}. Moreover, synthetic observations of millimeter-dust spirals formed via the two different mechanisms appear to be similar \citep{dong2015a,dong2015b,meru2017}. However, GI is expected to occur in young and cold sources \citep{kratter2009}, with likely high stellar accretion rates \citep{dong2015a}. WaOph 6 is indeed a young system, characterized by relatively low temperatures \citep[and this work]{andrews2009,law2022}, and a high stellar accretion rate ($\dot M=10^{-6.6\pm0.5}\ M_\odot \ yr^{-1}$, \citealt{eisner2005}). 
\citet{cadman2020} self-consistently modeled the spiral amplitudes for WaOph 6, managing to reproduce the observed fluxes, suggesting that GI may be the dominant mechanism driving the observed spiral structures.

On the other hand, a possible alternative mechanism to generate the observed substructures would be the disk interaction with a massive planetary companion \edit{(or binary interaction, but this source has no known binary companion)}. However, to trigger the two-armed symmetric spiral modes observed in WaOph 6, a massive planet of tens of Jupiter masses would be required, which should be visible at submillimeter/infrared frequencies \citep{meru2017,cadman2020}; yet, no planetary companion has been confirmed so far. 
The presence of an embedded planet was investigated by \citet{pinte2020}, supported by the potential kinematic detection of a kink in the CO emission, even though a larger-scale velocity structure could not be excluded due to the cloud contamination. \citet{speedie2022} confirmed that the observed millimeter-dust spirals in WaOph 6 do not align with the theoretical trajectory of the spiral that would originate at the suggested candidate planet location.
In case the planets responsible for the observed rings and gaps are less massive and fainter, thus non-detectable with the current instrumentation, GI could be driving the spiral structure, leading to a scenario of combined GI and planet-disk interactions at play.

However, the origin and the nature of the spiral structure in this disk is currently still highly debated.
Further considerations on the disk mass could help \edit{distinguish the two possibilities from each other} and constrain the more consistent one.


\section{Methodology and results}
\label{sec:4_methods_results}

In this section, we give an overview of the workflow of our analysis, describing in detail the methodology we followed. We then illustrate our main results.

\begin{itemize}
    \item 
    For each line, we modeled the observed line intensity in the continuum-subtracted channel maps using \texttt{discminer}\footnote{Publicly available on \url{https://github.com/andizq/discminer}} \citep{izquierdo2021}. From the best-fit model we obtained the disk geometric parameters (central offset coordinates $x_\mathrm{c}$ and $y_\mathrm{c}$, inclination $i$ and position angle PA) and the systemic velocity ($v_\mathrm{sys}$) that describe the observations. These parameters were used in the next step to extract the molecular emitting layers from the data cubes. At this point, we extracted the gas rotation curve with the \texttt{discminer} \radproffont routine. The details about the rotation curve extraction procedure depend on the specific source and the quality of the available data and are discussed in Sect. \ref{subsec:discminer}. To account for the beam smearing effect in the extracted rotation curves, for every line we applied a correction factor inferred from the convolved Keplerian model. The details of the beam smearing correction are discussed in Sect. \ref{subsec:beam_smearing}.
    
    \item 
    Using the disk geometrical parameters obtained in the previous step, with \texttt{disksurf}\footnote{Publicly available on \url{https://github.com/richteague/disksurf}} \citep{teague_disksurf} we extracted the molecular emitting layer for each line and the brightness temperature $T_\mathrm{B}$ along them \edit{(for optically thick lines, $T_\mathrm{B}$ traces the gas kinetic temperature)}. For each disk we obtained a collection of $(R,z)$ points with a known temperature.
    
    \item 
    We adopted a 2D parametric expression to describe the 2D $(R,z)$ disk thermal structure. For each source, we simultaneously fitted the temperature across the retrieved emission layers with the chosen parameterization, constraining the values for the thermal parameters that best describe the 2D thermal structure of the disk (Sect. \ref{subsec:pipeline}). 
    
    \item 
    With \texttt{DySc},\footnote{The code is publicly available at \url{https://github.com/crislong/DySc}. More details on the implementation can be found in \citet{exoalma_longarini}.} for each source we simultaneously fitted the rotation curves of all the considered tracers with a thermally stratified model \citep{martire}, including the 2D disk thermal structure found in the previous step. As the model includes contributions from the stellar gravity, the pressure gradients, and the disk self-gravity, we dynamically constrained the stellar and disk masses, and disk scale radius.
    
    \item 
    As the major contributions to the systematic uncertainties in this method are represented by both the disk geometry and thermal structure \citep{andrews2024}, we repeated the whole procedure 100 times from step 2 to step 5, bootstrapping over both geometrical and thermal parameters, as explained in Sect. \ref{subsec:bootstrapping}. In this way, we accounted for the major sources of systematic uncertainty and were able to provide a realistic, comprehensive estimate of the uncertainties on the obtained dynamical measurements. 
\end{itemize}

\subsection{Disk modeling and extraction of rotation curves}\label{subsec:discminer}
In the first step of our analysis, we model the line intensity of each molecular tracer with \texttt{discminer}, introduced by \citet{izquierdo2021}. We generate the Keplerian model channel maps that best reproduce the observed line intensity, constrained by the set of parametric prescriptions summarized in Table \ref{table:discminer_pars}, describing the disk geometry, the velocity, the shape and intensity of line emission, and the upper and lower parametric emitting surfaces. To sample the parameter space, we employed the Markov chain Monte Carlo random sampler \texttt{emcee} \citep{emcee}; for each analyzed tracer, we explored the parameter space with 150 walkers, and we ran the fit for 10000 steps. In Appendix \ref{app:discminer} we list the best-fit values from the \texttt{discminer} models of all the analyzed tracers in Table \ref{table:discminer_bestfit}.

\begin{table}[h!] 
    \centering
    \renewcommand\arraystretch{1.2} 
    \caption{\centering Attributes adopted in the \texttt{discminer} models.}
    \label{table:discminer_pars}
    \begin{adjustbox}{max width=\columnwidth}
    \begin{tabular}{c|c}
    \hline
     \textbf{Attribute} & \textbf{Prescription} \\
    \hline
        Orientation & \textit{i}, $\mathrm{PA}$, $x_\text{c}$, $y_\text{c}$ \\ 
        \hline
        Velocity & $v=\sqrt{\frac{GM_{\star,\mathrm{dm}}}{r^3}}R$, $v_\mathrm{sys}$ \\ 
        \hline
        Upper surface & $z_\mathrm{u}=z_0 (R/D_0)^p \exp[-(R/R_\mathrm{t})^q]$ \\
        Lower surface & $z_\mathrm{l}=-z_0 (R/D_0)^p \exp[-(R/R_\mathrm{t})^q]$ \\
        \hline
        Peak intensity & $I_\mathrm{p}=I_0 (R/D_0)^p (z/D_0)^q$ \\
        Line width & $L_\mathrm{w}=L_\mathrm{w0} (R/D_0)^p (z/D_0)^q$ \\
        Line slope & $L_\mathrm{s}=L_\mathrm{s0} (R/D_0)^p$ \\
        
    \hline
    \end{tabular}
    \end{adjustbox}
    \tablefoot{Here, $x_\text{c}$ and $y_\text{c}$ represent the central offset, $M_{\star,\mathrm{dm}}$ is the Keplerian mass, $G$ is the gravitational constant, $D_0=100$ au is a normalization factor, $z$ is the height above the midplane, $R$ is the cylindrical radius, and $r$ is the spherical radius. All the other variables are free parameters that are fitted independently.}
    
\end{table}

\begin{figure}
   \centering
    \caption*{\LARGE\textbf{HD 97048}}
   \includegraphics[width=\columnwidth]{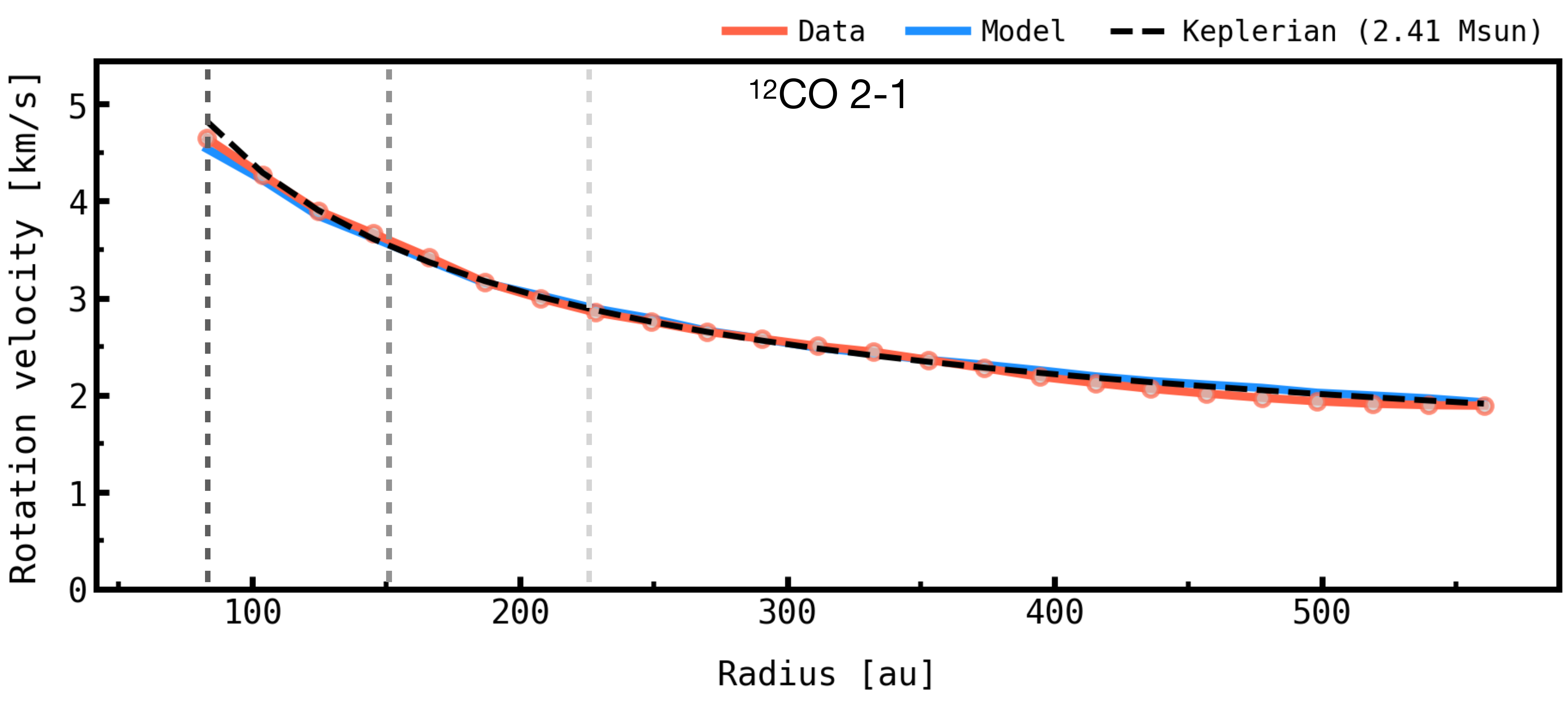}
   \includegraphics[width=\columnwidth]{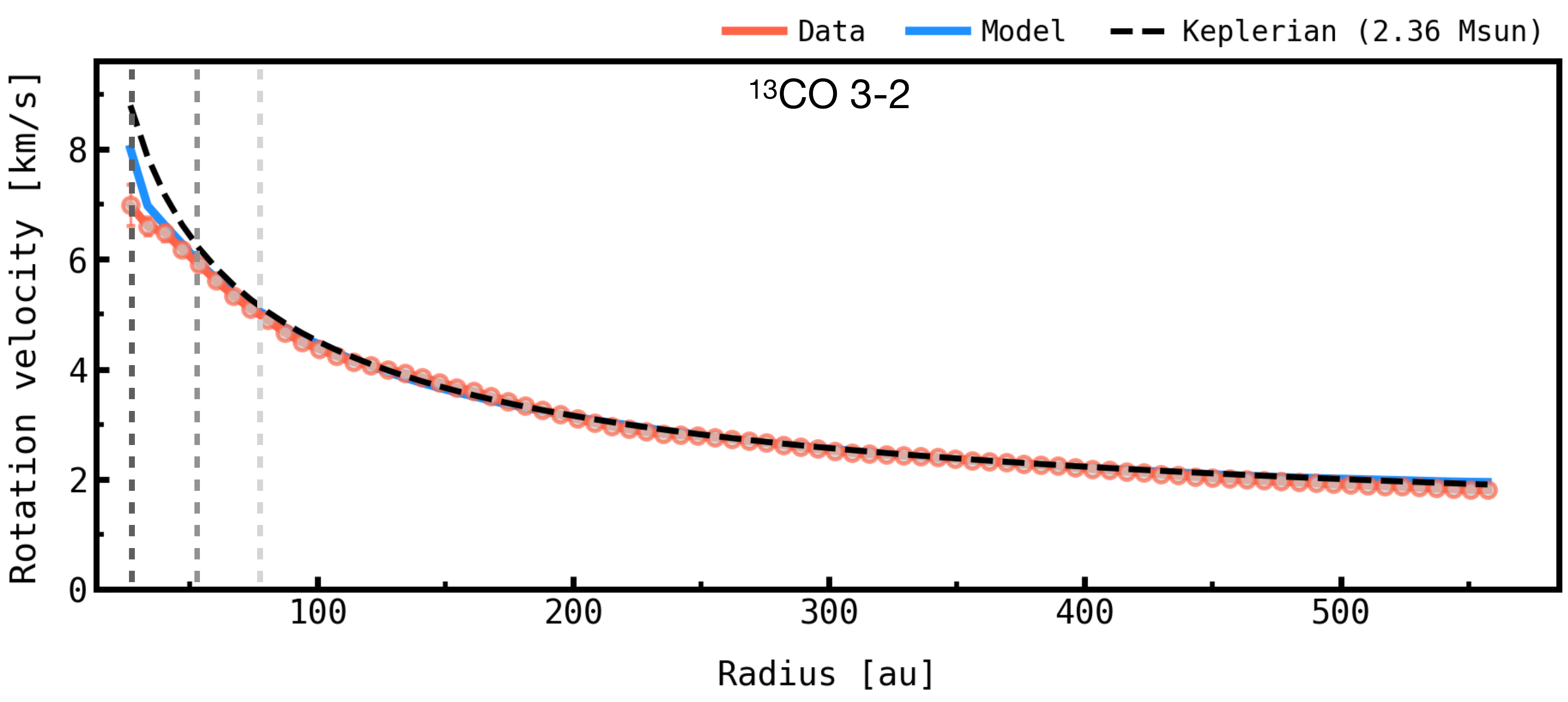}
   \includegraphics[width=\columnwidth]{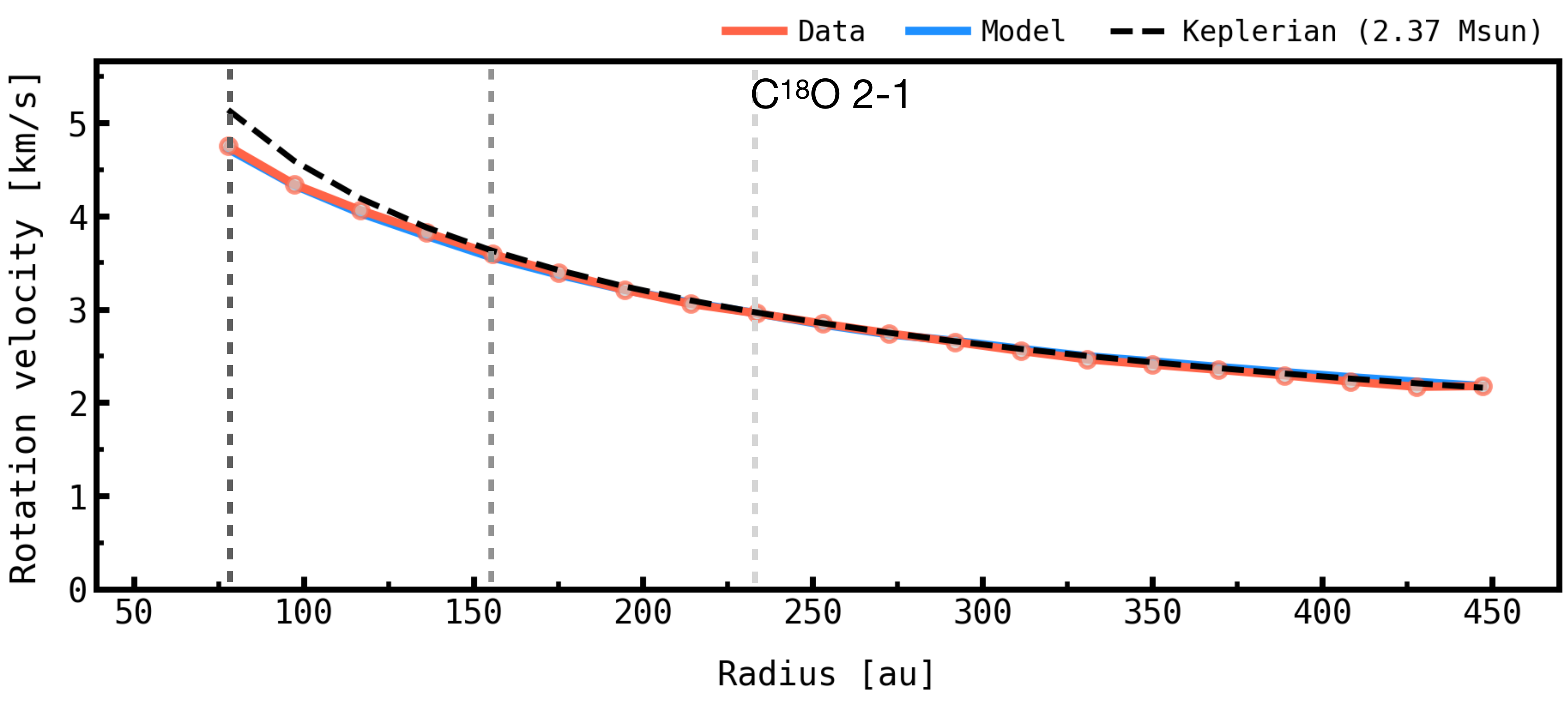}
   
      \caption{
      Rotation curves extracted with \texttt{discminer} for the HD 97048 disk: \ce{^12CO} 2 -- 1 (top left), \ce{^13CO} 3 -- 2 (top right), and \ce{C^18O} 2 -- 1 (bottom). The red line shows the data before applying the beam smearing correction, the black dashed line shows the pure Keplerian model, and the blue line shows the beam-convolved Keplerian model. \edit{Gray dashed lines represent one, two, and three beams from the central star. Velocity ranges on the y-axes are different for visualization purposes.}}
         \label{fig:rot_curves_dm_hd}
\end{figure}

\begin{figure}[h!]
   \centering
    \caption*{\LARGE\textbf{WaOph 6}}

   \includegraphics[width=\columnwidth]{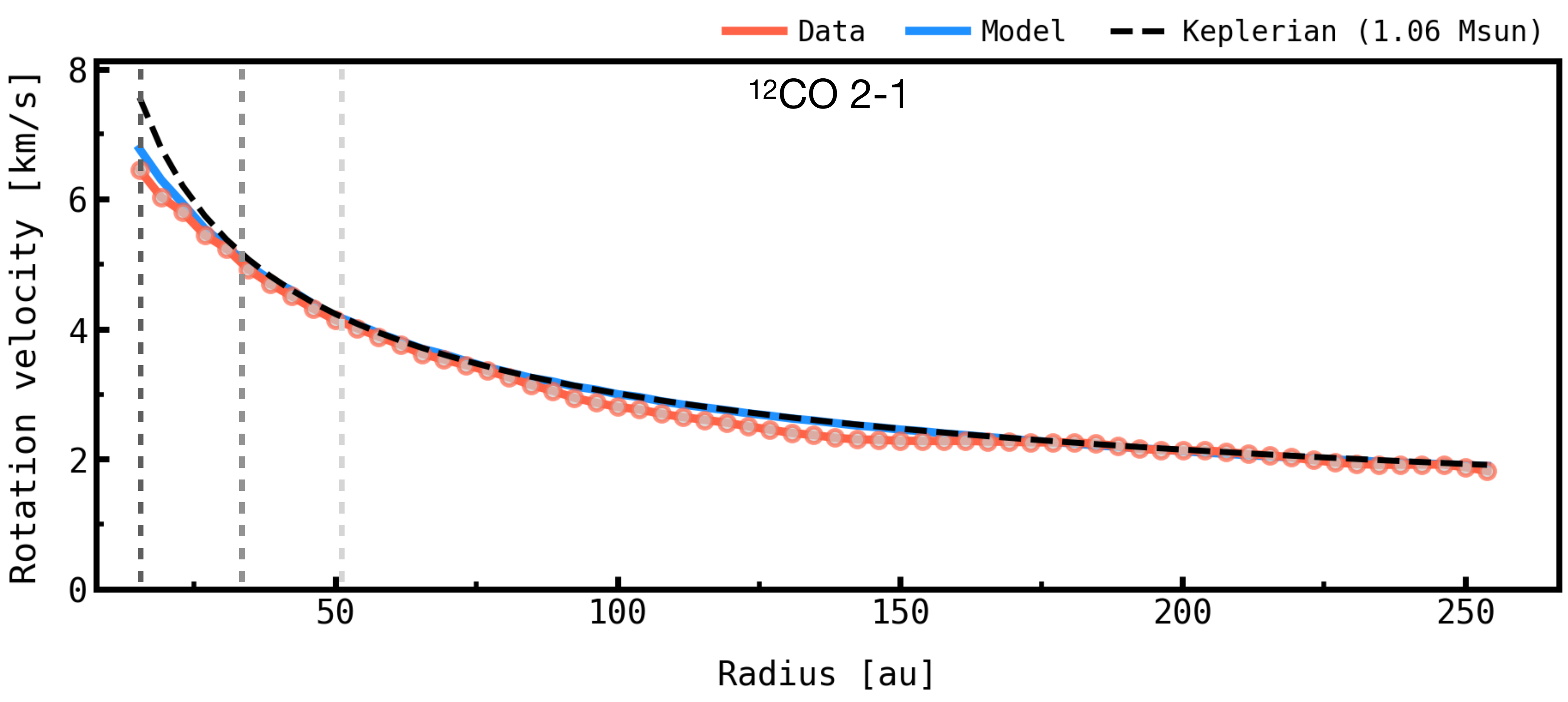} 
   \includegraphics[width=\columnwidth]{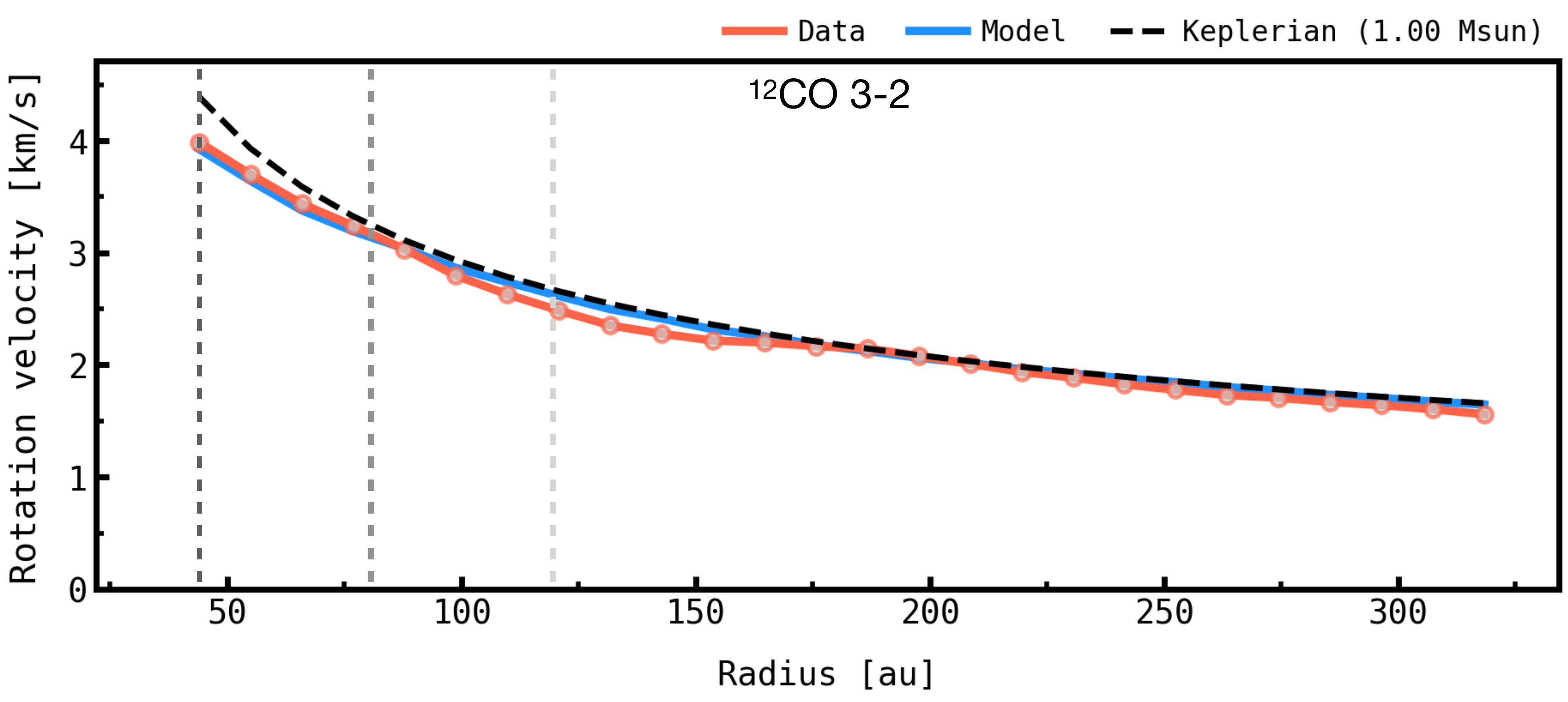} 
   \includegraphics[width=\columnwidth]{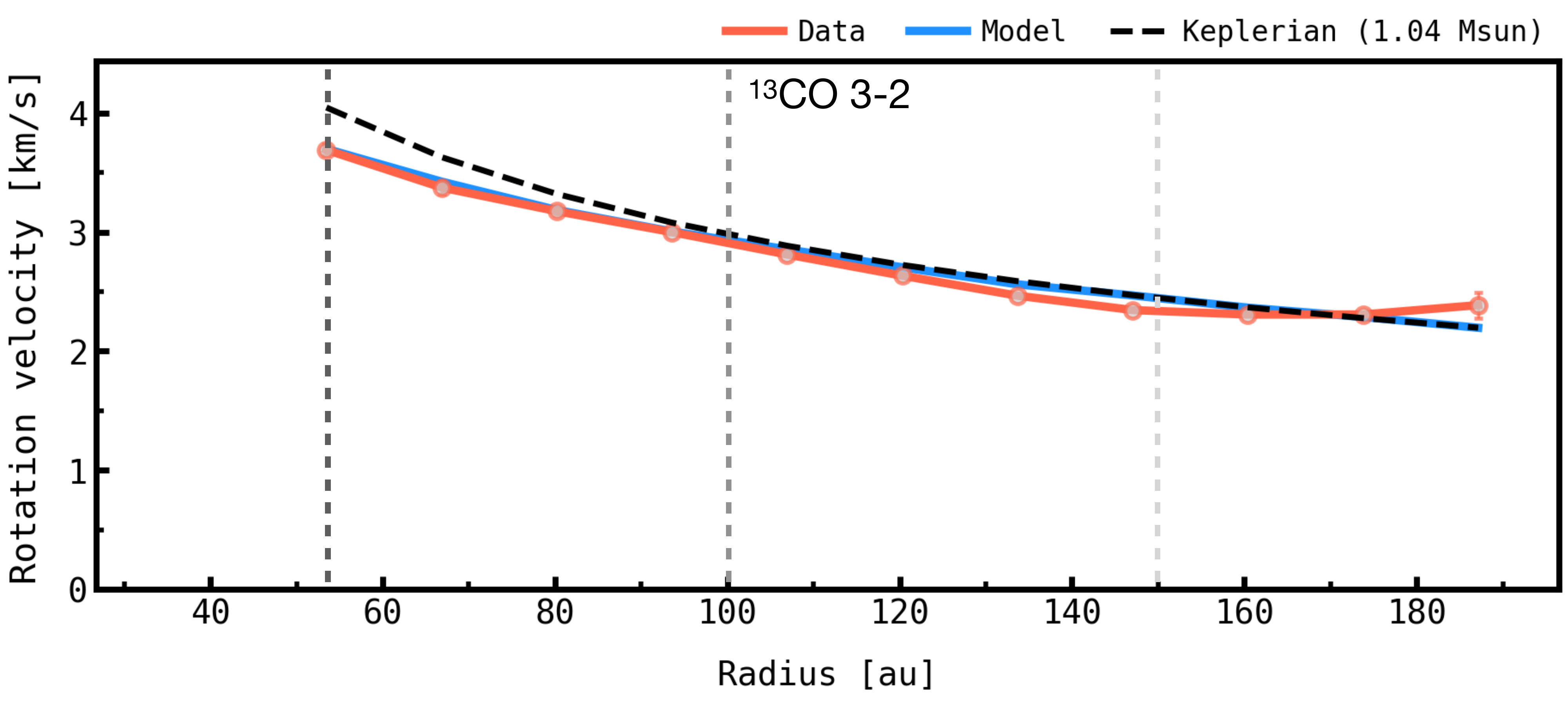} 
   
      \caption{
      Rotation curves extracted with \texttt{discminer} for the WaOph 6 disk: \ce{^12CO} 2 -- 1 (top left), \ce{^12CO} 3 -- 2 (top right), and \ce{^13CO} 3 -- 2 (bottom). The red line shows the data before applying the beam smearing correction, the black dashed line shows the pure Keplerian model, and the blue line shows the beam-convolved Keplerian model. \edit{Gray dashed lines represent one, two, and three beams from the central star. Velocity ranges on the y-axes are different for visualization purposes.}}
         \label{fig:rot_curves_dm_wa}
\end{figure}

After modeling the channel maps for each cube, with the \texttt{discminer} \radproffont routine we extract the rotation curves for all the tracers. 
We show the extracted rotation curves of all the lines for HD 97048 and WaOph 6, respectively in Figs. \ref{fig:rot_curves_dm_hd} and \ref{fig:rot_curves_dm_wa}. We notice that all the rotation curves display an internal flattening at smaller radii, which is due to beam smearing. In the next step of our analysis, we correct for this observational bias, as explained in Sect. \ref{subsec:beam_smearing}, and we use the corrected rotation curves to fit the disk masses.

\subsubsection{Accounting for absorption in the rotation curves extraction} \label{subsec:absorption}

\begin{figure*}
\centering

\begin{minipage}[t]{0.48\textwidth}
    \centering
    {\Large\textbf{HD 97048}\par\medskip}
    
    \begin{minipage}[c]{0.48\linewidth}
        \includegraphics[width=\linewidth]{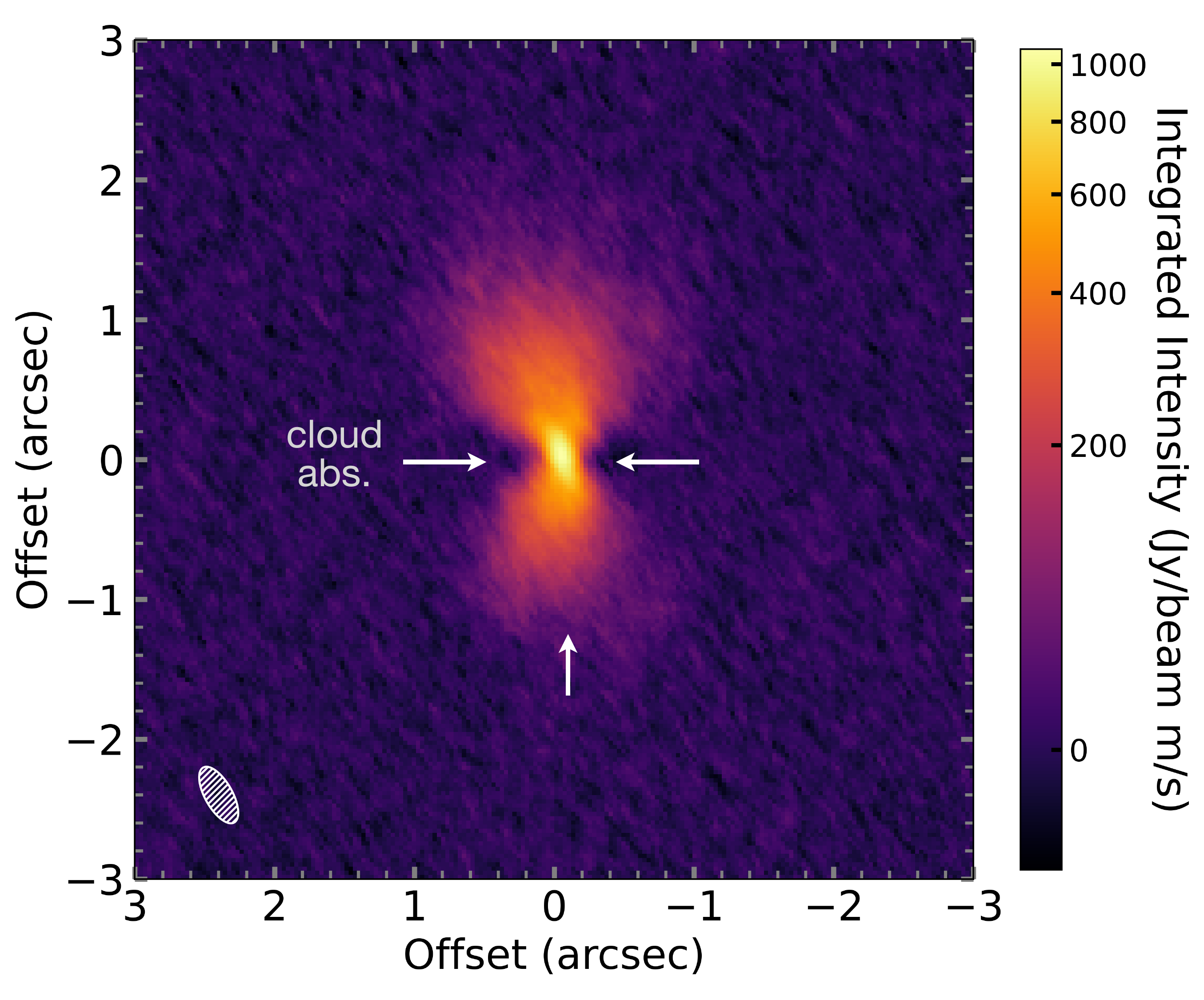}
    \end{minipage}\hfill
    \begin{minipage}[c]{0.48\linewidth}
        \includegraphics[width=\linewidth]{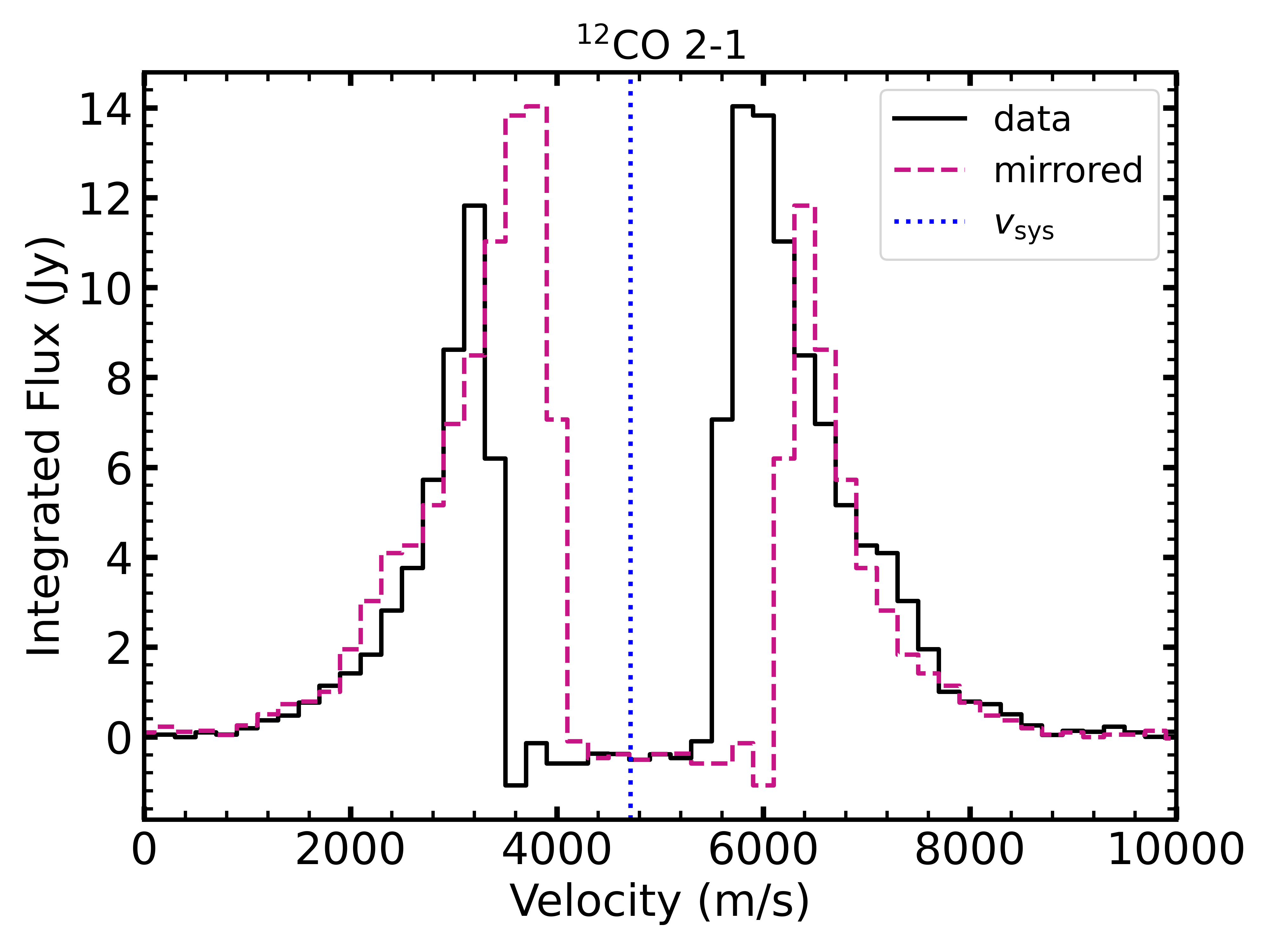}
    \end{minipage}

    \begin{minipage}[c]{0.48\linewidth}
        \includegraphics[width=\linewidth]{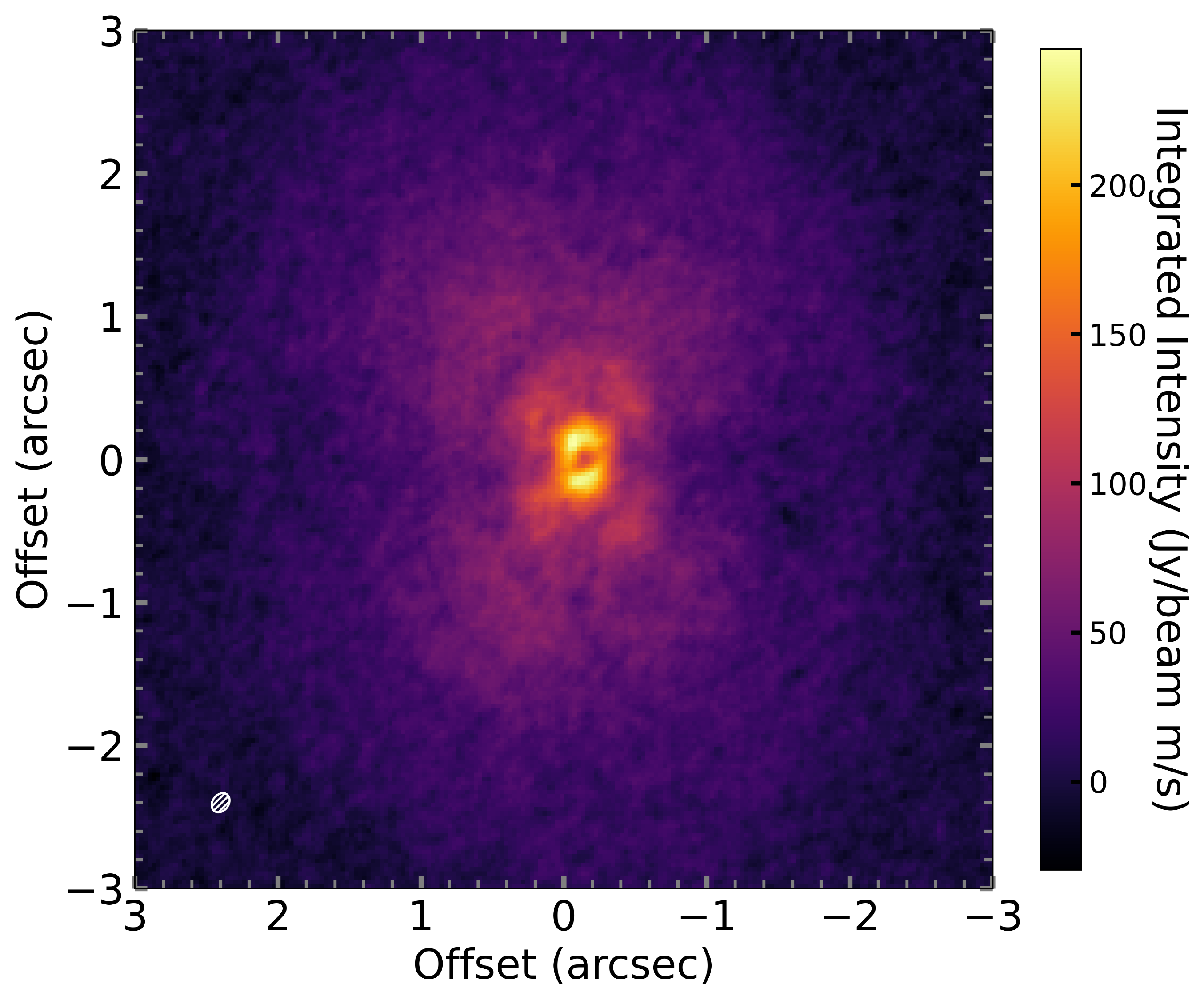}
    \end{minipage}\hfill
    \begin{minipage}[c]{0.48\linewidth}
        \includegraphics[width=\linewidth]{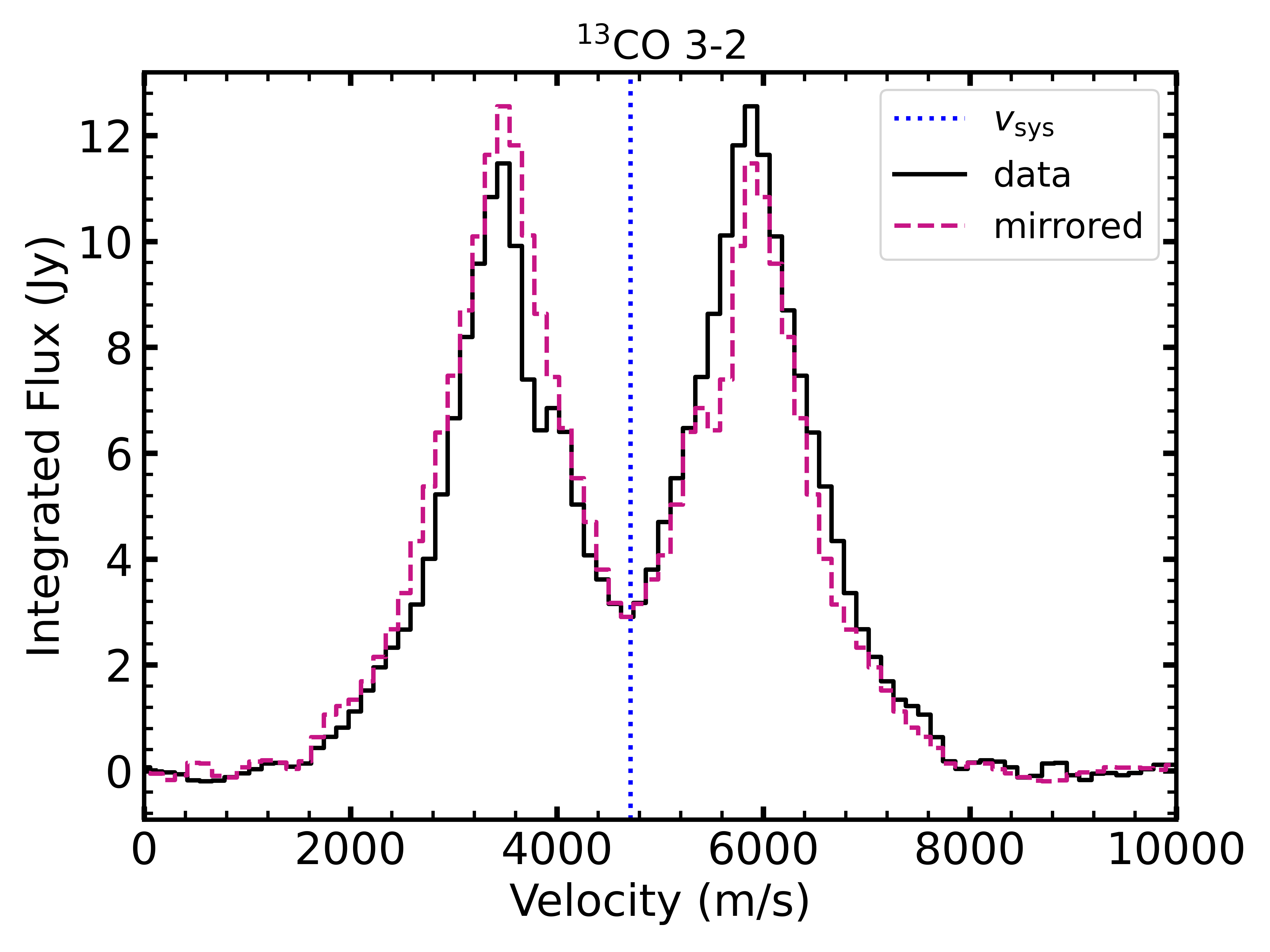}
    \end{minipage}

    \begin{minipage}[c]{0.48\linewidth}
        \includegraphics[width=\linewidth]{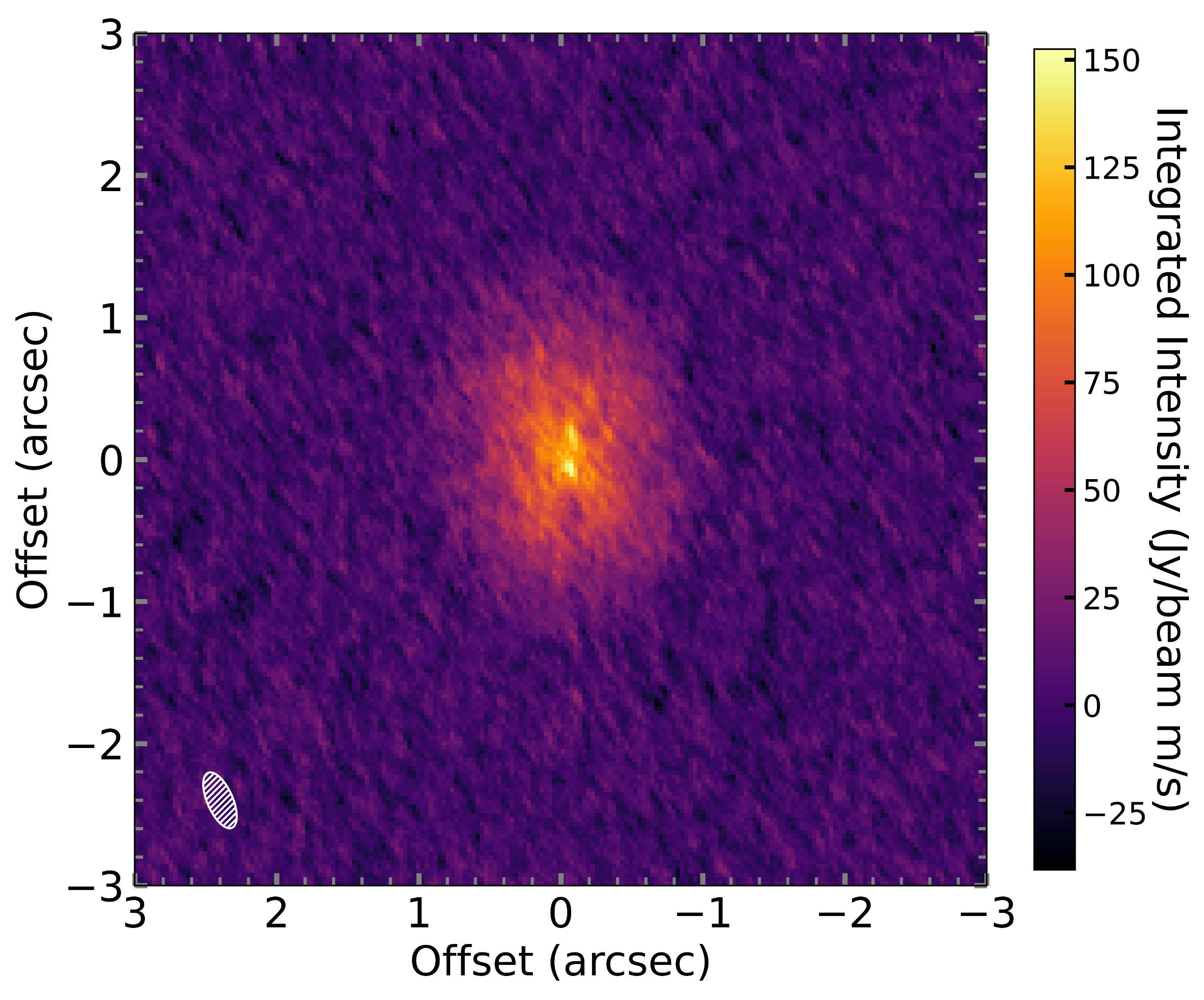}
    \end{minipage}\hfill
    \begin{minipage}[c]{0.48\linewidth}
        \includegraphics[width=\linewidth]{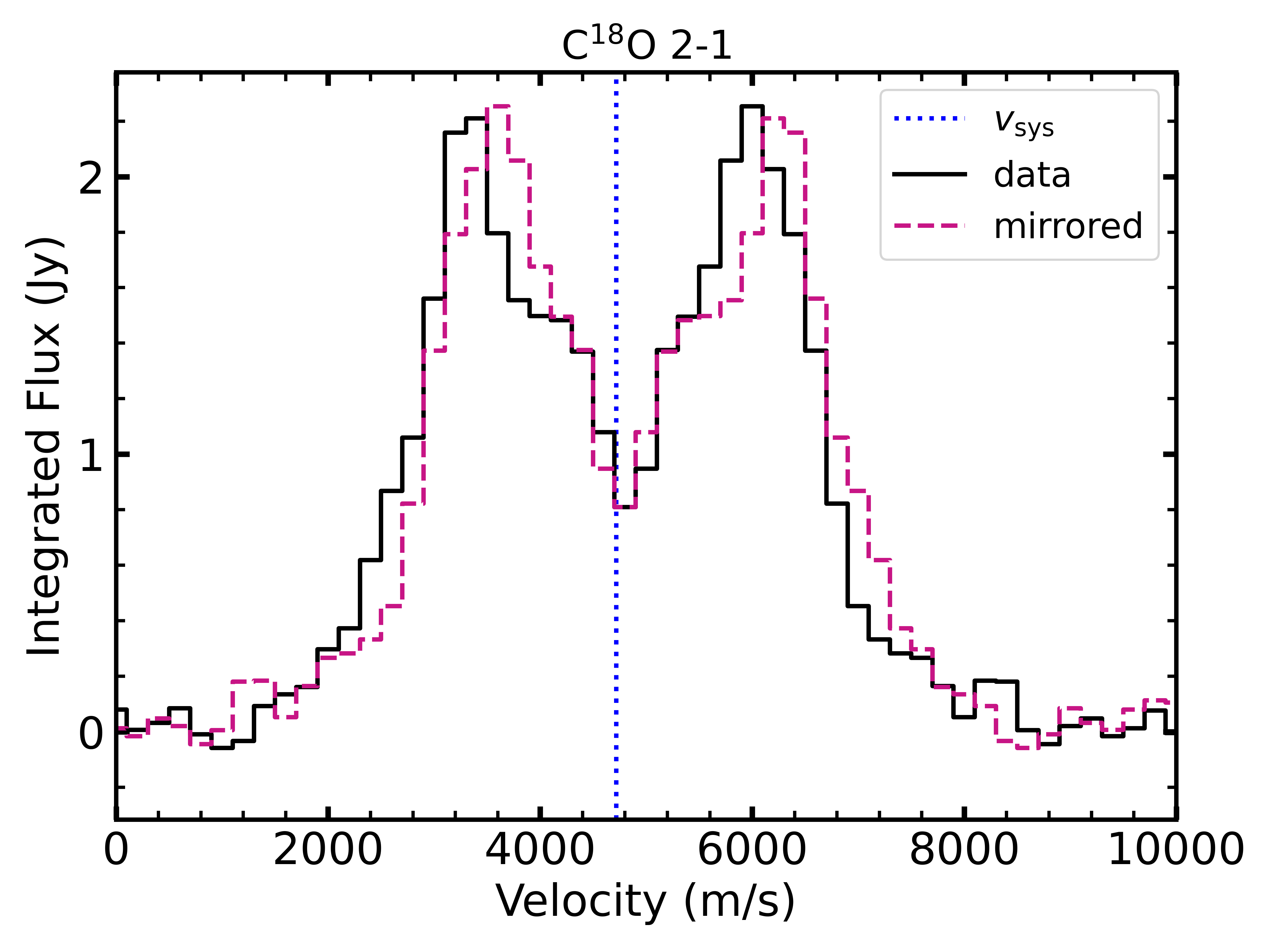}
    \end{minipage}
\end{minipage}
\hfill
\begin{minipage}[t]{0.48\textwidth}
    \centering
    {\Large\textbf{WaOph 6}\par\medskip}

    \begin{minipage}[c]{0.48\linewidth}
        \includegraphics[width=\linewidth]{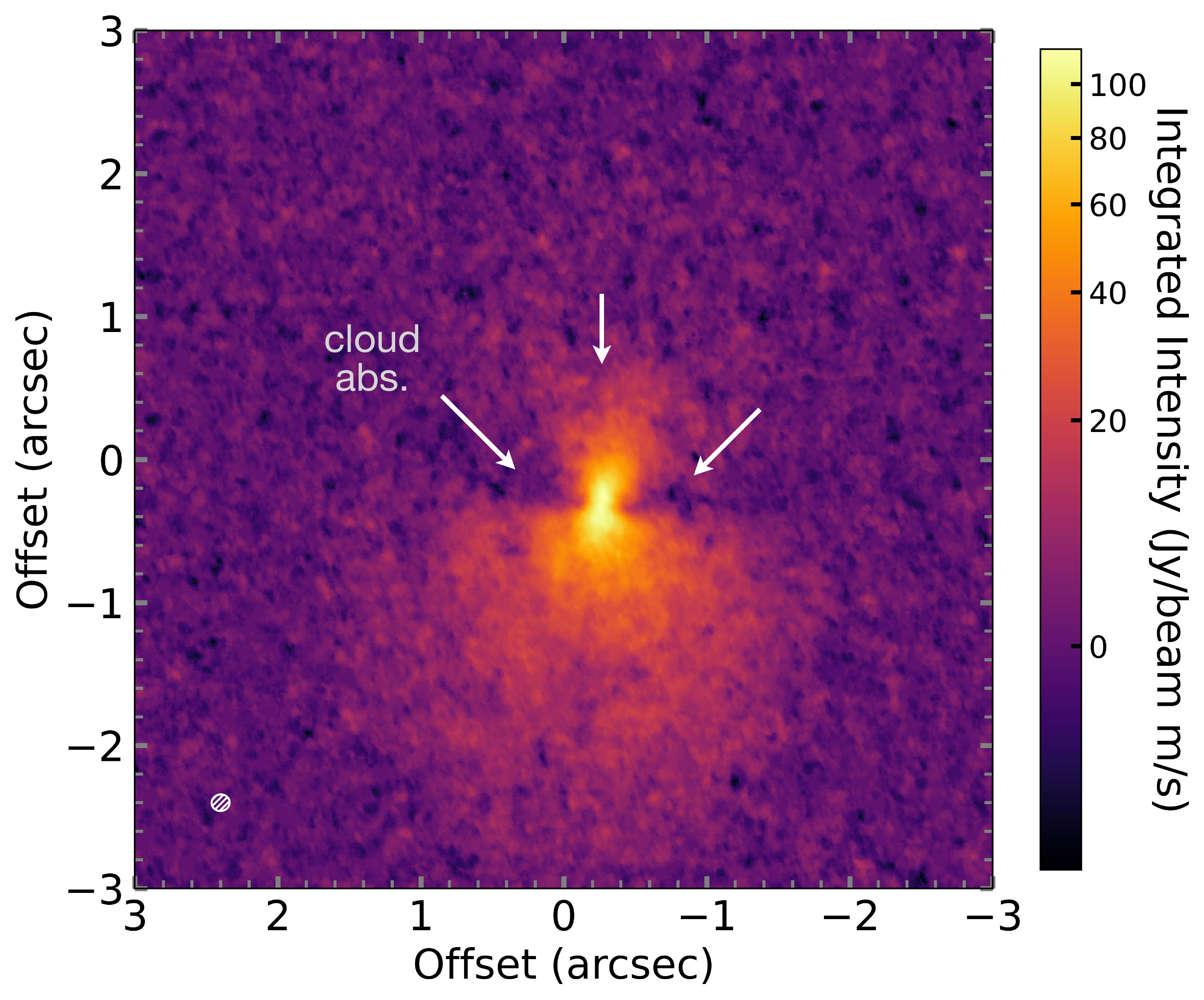}
    \end{minipage}\hfill
    \begin{minipage}[c]{0.48\linewidth}
        \includegraphics[width=\linewidth]{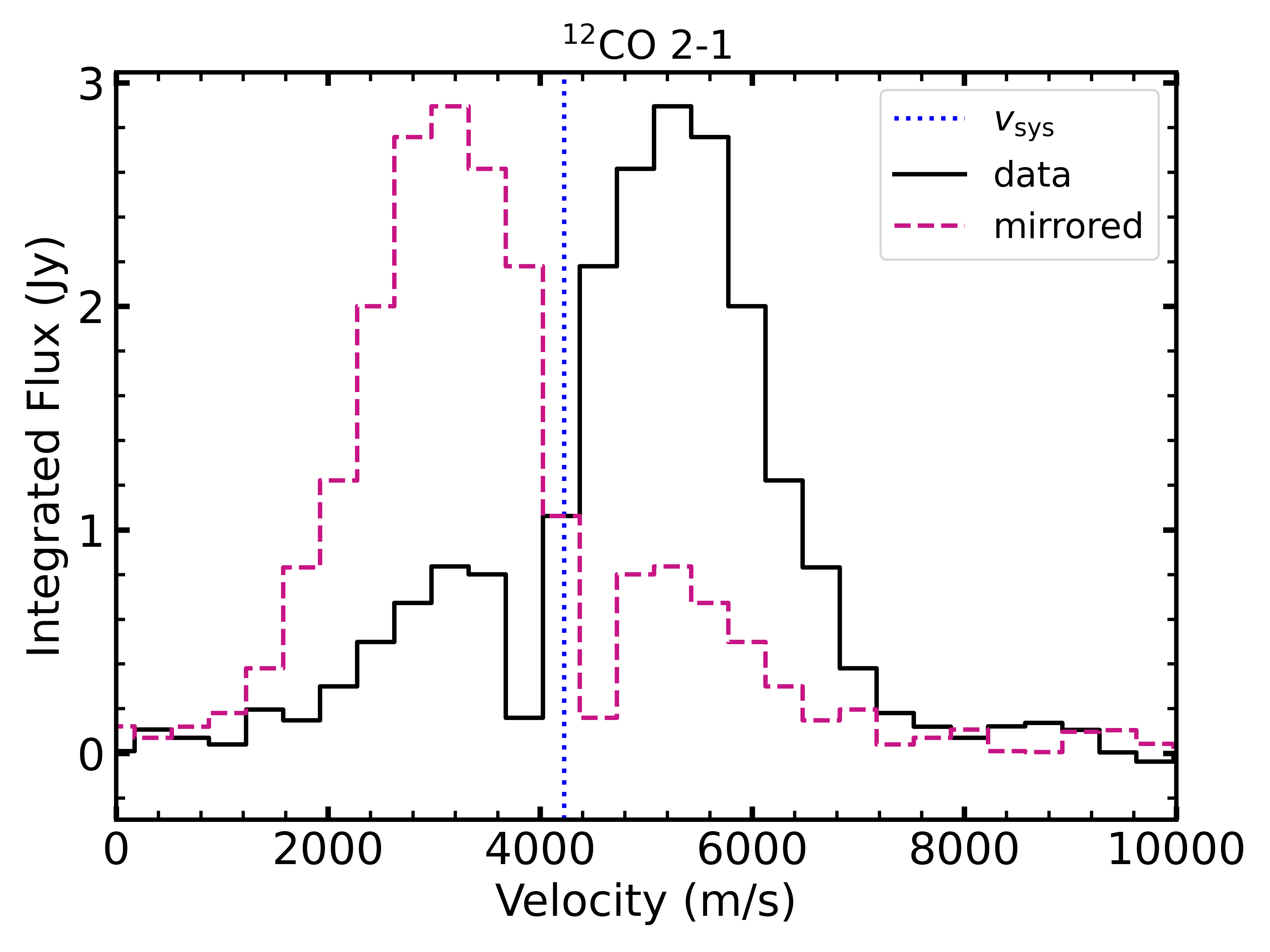}
    \end{minipage}

    \begin{minipage}[c]{0.48\linewidth}
        \includegraphics[width=\linewidth]{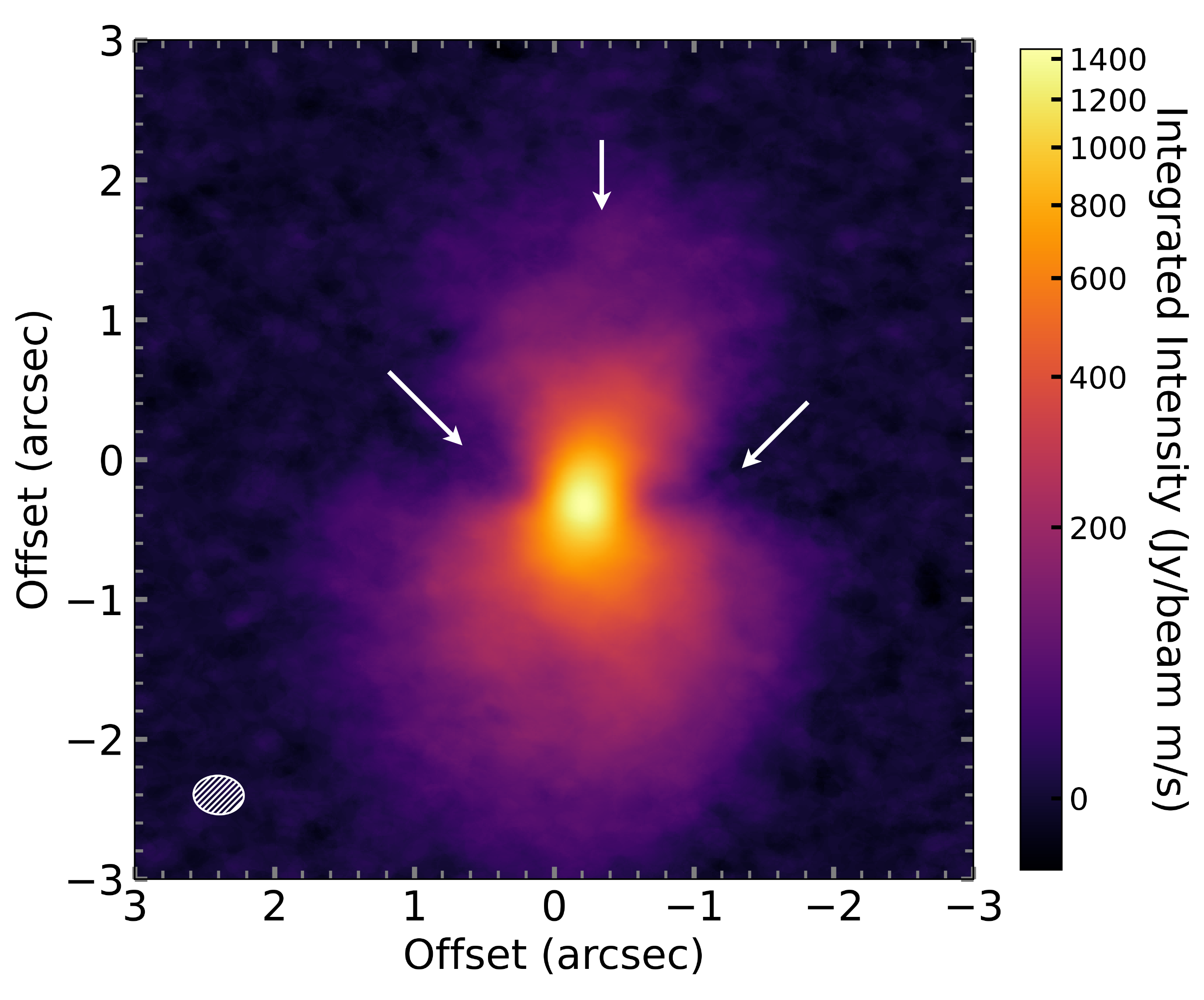}
    \end{minipage}\hfill
    \begin{minipage}[c]{0.48\linewidth}
        \includegraphics[width=\linewidth]{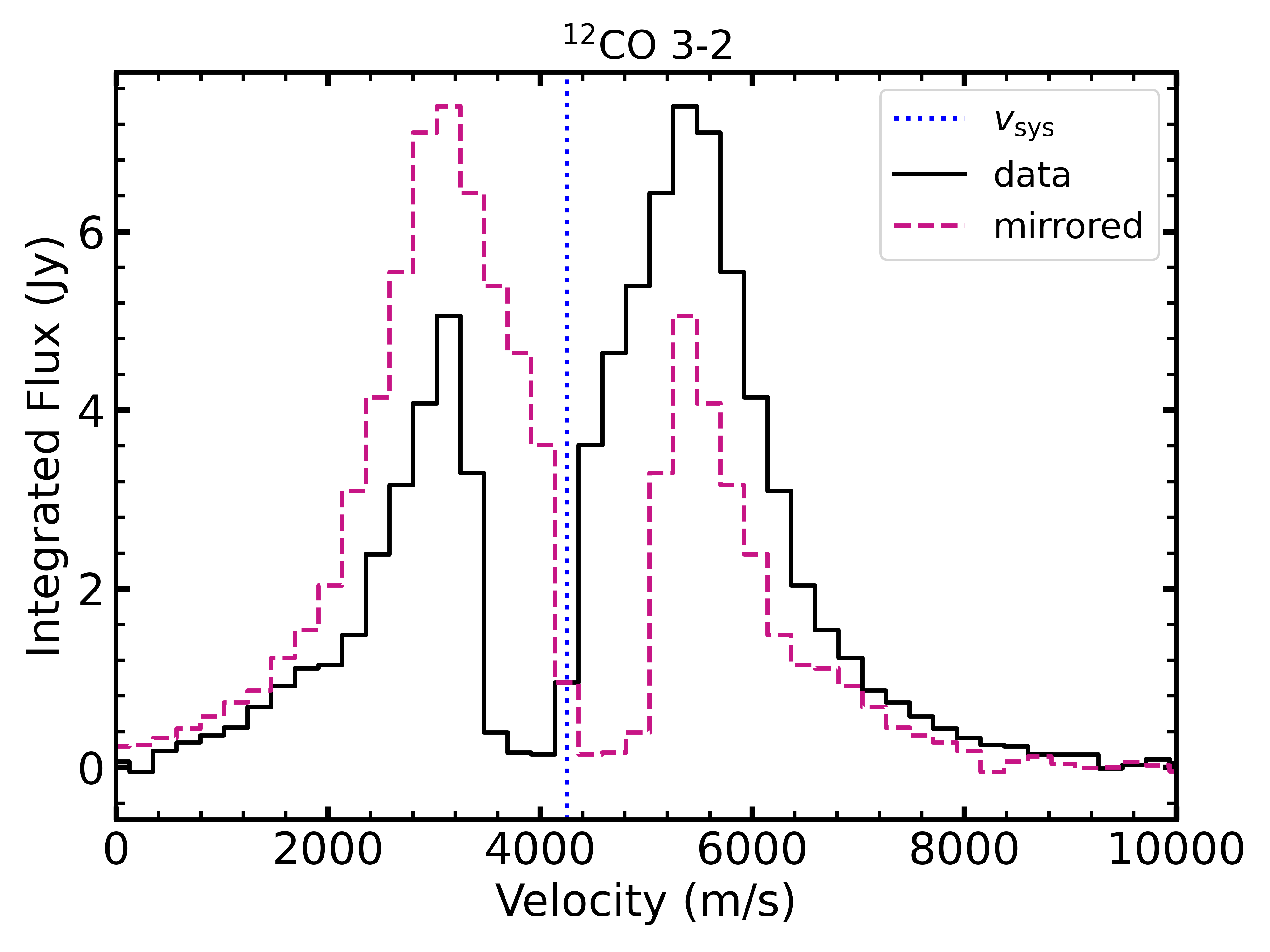}
    \end{minipage}

    \begin{minipage}[c]{0.48\linewidth}
        \includegraphics[width=\linewidth]{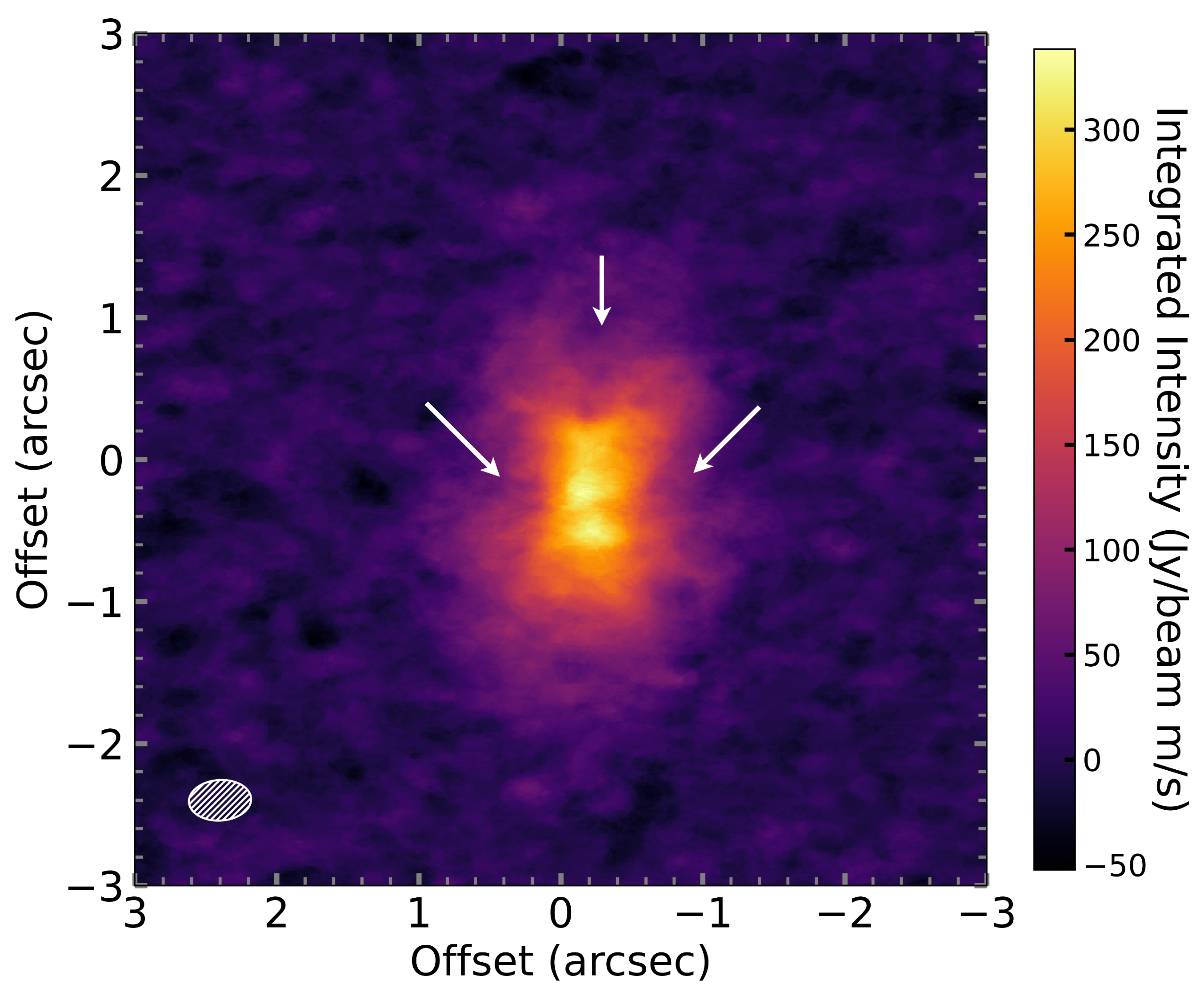}
    \end{minipage}\hfill
    \begin{minipage}[c]{0.48\linewidth}
        \includegraphics[width=\linewidth]{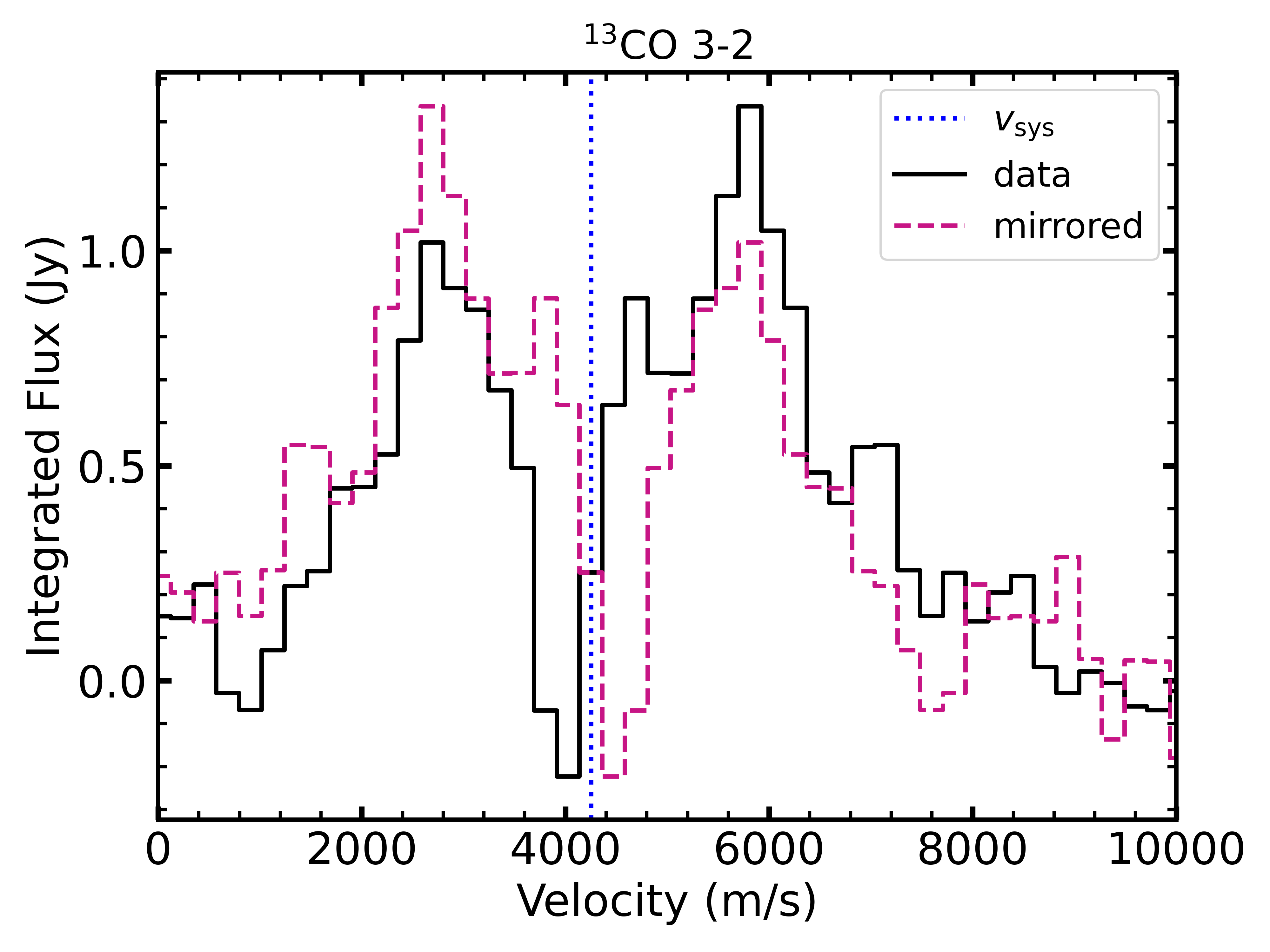}
    \end{minipage}
\end{minipage}

\caption{Integrated flux maps (left panels) and integrated spectra from 0.5 to 5 arcseconds (right panels) of the analyzed tracers for HD 97048 (left Col.) and WaOph 6 (right Col.). In the maps we apply an \textit{asinh} stretch for the \ce{^12CO} lines and a \textit{linear} stretch for \ce{^13CO} and \ce{C^18O}.
The synthesized beam is shown in the lower-left corner of each panel. We mark with arrows the regions where absorption is present. In the spectra, both the original and mirrored versions are shown to highlight absorption features.
}

\label{fig:abs_combined}
\end{figure*}

Both sources are embedded and thus affected by contamination. The CO cubes show significant absorption, especially in the channels close to the systemic velocity of the star but also in the blueshifted side of the disks. In this section, we describe in detail how we tackled the complications linked to the presence of absorption.

First, for all available tracers we obtained the integrated intensity maps and spectra, shown in Fig. \ref{fig:abs_combined}, highlighting the velocities at which total or partial absorption is present. Through an initial visual inspection of the data cubes and the inspection of the integrated spectra, we removed the fully absorbed channels close to the systemic velocity from the data cubes used to fit the channel maps with \texttt{discminer}. We performed this initial flagging on the data for all lines showing absorption at the systemic velocity: the \ce{^12CO} 2 -- 1 line for HD 97048, and all available tracers for WaOph 6. 

In addition to the strong features in the central channels, the \ce{^12CO} 2 -- 1 line for HD 97048 and all lines for WaOph 6 also show mild absorption in the blueshifted channels. From both integrated intensity maps and integrated spectra, weaker emission was observed in the blueshifted side of the disk with respect to the redshifted one. This discrepancy in the flux is clearly visible in the integrated spectra, mirrored with respect to the systemic velocity. 
For the \ce{^12CO} 2 -- 1 line of HD 97048, we excluded from the modeling procedure both the central totally absorbed channels and the mildly absorbed blueshifted channels, to avoid possible biases due to the fainter emission on one side of the disk. For this line, the high signal-to-noise ratio and the good spectral resolution of the archival data ensured that the number and quality of the remaining channels were sufficient to provide an efficient fit of the observed channel maps and an accurate disk model. Hence, for this line, our \texttt{discminer} best-fit model was extracted using the redshifted channels only. On the other hand, the signal-to-noise ratio for the WaOph 6 data is much worse than for HD 97048: for this reason, we considered all the available channels for the \texttt{discminer} model, excluding only the totally absorbed ones close to the systemic velocity. In Table \ref{tab:abs_channels} we list the excluded velocity ranges and corresponding portion of channels for all the lines.

\begin{table}[h!]\label{tab:abs_channels}
   \centering
    \renewcommand\arraystretch{1.2}
    \caption{Excluded velocity ranges in the \texttt{discminer} models.}
    \label{tab:abs_channels}  
    \begin{threeparttable}
    \begin{adjustbox}{max width=\columnwidth}
    \begin{tabular}{c|c|c|c}
    \hline
    \textbf{Source} & \textbf{Transition} & \textbf{Excluded $\upsilon$ range [km/s]} & \textbf{Excluded channels} \\ 
    \hline
       HD 97048 & \ce{^12CO} 2 -- 1  & < 5.4 & blueshifted and central \\
                & \ce{^13CO} 3 -- 2  & \ldots & none \\
                & \ce{C^18O} 2 -- 1  & \ldots & none \\
        WaOph 6 & \ce{^12CO} 2 -- 1  & 3.5 - 4.2 & central \\    
                & \ce{^12CO} 3 -- 2  & 3.582 - 4.251 & central \\ 
                & \ce{^13CO} 3 -- 2  & 3.582 - 4.251 & central \\ 
    \hline
    \end{tabular}
    \end{adjustbox}
    \tablefoot{For each tracer, we list the excluded velocity range and the corresponding portion of channels excluded from the model.}
   
    \end{threeparttable}
\end{table}

We highlight that the disk geometry (i.e., the orientation angles) and the systemic velocity of the disk should be fixed to the same values among the various tracers available for each source, to be fully consistent within our models.
As a second measure to further ease the modeling of the data cubes with partially absorbed emission \edit{from the foreground cloud}, we fixed the inclination $i$, position angle $\mathrm{PA}$, and systemic velocity $v_\mathrm{sys}$ in the fitting procedure to the best-fit values of the \ce{^13CO} 3 -- 2 \texttt{discminer} model. We selected this line as reference model because \ce{^13CO} is the least absorbed tracer for both sources, as clearly visible in Fig. \ref{fig:abs_combined}.

Additional measures were implemented within the procedure for the extraction of the molecular rotation curves, to minimize the impact of the presence of absorption in the archival data.
As many of the tracers show line absorption close to $v_\mathrm{sys}$, to obtain the rotation curves we computed the radial profiles of the averaged deprojected velocity within an azimuthal section of [-30\textdegree, 30\textdegree] along the disk major axis: by masking the azimuthal wedge around the disk minor axis, we ensured that no contribution from any absorption feature would bias the rotation curve retrieval. 

In addition, for the lines with clear absorption in the blueshifted side of the disk (\ce{^12CO} for HD 97048 and all tracers for WaOph 6) we extracted the rotation curves by computing the azimuthal velocities from the redshifted channels only, masking the partially absorbed side of the emission. The motivation for the selected approach is evident in the maps of the velocity residuals, presented in Figs. \ref{fig:map_vel_res_hd} and \ref{fig:map_vel_res_wa}), and are discussed in detail in Appendix \ref{app:vel_res_maps}.

\subsubsection{Beam smearing correction}\label{subsec:beam_smearing}
When extracting rotation curves, beam smearing \rr{(also referred to as "spatial resolution bias")} plays an important role in determining the gas rotation velocity, primarily affecting the innermost regions of the disk, where the rotation velocity varies most rapidly with radius.
Within a single resolution element, \rr{this steep variation causes slower-moving gas at larger radii -- arising from a larger emitting area -- to contribute more strongly to the intensity-weighted velocity field (i.e., modulated by underlying brightness gradients) than gas at smaller radii.} This leads to a systematic underestimate of the true rotation velocity and an apparent flattening of the rotation curve in the disk inner region, as observed in Figs. \ref{fig:rot_curves_dm_hd} and \ref{fig:rot_curves_dm_wa} \citep[see also][]{keppler2019,boehler2021,andrews2024,pezzotta2025}. 
It is not straightforward to determine the impact of this observational bias and to derive an analytical correction factor, as it may depend on the angular resolution, disk geometry, and gas velocity distribution itself. However, as this effect becomes significantly relevant at small radii, we adopted an empirical debiasing technique to correct our rotation curves, similar to the approach described in \citet{andrews2024}. 
\texttt{Discminer} fits the data with a beam-convolved model ($v_\mathrm{Kepl,model}(R)$), which is affected by the same beam smearing effects as the data, since the convolution with the beam provides information on both intensity and velocity gradients. We compare the rotation curve extracted by the beam-convolved model to the pure, analytical Keplerian curve ($v_\mathrm{Kepl,pure}(R)$) based on the \texttt{discminer} best-fit stellar mass $M_{\star,\mathrm{dm}}$ and emitting layers. In this way, we infer the correction factor $\xi$ to be applied to the original data ($v_\mathrm{data}(R)$) in order to obtain the unbiased velocities ($v_\mathrm{corr}(R)$), which are not affected by beam smearing. In particular, the correction factor $\xi$ was computed, at each radius, as the ratio between the pure analytical Keplerian rotation velocity $v_\mathrm{Kepl,pure}(R)$ and the biased beam-convolved one $v_\mathrm{Kepl,model}(R)$:
\begin{equation}
    v_\mathrm{corr}(R)=v_\mathrm{data}(R) \ \xi(R) = v_\mathrm{data}(R) \frac{v_\mathrm{Kepl,pure}(R)}{v_\mathrm{Kepl,model}(R)} .
\end{equation}

\begin{figure*}
   \centering
   \includegraphics[width=\columnwidth]{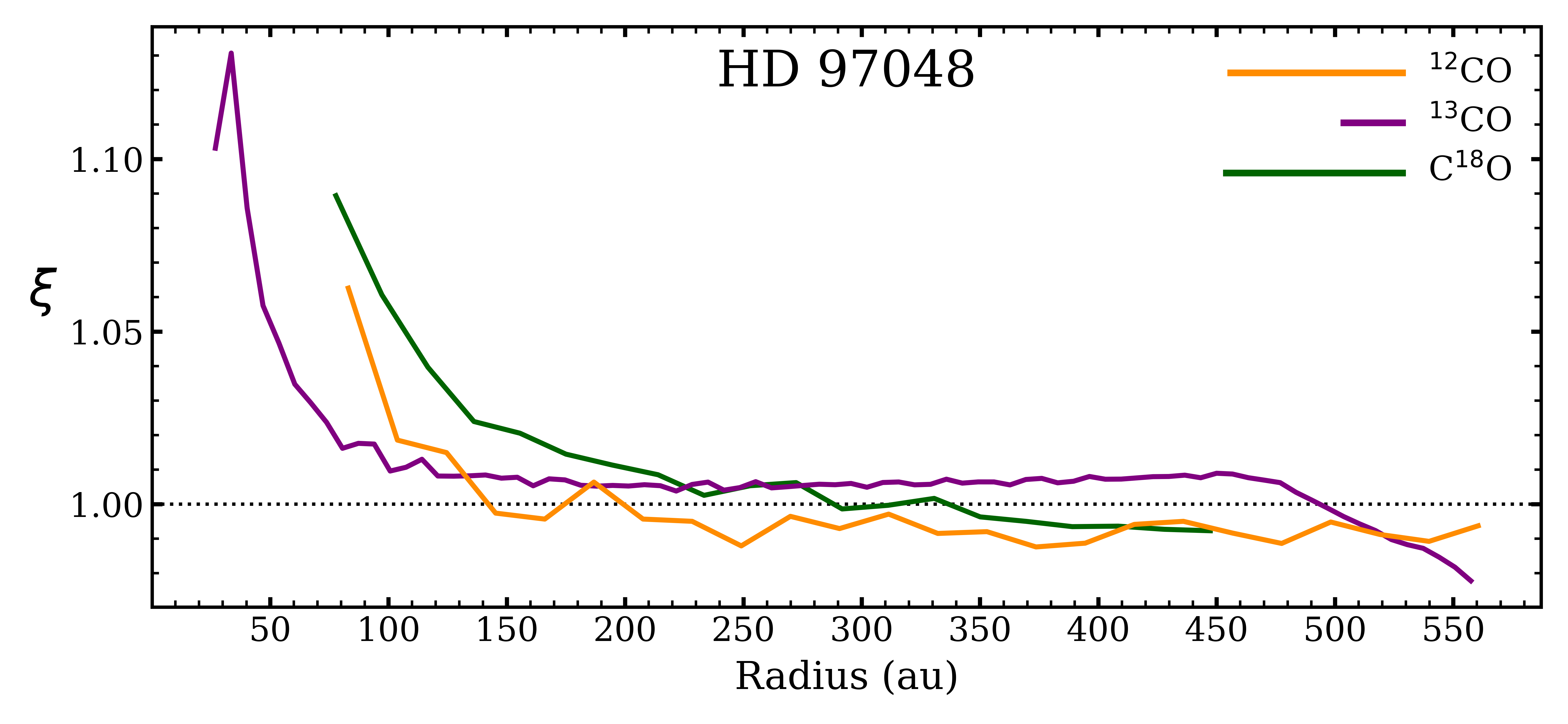}
   \includegraphics[width=\columnwidth]{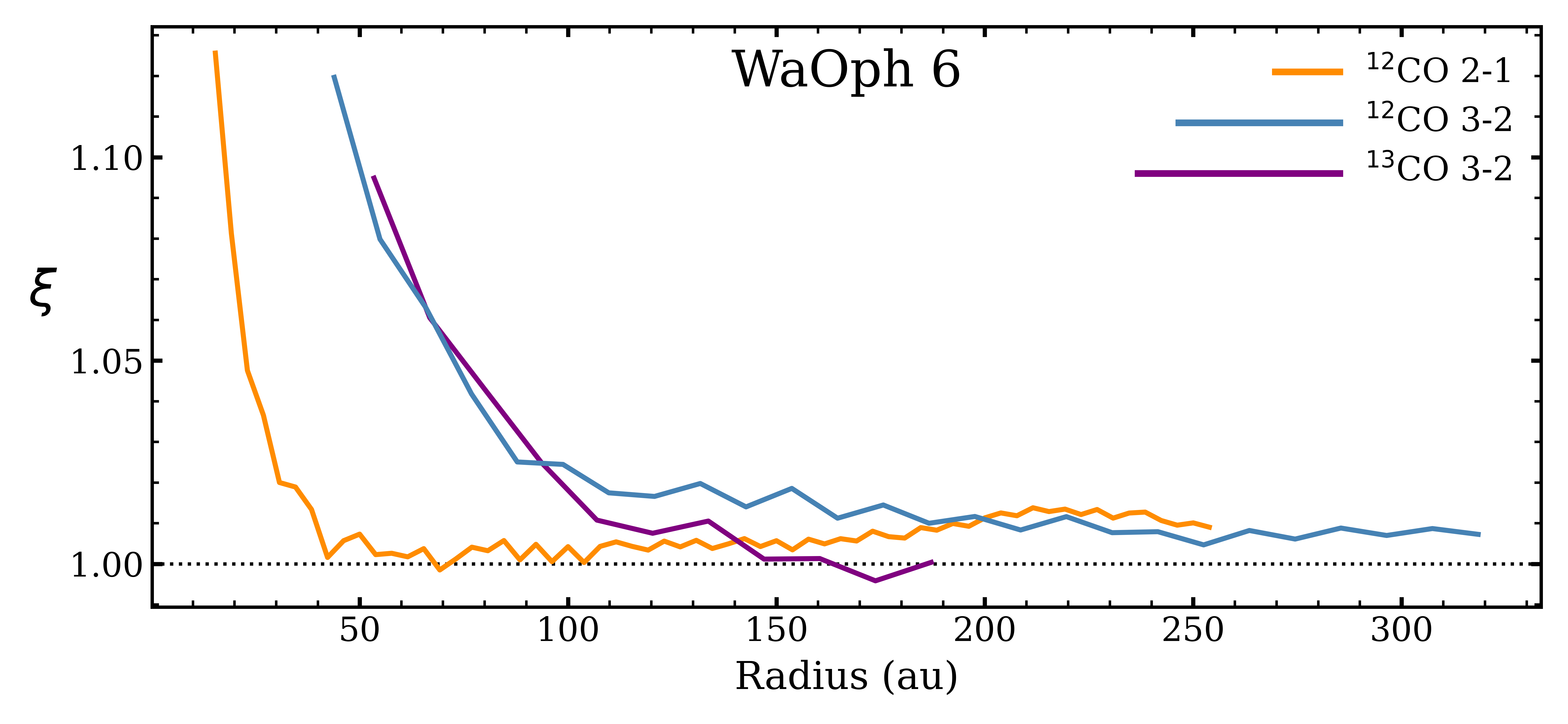}
      \caption{
      Radial profiles of the correction factor $\xi(R)$ for all tracers for HD 97048 (left) and WaOph 6 (right). For each tracer, the length of the bar in the legend represents the beam size.}
         \label{fig:corr_factors}
\end{figure*}

\begin{figure}
   \centering
   \includegraphics[width=\columnwidth]{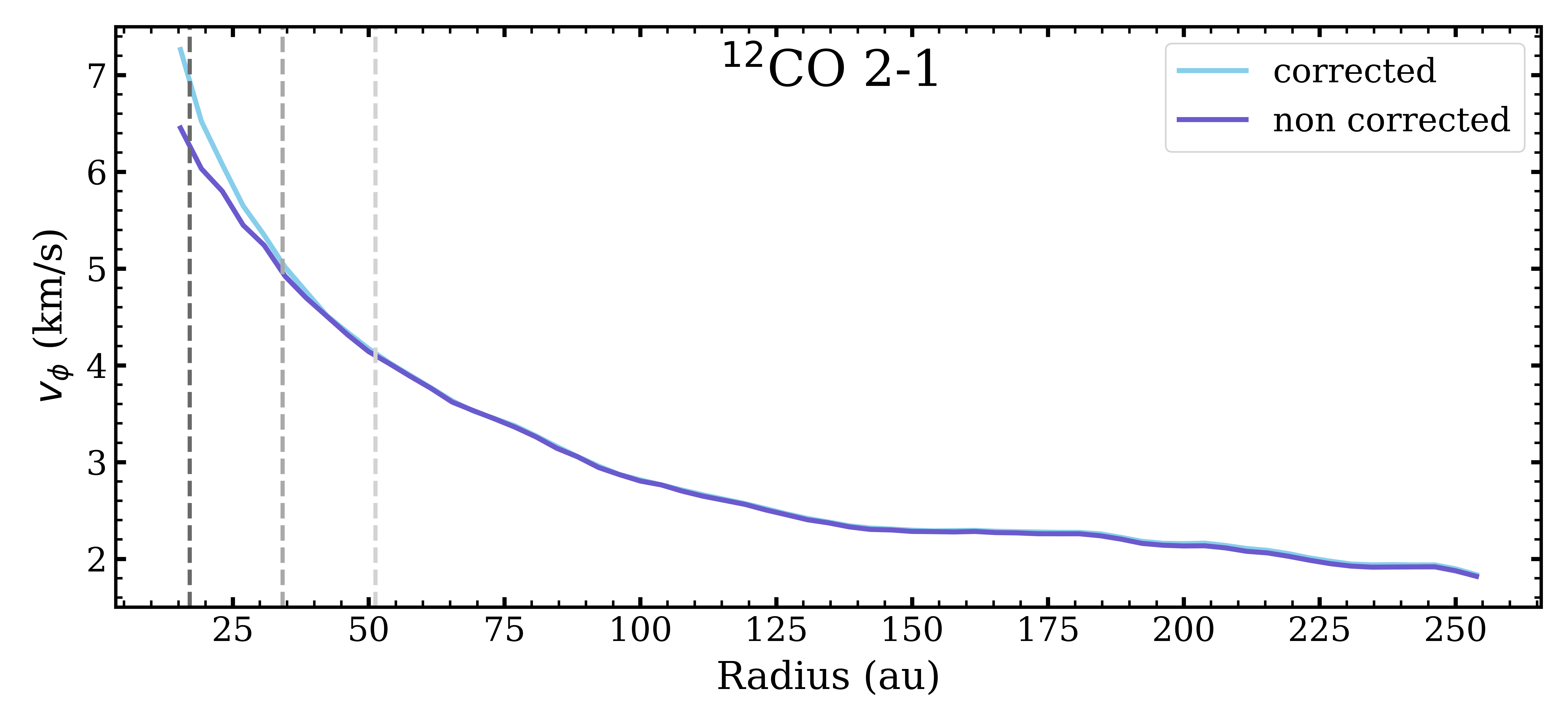}
      \caption{
      Rotation curve of the \ce{^12CO} 2 -- 1 line for WaOph 6 before (violet) and after (turquoise) beam smearing correction. Gray dashed lines represent one, two, and three beams from the central star.}
         \label{fig:12co21_corr_vs_noncorr}
\end{figure}

We show the radial profiles of the correction factors for both sources and all tracers in Fig. \ref{fig:corr_factors}. As an example, in Fig. \ref{fig:12co21_corr_vs_noncorr} we compare the corrected and non-corrected rotation curves for the \ce{^12CO} 2 -- 1 line of WaOph 6. We show the same plot for all the other tracers in Appendix \ref{app:corr_vs_noncorr}.

With the intention to show how important the beam smearing correction can be when fitting rotation curves, we repeated the whole analysis, which we describe in detail in Sects. \ref{subsec:pipeline} and \ref{subsec:bootstrapping}, using the non-corrected rotation curves. 
Without the correction, we underestimate the stellar mass by $\approx5.7\%$ for HD 97048 and $\approx8.3\%$ for WaOph 6; we overestimate the corresponding disk mass by $\approx46.3\%$ for HD 97048, and we underestimate it by $\approx38.2\%$ for WaOph 6. 
As expected, we get significant offsets in the derived estimates: the inner regions, where the Keplerian contribution is dominant, are fundamental to well constrain the stellar mass, and thereby the disk mass.
The obtained discrepancies between the corrected and non-corrected mass estimates are of the same order of magnitude as the typical systematic uncertainties for dynamical measurements \citep{veronesi2024,andrews2024}. Hence, including the correction to the beam smearing proves to be critical to obtain precise and reliable measurements through the dynamical method.

\subsection{Disk thermal structure and rotation curves fitting}\label{subsec:pipeline}

Assuming the systemic velocity, inclination and position angle of the disks as extracted from the \texttt{discminer} best-fit model of the \ce{^13CO} reference line, 
for each line we extract the molecular emitting layer with \texttt{disksurf}. For each data cube, we only select the frequency channels where the line emission is not contaminated by any gas absorption, and we discard the central channels centered around the systemic velocity, as they do not provide constraints on the emitting surfaces within the \texttt{disksurf} framework. We select the minimum signal-to-noise ratio requested for the surface points according to the data quality of the different cubes, ranging from 5 (for the less noisy cube) to 3.5 (for the most noisy cube). With the available data quality, for each tracer we only retrieve the front surface; if necessary, we manually mask the obtained surfaces to exclude visually identified noise.

To extract the emitting surfaces and consequently the gas temperature from line emission, we used the non-continuum-subtracted data cubes. Continuum subtraction can cause an underestimation of the gas temperature if the considered line is optically thick and absorbs the continuum background emission \citep{boehler2017,weaver2018}. 
We converted the measured line intensity $I_\nu$ along the emitting surfaces into a brightness temperature $T_\mathrm{B}$ according to the Planck law;
assuming that the considered CO lines are optically thick and in local thermodynamic equilibrium, the observed brightness temperature is then representative of the actual kinetic temperature of the gas. 
As we measured the brightness temperature tracing lines that probe different heights in the disk, for each source we obtained a collection of $(R,z)$ points where the disk temperature is known. In Fig. \ref{fig:layers} we show the retrieved emitting layers for both sources, and we show the corresponding brightness temperature with the color scale. The available data quality did not allow for the extraction of the WaOph 6 \ce{^13CO} surface, which was thus not considered for the retrieval of the 2D thermal structure.

At this point, we assumed a 2D parameterization\footnote{We also tested the hyperbolic tangent \citep{dullemond2020} and a simple power law as parametric expressions, but as in \citet{exoalma_maria} the sinusoidal function in Eq. \ref{eq:sinusoidal_T} provided the best fits to our data.} to describe the disk thermal structure. We adopted the two-layer model introduced by \citet{dartois}, where the atmospheric temperature $T_\mathrm{atm}$ is connected to the midplane temperature $T_\mathrm{mid}$ with a sinusoidal function 
\begin{equation}\label{eq:sinusoidal_T}
    T(z) =
        \begin{cases}
        T_{\text{atm}} + (T_{\text{mid}} - T_{\text{atm}}) \cos^{2}\!\left( \frac{\pi}{2} \frac{z}{z_q} \right) & \text{if } z < z_q, \\
        T_{\text{atm}} & \text{if } z \geq z_q.
    \end{cases}
\end{equation}

In Eq. \ref{eq:sinusoidal_T}, $T_\mathrm{atm}$, $T_\mathrm{mid}$, and $z_q$ are described by power laws with respective indexes $q_\text{atm}$, $q_\text{mid}$, and $\beta$, and with respective normalization factors $T_\text{atm,0}$, $T_\text{mid,0}$ and $z_0$ at the reference radius $R_0=100$ au.

For each source we fitted the non-parametric temperature measured in all the available $(R,z)$ points with Eq. \ref{eq:sinusoidal_T}, determining the best-fit estimates for the six free thermal parameters: $T_\text{atm,0}$, $T_\text{mid,0}$, $z_0$, $q_\text{atm}$, $q_\text{mid}$, and $\beta$. For this fit, we only consider the data points with brightness temperatures above the CO freeze-out temperature (20 K): points with $T_\mathrm{B}<20$ K could be tracing optically thin emission, and thus would not probe the real gas temperature.
To ease the convergency of the fit, we set some broad \rr{boundaries} on the allowed values for the different parameters, based on previous literature (see \citealt{exoalma_maria} and references therein), exploring this portion of the parameter space: $T_\text{mid,0}<T_\text{atm,0}<150$ K, $T_\mathrm{CMB}$ $<T_\text{mid,0}<60$ K, $-2<q_\text{atm}<1$, $-2<q_\text{mid}<0$, $0<\beta<3$, $0<z_0<r_\text{max}$, where $r_\text{max}$ is the largest radial point.

To fit the rotation curves, we followed the same procedure adopted in \citet{pezzotta2025}, which we briefly summarize here. For each source, we simultaneously fitted the beam smearing corrected rotation curves of the analyzed tracers with a thermally stratified model (see \citealt{martire} for details), 
which includes contributions to the gas rotation velocity from stellar gravity, pressure gradients, thermal stratification, and disk self-gravity. The various terms are evaluated along the molecular emitting surface extracted by \texttt{discminer}.
\rr{The adopted model assumes a self-similar prescription for the disk surface density \citep{lyndenbell}, and describes the rotating gas as a barotropic fluid in vertical hydrostatic equilibrium.}
To perform the simultaneous fit of the rotation curves, we extended the code \texttt{DySc} to fit more than two lines together; all fits were performed using ten walkers, 1000 steps of burn-in and 5000 further steps. From the rotation curves fit, we obtain accurate dynamical estimates of the three fundamental parameters the model depends on: the stellar mass $M_\star$, the disk mass $M_\text{d}$, and the disk scale radius $ R_\text{c}$.

\subsection{Bootstrapping over disk geometry and thermal structure: A more comprehensive view of the systematic uncertainties}\label{subsec:bootstrapping}

The fits to the rotation curves presented here involve using a set of geometrical parameters determining the disk geometry and systemic velocity, and a set of thermal parameters describing the 2D thermal structure.
These two sets of variables both represent an important contribution to the systematic uncertainties on the final best-fit values for the stellar mass, disk mass, and disk scale radius, as discussed in \citet{andrews2024}. \edit{The systematic errors intrinsic to the described dynamical method are typically comparable to (or even larger than) the statistical ones, making it essential to account for how uncertainties on the disk geometry and temperature affect the inferred dynamical estimates.}
To do so, we performed the whole procedure described in Sect. \ref{subsec:pipeline} 100 times, first bootstrapping over the geometrical parameters we obtain from the \texttt{discminer} model ($\{i$, $\mathrm{PA}$, $x_\mathrm{c}$, $x_\mathrm{c}$, $v_\text{sys}\}_\text{best}$), and then over the thermal parameters. We start extracting a random value for each geometrical parameter, sampling a normal distribution centered on the best-fit value obtained by the \texttt{discminer} model, and with a fixed standard deviation. The width of the Gaussian is set to $1/5$ of the beam major axis for the central offsets $x_\mathrm{c}$ and $y_\mathrm{c}$, $1$\textdegree $\ $ for the angles ($i$, PA) and $3$ m/s for $v_\text{sys}$; these values allowed us to adequately explore the parameter space and take into account the systematic uncertainty on the extracted disk geometry. 
For each set of new randomly extracted geometrical parameters $\{i$, $\mathrm{PA}$, $x_\mathrm{c}$, $y_\mathrm{c}$, $v_\text{sys}\}_\text{random}$, we repeat the extraction of the emitting surfaces and the brightness temperature along them. At this point, for each set of emitting layers we fitted the 2D thermal structure of the disk, obtaining a total of 100 possible thermal structures. Each one of these thermal structures is described by six best-fit thermal parameters ($\{T_\text{atm,0}$, $T_\text{mid,0}$, $q_\text{atm}$, $q_\text{mid}$, $z_0$, $\beta\}_\text{best}$), with their corresponding posterior distributions. At this point, instead of adopting the best-fit values (i.e., the $50^\text{th}$ percentiles of the distributions), to describe the disk thermal structure, we bootstrap over the thermal parameters. For 100 times, we extract a random value for each parameter, sampling a Gaussian distribution which is centered on the best-fit value we found at the previous step ($\{T_\text{atm,0}$, $T_\text{mid,0}$, $q_\text{atm}$, $q_\text{mid}$, $z_0$, $\beta\}_\text{best}$), and with a standard deviation given by the $16^\text{th}$ and $84^\text{th}$ percentiles of the original distribution of the considered parameter. 
At this point, for each of the 100 thermal structures, we performed the rotation curves fit with \texttt{emcee}. We concatenated the Markov chains obtained from the 100 iterations for the three parameters of interest: $M_\star$, $M_\text{d}$, and $ R_\text{c}$. The final best-fit dynamical estimates of these quantities are given by the $50^\text{th}$ percentiles of the overall distributions, and the corresponding errors are described by the $16^\text{th}$ and $84^\text{th}$ percentile range.

The parameters describing the final thermal structures were extracted by concatenating the Markov chains obtained from the 100 iterations, for the thermal parameters randomly extracted.
We adopt the $50^\text{th}$ percentiles of the overall distributions as the final best-fit estimates of these parameters, while uncertainties are given by the $16^\text{th}$ and $84^\text{th}$ percentiles, \edit{corresponding to $1\sigma$ credible intervals ($\approx68\%$)}.
In Table \ref{table:final_T_estimates} we report for both sources the best-fit estimates for the thermal parameters, with their corresponding errors.
In Fig. \ref{fig:2D_T_fits} in Appendix \ref{app:2D_T} we show the 2D thermal structures of the two disks, after the bootstrapping procedure. \edit{The disk around HD 97048 (a massive, hot Herbig star) is particularly warm, especially in the inner regions, while WaOph 6 (colder T Tauri star with lower mass) hosts a cooler disk.}

\begin{table}[h!] 
    \centering
    \renewcommand\arraystretch{1.2} 
    \caption{\centering Best-fit estimates for the thermal parameters \edit{with their corresponding $1\sigma$ uncertainties.}}
    \label{table:final_T_estimates}
    \begin{adjustbox}{max width=\columnwidth} 
    \begin{tabular}{c|c|c|c|c|c|c}
    \hline
     \textbf{Source}      & \textbf{$T_\text{atm,0}$} [K] & \textbf{$T_\text{mid,0}$} [K] & \textbf{$q_\text{atm}$} & \textbf{$q_\text{mid}$} & \textbf{$z_0$} [au]       & $\beta$ \\
     \hline
        \textbf{HD 97048} & $77.37^{+15.84}_{-5.48}$      & $59.96^{+0.03}_{-0.57}$       & $-0.21^{+0.14}_{-0.30}$   & $-0.32^{+0.05}_{-0.02}$   & $0.03^{+0.05}_{-0.01}$    & $2.99^{+0.01}_{-2.50}$ \\ 
    \hline
        \textbf{WaOph 6}  & $36.26^{+4.23}_{-0.77}$       & $35.19^{+0.80}_{-2.71}$       & $0.08^{+0.14}_{-0.40}$    & $-0.32^{+0.14}_{-0.11}$   & $0.49^{+0.22}_{-0.25}$    & $1.51^{+1.06}_{-1.27}$ \\ 
    \hline
        
    \end{tabular}
    \end{adjustbox}
    
\end{table}

\begin{figure*}
   \centering
   \includegraphics[width=\textwidth]{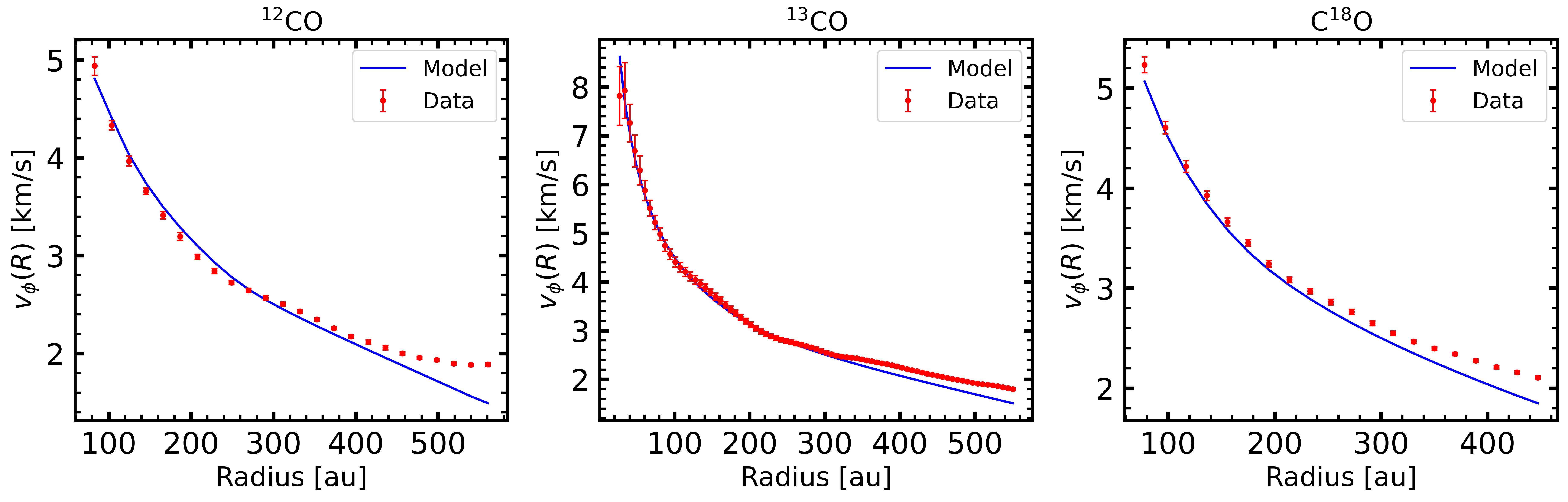} 
   \includegraphics[width=\textwidth]{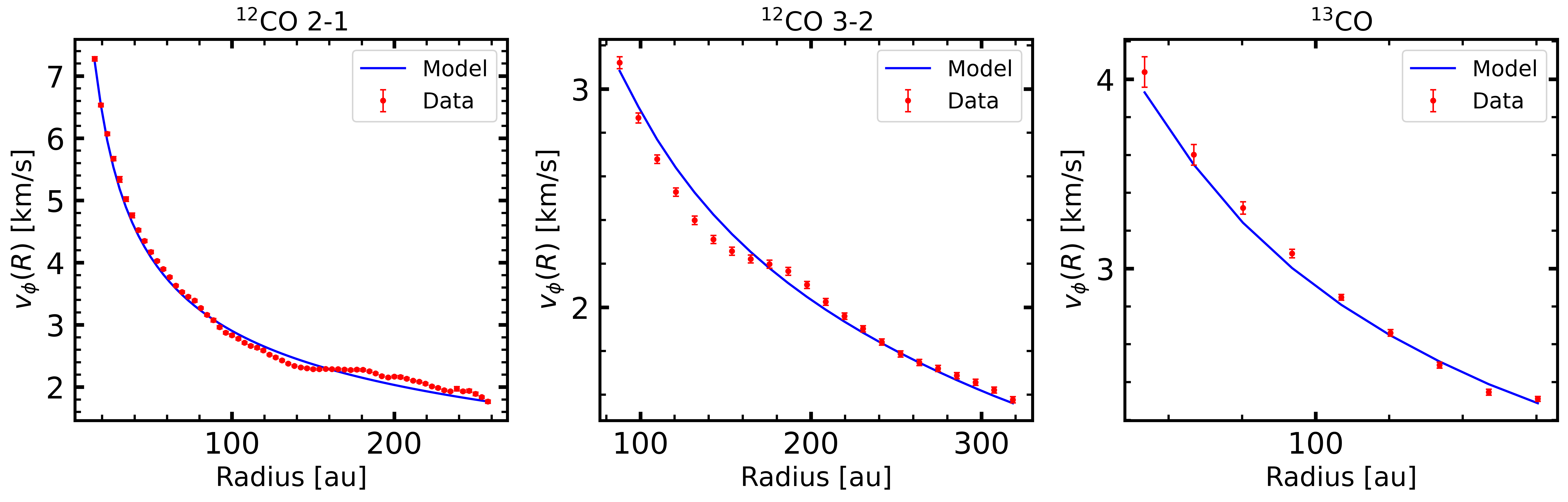} 
      \caption{
      Best-fit of the beam smearing corrected rotation curves of all the considered lines for HD 97048 (\rr{top row}) and WaOph 6 (\rr{bottom row}). The red dotted line shows the data, while the blue solid line represents the best-fit model.}
         \label{fig:final_curves_fit}
\end{figure*}

In Fig. \ref{fig:final_curves_fit} we show the corrected rotation curves best-fit for all the considered tracers, while in Table \ref{table:final_estimates} we report for both sources the best-fit dynamical estimates for the stellar mass, the disk mass, and the disk scale radius, with their \rr{uncertainties corresponding to the $16^\text{th}$ and $84^\text{th}$ percentiles of the distributions}. In Appendix \ref{app:cornerplots}, we show in Fig. \ref{fig:total_bootstrapping} the final corner plots of the rotation curves fits of both sources, after the bootstrapping procedure.

\begin{table}[h!] 
    \centering
    \renewcommand\arraystretch{1.2} 
    \caption{\centering Best-fit dynamical estimates for the stellar mass, the disk mass, and the disk scale radius\edit{with their corresponding $1\sigma$ uncertainties.}}
    \label{table:final_estimates}
    \begin{adjustbox}{max width=\columnwidth}
    \begin{tabular}{c|c|c|c}
    \hline
        \textbf{Source}   & \textbf{$M_\filledstar$} [$M_\odot$] & \textbf{$M_\mathrm{d}$} [$M_\odot$] & \textbf{$R_\mathrm{c}$} [au] \\
    \hline
        \textbf{HD 97048} & $2.226^{+0.054}_{-0.049}$            & $0.300^{+0.055}_{-0.061}$           & $172^{+24}_{-14}$ \\ 
    \hline
        \textbf{WaOph 6}  & $0.956^{+0.006}_{-0.006}$            & $0.210^{+0.045}_{-0.038}$           & $647^{+193}_{-155}$ \\ 
    \hline
        
    \end{tabular}
    \end{adjustbox}
    
\end{table}

\edit{Through this bootstrapping approach, we obtain uncertainties of $\approx 2.3\%$ for $M_\star$, $\approx19\%$ for $M_\mathrm{d}$ and $\approx11\%$ for $R_\mathrm{c}$ for HD 97048, while $\approx0.6\%$ for $M_\star$, $\approx20\%$ for $M_\mathrm{d}$ and $\approx27\%$ for $R_\mathrm{c}$ for WaOph 6.
These values take into account the systematic uncertainties deriving from the disk thermal structure and geometry. Instead, without the bootstrapping procedure we obtain underestimated uncertainties for almost all the quantities: $\approx1.8\%$ for $M_\star$, $\approx13\%$ for $M_\mathrm{d}$ and $\approx6\%$ for $R_\mathrm{c}$ for HD 97048, while $\approx0.3\%$ for $M_\star$, $\approx18\%$ for $M_\mathrm{d}$ and $\approx26\%$ for $R_\mathrm{c}$ for WaOph 6. }
Systematic uncertainties are generally slightly lower for the parameters of WaOph 6, with respect to HD 90748. This trend depends on the more prominent flaring of the HD 97048 disk: as the emitting surfaces are more flared, a small difference in one of the geometrical parameters can result in a large difference when extracting the emitting layers, and consequently when deriving the 2D thermal structure. 

We were also able to obtain the posterior distribution of the total dynamical mass of the two systems by computing, for each estimate of the stellar and the disk mass, their sum. The final best-fit estimate of the total mass is given by the $50^\text{th}$ percentile of the overall distribution, while the uncertainties are given by \edit{the $16^\text{th}$ and $84^\text{th}$ percentiles.
We obtained $M_\mathrm{tot}=2.524 ^{+0.014}_{-0.013}$ $M_\odot$ for HD 97048, and $M_\mathrm{tot}=1.166 ^{+0.045}_{-0.038}$ $M_\odot$ for WaOph 6.}
The uncertainties on the total dynamical masses do not correspond to the sum of the uncertainties on the stellar and disk masses, as the two estimates from the fit are not independent. As the two parameters are anticorrelated (lower values for the stellar mass are compensated by higher disk mass estimates), the corresponding uncertainties on the best-fit estimate of their sum are lower with respect to the errors we find on the single parameters.



\section{Analysis and discussion}
\label{sec:5_discussion}

\subsection{Comparisons with the literature} \label{subsec:comparisons}


The stellar mass of HD 97048 was determined by \citet{vioque2018} after deriving the luminosity of the source and locating it in the Hertzsprung–Russell diagram, \rr{using the PARSEC isochrones with solar metallicity \citep{bressan2012,marigo2017}}: the authors obtained $M_\star = 2.252^{+0.113}_{-0.135}\ M_\odot$, which is in perfect agreement with \edit{our dynamical estimate of $M_\star = 2.226^{+0.054}_{-0.049}\ M_\odot$.}
\citet{stapper2024} determined the disk mass of HD 97048 using the thermo-chemical code Dust And LInes \citep[DALI, ][]{bruderer2012,bruderer2013}, which is able to match the model line luminosities to the observed \ce{^13CO} and \ce{C^18O} ones. The authors obtained a best-fit disk mass of $M_\mathrm{d}=0.1\ M_\odot$, with corresponding uncertainties that cover the mass range from $M_\mathrm{d}=0.03\ M_\odot$ up to $M_\mathrm{d}=0.3\ M_\odot$; \edit{our estimate of $0.300^{+0.055}_{-0.061} \ M_\odot$} is compatible within the error bars. 
\edit{An upper limit of $M_\mathrm{d}\leq 9.4\times10^{-2}M_\odot$ on the disk mass was suggested by \citet{kama2020} from HD 1-0 \textit{Herschel/}PACS archival observations. However, an accurate knowledge of the vertical temperature structure is necessary to correctly estimate disk masses through this method: the uncertainty on the mass estimate can exceed one order of magnitude for more massive disks \citep{trapman2017}.}
\edit{The disk mass can also be estimated by tracing the location of the CO emitting surface back to the CO column density and, assuming a CO abundance, to the disk surface density \citep{giovanni_exoalma}. Both \citet{giovanni_exoalma} and \citet{paneque2025} find systematically lower disk masses compared to literature values derived using methods that are independent of the CO abundance. These results suggest that carbon depletion must be taken into account in order for height-based disk mass estimates to be consistent with those obtained through alternative approaches.
To reconcile the disk mass estimate of HD~97048 reported by \citet{paneque2025} with the dynamical measurement obtained in this work, a carbon depletion factor of approximately ten relative to the interstellar medium (ISM) value is required. This value is consistent with the CO depletion factors inferred by \citet{giovanni_exoalma} to bring height-based disk masses into agreement with the dynamically inferred masses for the exoALMA sample \citep{exoalma_longarini}.}
Finally, \citet{law2022} obtained a dynamical measurement for the stellar mass of $M_\star=2.7^{+0.015}_{-0.015}\ M_\odot$, fitting the CO rotation maps with \texttt{eddy} \citep{teague_eddy}, assuming that the velocity field is only shaped by the simple Keplerian contribution.
\edit{However, this method neglects the pressure gradient term and does not disentangle the two gravitational contributions of the star and the disk self-gravity.} Our dynamical measurement of the total mass of the system, \edit{equal to $M_\mathrm{tot}=2.524 ^{+0.014}_{-0.013}\ M_\odot$}, is in overall agreement with \citet{law2022}, especially considering that their error bars on the mass estimate are only statistical and hence underestimated \citep{andrews2024,pezzotta2025}. 

Concerning WaOph 6, there is a strong discrepancy in the literature estimates for the stellar mass. Within the DSHARP Large Program, \citet{andrews2018} obtained a best-fit estimate mass of $M_\star=0.676^{+0.324}_{-0.126}\ M_\odot$ using stellar evolutionary models.
\citet{brown-sevilla2021} measured a very similar stellar mass of $M_\star=0.7^{+0.1}_{-0.1}\ M_\odot$, from pre-main-sequence tracks. On the other hand, performing the velocity map fitting with a simaple Keplerian model, \citet{law2022} obtained $M_\star=1.12^{+0.008}_{-0.008}\ M_\odot$, a very different result which is not compatible with the previous literature values. This discrepancy has been traced back to the likely non ideal efficiency of stellar evolutionary models in inferring the masses of low-mass pre-main-sequence stars, especially if embedded, or to a possible underestimate of the spectral type of the star. 
However, with our measurement we are able to solve this discrepancy: by disentangling the gravitational contributions from the star and the disk, we obtain a \edit{stellar mass of $M_\star = 0.956^{+0.006}_{-0.006}\ M_\odot$}, which \edit{is within the uncertainties of} the previous estimate by \citet{andrews2018}, and \edit{a total mass of $M_\mathrm{tot}=1.166 ^{+0.045}_{-0.038}\ M_\odot$} which is in line with the measurement by \citet{law2022}. Hence, the presumed discrepancy is well explained if we take into account the disk mass: part of the dynamical mass measured by \citet{law2022} actually belongs to the disk, which according to our measurements (and previous estimates based on semi-analytic models, e.g., $M_\mathrm{d}=0.16 \ M_\odot$ by \citealt{cadman2020}) is quite massive and whose gravitational contribution cannot be neglected when analyzing its kinematical pattern.
\edit{Concerning the disk mass of WaOph 6, the comparison between the dynamical estimate derived in this work and the height-based disk mass inferred by \citet{paneque2025} suggests a depletion factor of $\approx100$ with respect to the ISM value, which is consistent with the CO depletion levels reported for some disks in the exoALMA sample \citep{giovanni_exoalma}.}

\subsubsection{Gas-to-dust mass ratios}

For both disks, we computed the gas-to-dust mass ratios employing our dynamical gas masses and dust masses that were extracted by previous studies, assuming optically thin emission from the millimeter-dust. 
For HD 97048, we consider the dust mass $M_\mathrm{dust}=4.6\times10^{-4}\ M_\odot$ \rr{computed from continuum emission at $\lambda=1.28$ mm} by \citet{stapper2022}, and \edit{we obtain $M_\mathrm{gas}/M_\mathrm{dust}\approx640$}. Our measurement is in agreement with the values recently found within the exoALMA Large Program by \citet{exoalma_longarini}, which are considerably higher than the canonical ISM value of 100 (the average mass ratio in their sample is $\approx400$). In addition, \citet{stapper2024} observed that the warm Herbig disks have significantly larger gas-to-dust mass ratios than the colder T Tauri stars, often exceeding the standard ISM value. 

For WaOph 6, we consider the dust mass $M_\mathrm{dust}=1.4\times10^{-4}\ M_\odot$ \rr{computed from continuum emission at $\lambda=1.25$ mm} by \citet{brown-sevilla2021}, and we obtain $M_\mathrm{gas}/M_\mathrm{dust}\approx1500$. Our value is approximately one order of magnitude larger than the typical ISM value; however, we need to take into account that this disk is massive with respect to typical disks, and the dust mass is likely underestimated due to the many assumptions that were made in order to extract it, as explained in \citet{brown-sevilla2021}. In particular, the optically thin emission hypothesis for the millimeter-dust is not always representative of the actual disk conditions, as already observed in many previous studies \citep{brown-sevilla2021,stapper2022,curone2025}. 
\citet{stapper2024} also observed this discrepancy, comparing the gas masses extracted through chemical modeling with DALI to those obtained through the simple assumption of an optically thin relation between line flux and column density.
Our value is also in line with the results by \citet{trapman2025} within the AGE-PRO Large Program \citep{zhang2025}, who found high gas-to-dust ratios ($\approx 1000$ or slightly higher) for some disks located in the Ophiuchus region, as the source we are considering. They also highlight the retrieved dust masses are likely underestimated, especially for the more massive disks, due to the optically thin dust assumption.

\subsubsection{Disk-to-star mass ratios}

We measure quite high values for the \edit{disk-to-star mass ratios in both disks: $M_\mathrm{d}/M_\star=0.13 ^{+0.03}_{-0.03}$ for HD 97048 and $M_\mathrm{d}/M_\star=0.22 ^{+0.05}_{-0.04}$ for WaOph 6.}
Our estimates are in line with the expectations: HD 97048 is a very massive source surrounded by a hot and extended disk, and WaOph 6 shows a clear spiral morphology, possibly hinting at an ongoing GI, which can more easily occur in sufficiently massive disks. 
Our value for the WaOph 6 mass ratio is in agreement with the estimate of $\approx 0.24$ by \citet{cadman2020}. 
The estimate for HD 97048 is compatible with the results for most of the exoALMA sources \citep{exoalma_longarini}, while the estimate for WaOph 6 is higher than their average $M_\mathrm{d}/M_\star$. This is consistent, as the exoALMA sample does not include any sources displaying a spiral morphology in the millimeter-dust continuum, which are expected to have a higher mass ratio if undergoing GI. Both our measurements are in line with the disk-to-star mass ratios computed \edit{for Elias 2-27 ($M_\mathrm{d}/M_\star\approx 0.17$, \citealt{veronesi2021}) and for \rr{IM Lup ($M_\mathrm{d}/M_\star\approx 0.09$}, \citealt{martire}), two disks with millimeter dust spiral structures.}
In Sect. \ref{subsec:Q}, we discuss in detail the possible interpretations of our results and the implications about the potential presence of ongoing GI in these systems.
Finally, we highlight that our constrained values are higher than the $5\%$ threshold needed for dynamical disk mass measurements \citep{veronesi2024,andrews2024}.

\subsection{Disk stability} \label{subsec:Q}

Precise measurements of stellar masses, disk masses and scale radii are also fundamental requirements to estimate the likelihood for a disk to be gravitationally unstable. For both our sources we computed the Toomre $Q$ parameter \citep{toomre1964}, defined as

\begin{equation}
    Q \simeq \frac{c_\mathrm{s,mid}\Omega_\mathrm{k}}{\pi G\Sigma} = 
             2\frac{H}{R}\bigg|_\mathrm{mid} \bigg(\frac{M_\star}{M_\mathrm{d}}\bigg) \bigg(\frac{R}{R_\mathrm{c}}\bigg)^{-1} \exp \bigg[\frac{R}{R_\mathrm{c}}\bigg].
\end{equation}

This parameter quantifies the significance of the stabilizing contributions of pressure ($c_\mathrm{s,mid}$) and rotation ($\Omega_\mathrm{k}$) in the force balance of a disk, with respect to the non-stabilizing disk self-gravity ($\Sigma$). \rr{By explicitly expressing $\Omega_\mathrm{k}=\sqrt{GM_\star/R^3}$, $c_\mathrm{s,mid}=\Omega_\mathrm{k}H_\mathrm{mid}$ and the assumed self-similar solution for the surface density \citep{lyndenbell}}, we highlight that this quantity depends on the three main parameters we obtain through our rotation curves fitting \citep{kratter2016}. In disks with a lower $Q$ (ideally close to $\approx 1$), the destabilizing effect of the disk mass itself dominates over the other balancing contributions, leading to the possible trigger and development of gravitational instabilities within the disk. 
\edit{We obtained the radial profiles of the Toomre parameter for HD 97048 and WaOph 6, from which we extracted its minimum value along the disk radial extent ($Q_\mathrm{min}$), as that would be the location where any GI would potentially be triggered. From our analysis, we obtained $Q_\mathrm{min}\approx4.7$ for HD 97048 and $Q_\mathrm{min}\approx5.4$ for WaOph 6. }

\edit{We compared the retrieved $Q_\mathrm{min}$ values for HD 97048 and WaOph 6 with the values for all disks whose masses have been dynamically constrained. In addition to our two sources, the sample includes Elias 2-27, and the disks studied by the MAPS and the exoALMA Large Programs, for a total number of 15 sources. We did not include AS 209 and AA Tau, whose mass determinations were each compatible with zero and highly uncertain.  The value of 
$Q_\mathrm{min}$ depends on some parameters: $M_\star$, $M_\mathrm{d}$, $R_\mathrm{c}$, $T_\mathrm{mid,0}$, and $q_\mathrm{mid,0}$. Adopting the literature values for these parameters,\footnote{The best-fit values are from this work for HD 97048 and WaOph 6, \citet{exoalma_longarini} for the exoALMA sources, \citet{martire} for the MAPS sources, and \citet{veronesi2021} for Elias 2-27.} we computed $Q_\mathrm{min}$ for all 15 sources. We accounted for the uncertainties on the input physical parameters through a bootstrap resampling procedure with $N_\mathrm{iter}=10000$: for each iteration, all parameters were randomly drawn from Gaussian distributions centered on their best–fit values, with standard deviations equal to the corresponding uncertainties. This procedure yielded a distribution of $N_\mathrm{iter}=10000$ values of $Q_\mathrm{min}$ for each source.
\rr{As the presence of millimeter-dust spiral morphologies can be induced by ongoing GI \citep{dipierro}, in order to investigate the relation between the appearance of spirals and disk stability} we then distinguished two subsamples: disks with observed spiral structures in the millimeter-dust continuum emission (Elias 2-27, WaOph 6 and IM Lup; none of them is known to host outer binary companions), and disks without such spirals.
For each bootstrap iteration, we constructed the empirical cumulative distribution function (ECDF) of $Q_\mathrm{min}$ separately for the two subsamples.}

\begin{figure*}
   \centering
   \includegraphics[width=\textwidth]{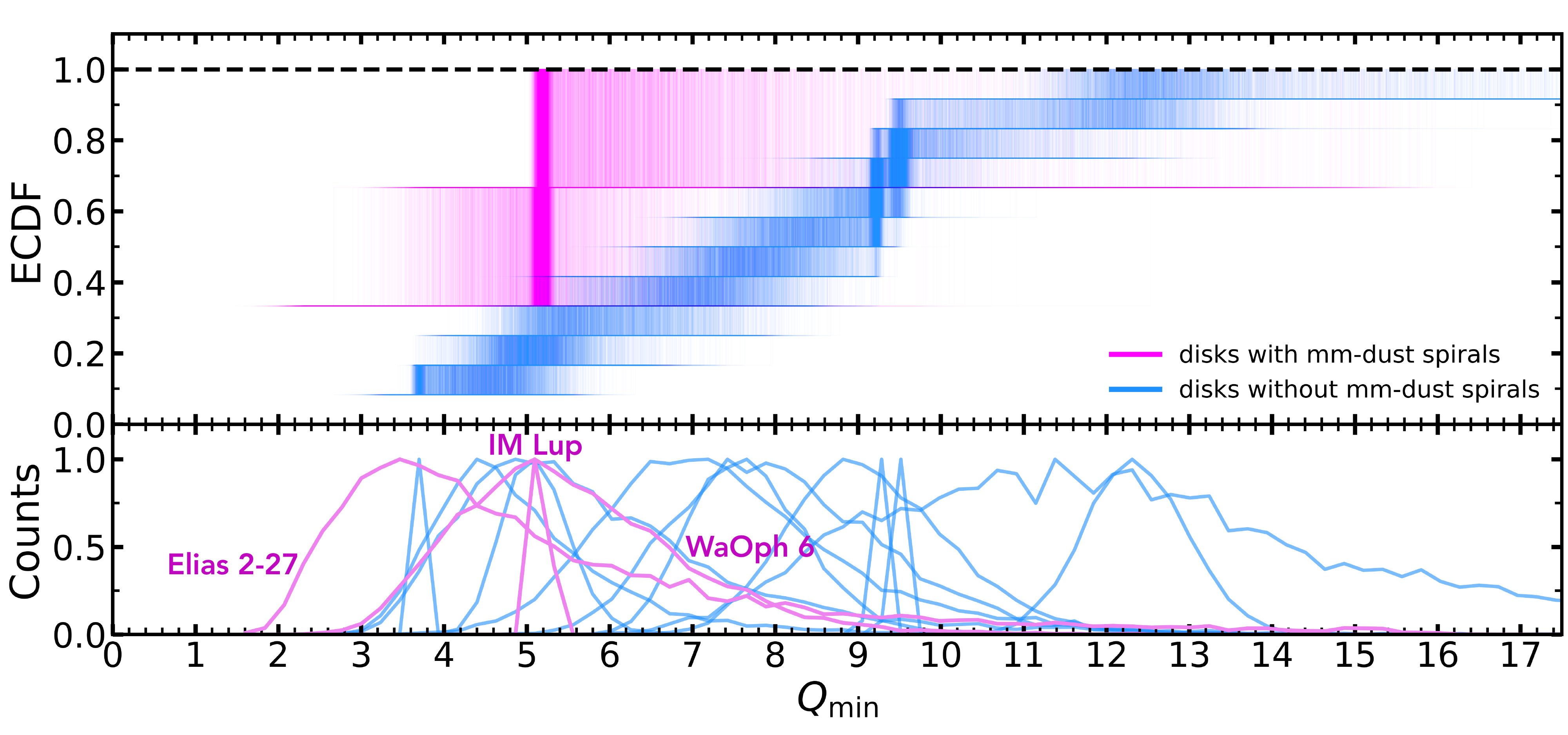}
   
      \caption{
      \edit{Obtained $Q_\mathrm{min}$ distributions for all disks with dynamical mass estimates (lighter lines) together with the corresponding cumulative distributions (darker lines), from $N_\mathrm{iter}=10000$ bootstrap realizations. Disks hosting millimeter-dust spirals are shown in pink, while disks without spirals are shown in blue. The distributions of individual sources have been normalized to the same maximum value for visualization purposes.}
      }
      
         \label{fig:toomre}
\end{figure*} 

\edit{In Fig. \ref{fig:toomre} we show the obtained distributions for $Q_\mathrm{min}$ for all disks with known dynamical disk mass estimates, and the corresponding cumulative distributions, separating disks with millimeter-dust spirals and without such spirals. \rr{The whole $Q(R)$ radial profiles are reported in Appendix \ref{app:Q_profiles}.}
We observe that the two distributions are significantly displaced: for disks with millimeter-dust spirals (pink line), the cumulative distribution rapidly reaches unity with a steep slope, implying that all considered disks with millimeter-dust spiral morphology have lower $Q_\mathrm{min}$ values (between 4 and 6). For the sources without millimeter-spirals (blue line), the cumulative distribution for $Q_\mathrm{min}$ has a much shallower slope, slowly reaching unity. 
Both the bootstrapped distributions of $Q_\mathrm{min}$ and the cumulative functions show that disks hosting millimeter-dust spirals are systematically shifted toward lower $Q_\mathrm{min}$ values compared to disks without spirals. 
To quantitatively assess whether the pink and blue empirical samples are consistent with being drawn from the same parent distribution, for each iteration we performed a two-sample Anderson–Darling (AD) statistical test \citep{anderson1952}.
Across the ensemble of bootstrap realizations, the Anderson–Darling test consistently rejected the null hypothesis of a common underlying distribution, yielding an average p-value of 0.001 (corresponding to a Gaussian two-sided significance of $>3\sigma$).}

The first main outcome of this quantitative test is to show, by leveraging precise dynamical measurements of the stellar and disk masses, that disks hosting spiral structures in the millimeter continuum emission generally have statistically significant lower $Q_\mathrm{min}$ values. 
This result suggests that the observed millimeter-dust spirals are likely driven by GI, and may be regarded as signatures of ongoing GI. This interpretation is in agreement with recent studies by \citet{yoshida2025}, tracing back the millimeter-dust spirals in the IM Lup disk to the presence of gravitational instabilities, based on the Keplerian motion of the spiral arms.

Despite not showing any spiral morphologies in the millimeter continuum emission, \edit{HD 97048, SY Cha, and GM Aur also} have low $Q_\mathrm{min}$ values: their measured $Q_\mathrm{min}$ is comparable with the results for disks with observed spirals ($Q_\mathrm{min}\approx4-6$). Within the sample of sources with dynamical disk mass measurements available, \edit{these three outliers have prominent} non-Keplerian motions detected in the gas kinematical pattern. 
HD 97048 and SY Cha have evident localized kinks visible in some velocity channels \citep{pinte2019,exoalma_pinte}, and diffuse deviations from the Keplerian background up to $\approx 0.3$ km/s in the velocity residuals maps (as clear from Fig. \ref{app:vel_res_maps} for HD 97048). For both disks, the presence of an embedded planetary companion was suggested to explain the pronounced kinks in the CO emission; however, despite the presence of rings and annular gaps, no planet has been detected so far. On the other hand, the low $Q_\mathrm{min}$ values and the presence of more extended kinematic perturbations visible in multiple locations in the velocity residuals may hint at the presence of a more global mechanism at play in these disk, such as GI; however, the two disks do not exhibit any spiral structure in the continuum emission \citep{walsh2016,vanderplas2017,speedie2022}. 
A possible scenario that could explain both the observations and the retrieved $Q_\mathrm{min}$ values for these two sources is the combined effect of the perturbations induced by GI and by an embedded planetary companion. The observability of GI-driven large-scale spirals in the millimeter-dust emission can be weakened or even suppressed by the interaction with a planet, depending on the ratio between the mass of the disk and the embedded companion: with a slightly unstable disk and a massive planet, the dust spiral can be washed out and any signature of GI in the continuum emission can be dynamically hidden by the perturbations driven by the planet \citep{rowther2020,rowther2023}. Similarly, either GI or planet-disk interaction signatures can be detected in the gas kinematics, according to which term is dominant in the overall disk dynamics. Hence, the observations for these two sources are compatible with the presence of a massive planet suppressing the signatures of GI. 
\edit{On the other hand, GM Aur exhibits large-scale, non-axisymmetric gas structures in the \ce{^12CO} emission, in striking contrast with the axisymmetric and multi-ringed millimeter continuum emission. \ce{^12CO} line emission reveals a complex morphology, consisting of spiral arms up to $\approx1200$ au, an extended tail, and diffuse structures, up to $\approx1900$ au from the central star. The complex morphology and large-scale non-Keplerian motions in GM Aur were interpreted as the result of late infall of surrounding material onto the disk \citep{huang2021}. The overall complexity of gas emission made it challenging to confidently determine the Toomre parameter and assess the disk stability \citep{mcclure2016,schwarz2021,lodato}. Thus, any conclusion about ongoing gravitational instabilities in the GM Aur disk should be regarded with some caution.}

\subsection{Pressure substructures} \label{subsec:Psubs}
 
While the stellar and disk gravitational fields influence the rotational velocity on a large scale, small-scale modulations in the rotation curve can be directly traced back to pressure gradient variations \citep{teague2018b,pezzotta2025,exoalma_stadler}.
We can subtract the Keplerian contribution from the extracted rotation curves to obtain the observed deviation from the Keplerian motion $\delta v_\phi$. We expect $\delta v_\phi=0$ in correspondence of pressure minima and maxima (i.e., rings and gaps), where the pressure gradient is zero \citep[e.g.,][]{rosotti2020}.
Any inaccuracy in the subtracted Keplerian contribution due to the uncertainty on the retrieved stellar mass, or the presence of the disk self-gravity, whose contribution is generally lower than the pressure gradient \citep{andrews2024,exoalma_longarini}, would only induce a vertical shift in $\delta v_\phi$, without changing the small-scale oscillations. To circumvent this possible problem, it is sufficient to determine the sign of the radial derivative of $\delta v_\phi$, which is expected to be positive in proximity of a pressure minimum and be negative around the location of pressure maxima \citep[e.g.,][]{izquierdo2023,exoalma_stadler}. 

\begin{figure}
   \centering
   \includegraphics[width=.5\textwidth]{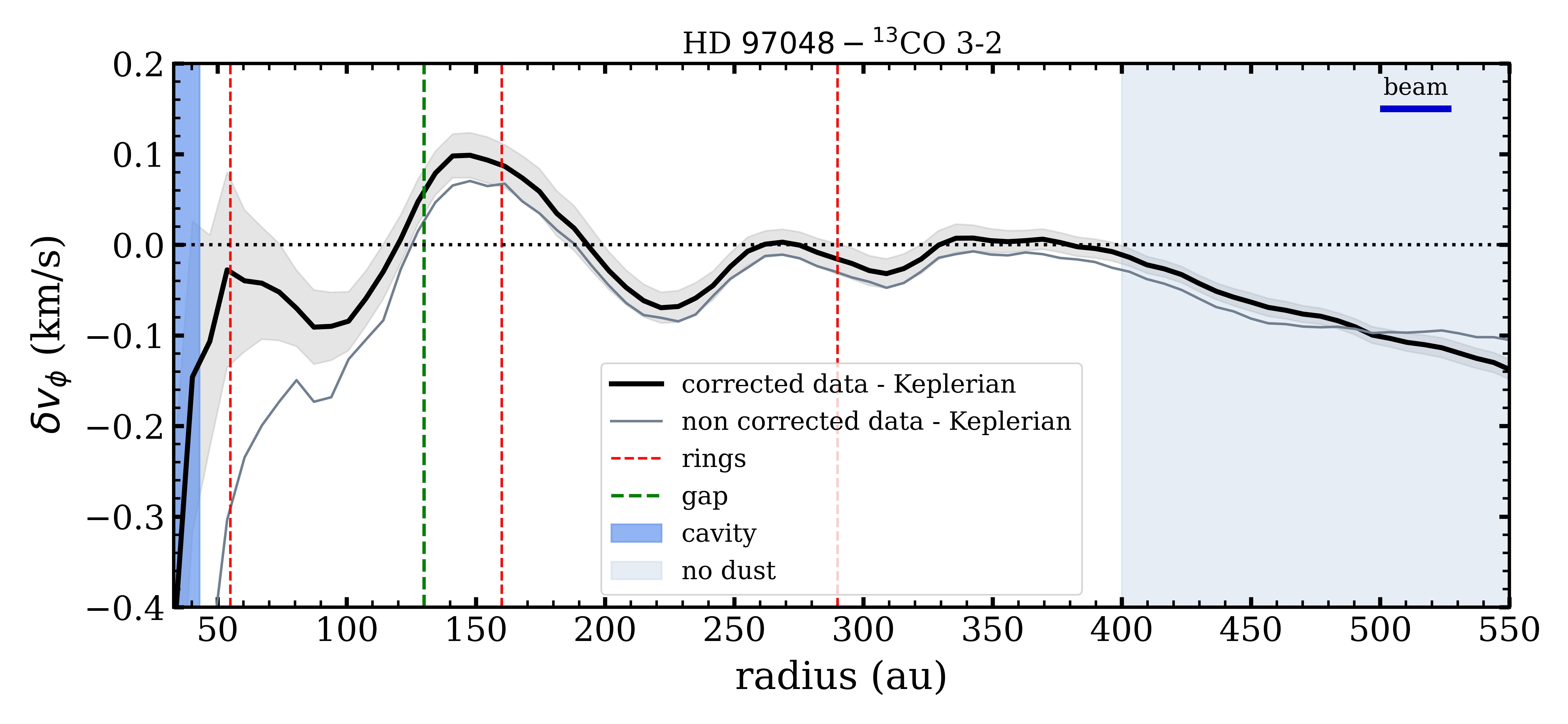}
   \includegraphics[width=.5\textwidth]{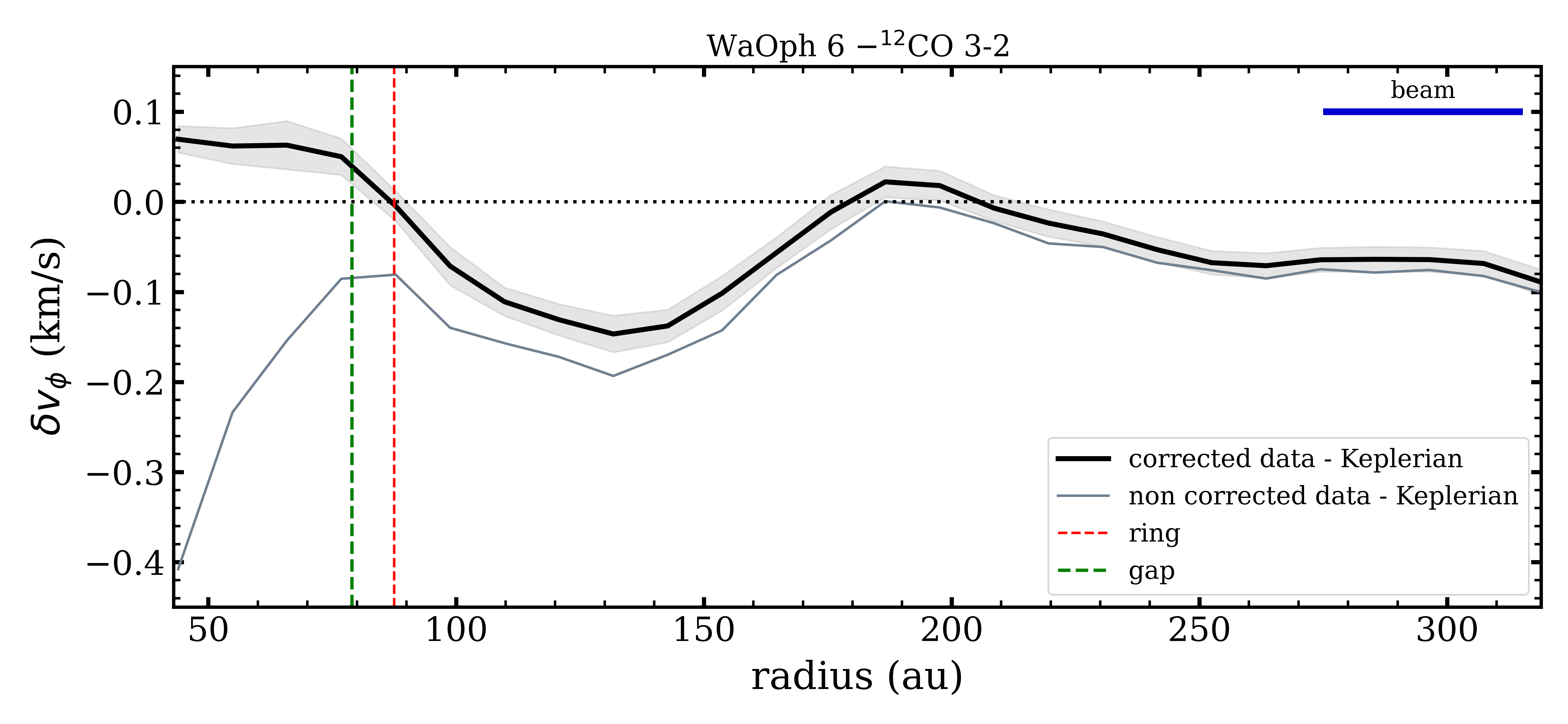}
      \caption{
      Radial profiles of the azimuthal velocity residuals $\delta v_\phi$ (data-Keplerian) of the \ce{^13CO} 3 -- 2 line for HD 97048 (top) and the \ce{^12CO} 3 -- 2 line for WaOph 6 (bottom), with the corresponding uncertainties. The thick black line was obtained using beam smearing corrected rotation curves, while the thin gray line was obtained with the non-corrected curves. Red lines show the locations of millimeter-dust rings, while green lines mark the millimeter-dust gaps (as reported by \citealt{pinte2019}). The horizontal blue line in the upper-right corner of each panel shows the beam size of the observations.
      }
      
         \label{fig:deltavphi}
\end{figure}

For one line per source, in Fig. \ref{fig:deltavphi} we show the radial profiles of the azimuthal velocity residuals $\delta v_\phi$, that we obtained by subtracting the \texttt{discminer} Keplerian profile 
from the extracted rotation curves. 
We can clearly see the presence of pressure substructures in both sources, with deviations from Keplerian motion up to $\approx 0.1$ km/s. The observed oscillations in the radial profile of the azimuthal velocity residuals for HD 97048 are in agreement with the expected deviations induced by the presence of rings and gaps. The radial derivative of $\delta v_\phi$ is positive in the gap where the planet candidate suggested by \citet{pinte2019} would be located, while it is negative at all ring locations. A similar behavior is also visible for the other available tracers, whose plots are shown in Fig. \ref{fig:deltavphi_others} in Appendix \ref{app:vel_azim_res}. In the case of WaOph 6, the locations of the millimeter-dust ring and gap are radially very close \edit{($\leq10$ au)}, according to \citet{huang2018_2}; moreover, the data quality for this disk is significantly reduced by cloud absorption and lower signal-to-noise ratio. Hence, clearly detecting fine substructures within azimuthally averaged quantities is a complicated task. However, we can still notice some deviations from the Keplerian background motion, which can be traced back to the effect of pressure gradients. In particular, we measure exactly $\delta v_\phi=0$ where the dust ring is located, as expected if it corresponds to a region of pressure maximum.

\section{Conclusions}
\label{sec:6_conclusions}

In this work, we have explored the kinematics of the emission lines of CO isotopologs in the disks of HD 97048 and WaOph 6, two sources exhibiting features of potential planet-disk interaction and GI, to dynamically retrieve their stellar masses, disk masses, and scale radii. We followed the pipeline adopted by the exoALMA collaboration \citep{exoALMA_1} to model the line emission \citep{exoalma_izquierdo}, extract the 2D thermal structure \citep{exoalma_maria}, and fit the rotation curves \citep{exoalma_longarini}.
Our results show the following:

\begin{itemize}
    \item[$-$] The pipeline to retrieve dynamical measurements can be efficiently applied to disks with significant absorption features. A detailed kinematical analysis and some careful technical measures \edit{in the modeling and curve extraction procedures} are required to analyze the effects of cloud contamination and avoid consequent biases in the retrieved line centroids.
    \item[$-$] Beam smearing can systematically bias the rotation curves, affecting the retrieved dynamical masses, with discrepancies of $\approx45\%$. For each tracer, we applied a correction factor to fix this bias. 
    \item[$-$] By bootstrapping over both the geometrical and thermal parameters of the disk, we have provided more reliable estimates of the uncertainties on dynamical disk masses, which we estimate \edit{to be $\approx20\%$.}
    \item[$-$] From our analysis we obtained \edit{$M_\star=2.226^{+0.054}_{-0.049} \ M_\odot$, $M_\mathrm{d}=0.300^{+0.055}_{-0.061}\ M_\odot$, and $R_\mathrm{c}=172^{+24}_{-14}$ au for HD 97048 and $M_\star=0.956^{+0.006}_{-0.006} \ M_\odot$, $M_\mathrm{d}=0.210^{+0.045}_{-0.038}\ M_\odot$, and $R_\mathrm{c}=647^{+193}_{-155}$ au for WaOph 6}. We measured higher gas-to-dust mass ratios than the canonical ISM value, implying that dust masses computed from the continuum flux under the assumption of optically thin emission are often underestimated. For both disks we extracted high disk-to-star mass ratios that are in line with the literature values obtained for disks displaying spiral morphology in the millimeter continuum emission. 
    \item[$-$] Precise mass measurements allowed us to further investigate the stability of our sources by evaluating the minimum Toomre parameter $Q_\mathrm{min}$. We obtained $Q_\mathrm{min}\approx4.7$ for HD 97048 and $Q_\mathrm{min}\approx5.4$ for WaOph 6. Considering all disks with available dynamical mass measurements, we observed a \edit{systematic difference between the values of the minimum Toomre parameter for disks with and without millimeter-dust spirals}. 
    \item[$-$] The only sources with a lower $Q_\mathrm{min}$ value but no observed millimeter-dust spirals are HD 97048, SY Cha, \edit{and GM Aur. However, these sources have extended non-Keplerian motions in the gas kinematics, and the possible presence of ongoing GI cannot be ruled out.} 
    \item[$-$] In the disks of HD 97048 and WaOph 6, the observed pressure modulations are generally consistent with the locations of the millimeter-dust substructures, confirming the correspondence between pressure maxima (minima) and the existence of rings (gaps). We stress that including the beam smearing correction in the rotation curves allowed us to efficiently measure the deviations from the Keplerian background up to the innermost regions of the disks.
\end{itemize}

\begin{acknowledgements}
This paper makes use of the following ALMA data: ADS/JAO.ALMA\#2015.1.00168.S, ADS/JAO.ALMA\#2015.1.00192.S, ADS/JAO.ALMA\#2016.1.00484.L,  ADS/JAO.ALMA\#2016.1.00826.S.
ALMA is a partnership of ESO (representing its member states), NSF (USA) and NINS (Japan), together with NRC (Canada), MOST and ASIAA (Taiwan), and KASI (Republic of Korea), in cooperation with the Republic of Chile. The Joint ALMA Observatory is operated by ESO, AUI/NRAO and NAOJ. 
We acknowledge Felipe Alarcón and Swastik Chowbay for useful discussions.
V.P. and S.F. are funded by the European Union (ERC, UNVEIL, 101076613). Views and opinions expressed are however those of the author(s) only and do not necessarily reflect those of the European Union or the European Research Council. Neither the European Union nor the granting authority can be held responsible for them. 
S.F. acknowledges financial contribution from PRIN-MUR 2022YP5ACE. 
Support for AFI was provided by NASA through the NASA Hubble Fellowship grant No. HST-HF2-51532.001-A awarded by the Space Telescope Science Institute, which is operated by the Association of Universities for Research in Astronomy, Inc., for NASA, under contract NAS5-26555.
GL acknowledges support by PRIN-MUR 20228JPA3A and by the European Union Next Generation EU, CUP: G53D23000870006.
CL has been supported by the UK Science and Technology Research Council (STFC) via the consolidated grant ST/W000997/1. 
Support for C.J.L. was provided by NASA through the NASA Hubble Fellowship grant No. HST-HF2- 859 51535.001-A awarded by the Space Telescope Science Institute, which is operated by the Association of Universities for Research in Astronomy, Inc., for NASA, under contract NAS5-26555.
T.P-C. acknowledges support from the Michigan Society of Fellows and has received funding by the Heising-Simons foundation through the 51 Pegasi B Fellowship.

\end{acknowledgements}
\bibliographystyle{aa}
\bibliography{bibliography}

\begin{appendix}

\section{Requirements for dynamical measurements: 2D disk temperature versus number of fitted rotation curves} \label{app:requirements}

While the importance of including the disk thermal structure has previously been verified, the impact of the number of fitted lines on the final disk mass estimate is still unknown: in particular, we here investigate \edit{if it is possible to obtain precise disk mass estimates fitting only one rotation curve, if we already know the underlying 2D thermal structure. This would represent a further step to extend the applicability of the dynamical method to larger samples. To tackle this question, we ran tests on the exoALMA sample \citep{exoalma1}, consisting of bright disks, radially extended beyond $1^{\prime\prime}$, with an inclination between $5-60$\textdegree, and no signs of envelope contamination}. Specifically, we considered the ten sources for which there are reference dynamical measurements by \citet{exoalma_longarini}: these dynamical estimates were obtained by simultaneously fitting two rotation curves (\ce{^12CO} and \ce{^13CO} 3 -- 2 lines) with the vertically stratified model by \citet{martire}, taking into account the 2D thermal structures by \citet{exoalma_maria}. \edit{In this test, the dynamical measurements reported by \citet{exoalma_longarini} were adopted as reference values, and our results were compared against them to assess consistency.}
We computed dynamical disk mass estimates for the same disks, this time fitting only one rotation curve (the \ce{^12CO} 3 -- 2 line) with the code \texttt{DySc}, in line with recent works \citep{martire,pezzotta2025,exoalma_longarini}. 
\edit{Given the curves, the 2D thermal structure of the disk and the molecular emitting layers, the code explores the parameter space with Monte Carlo Markov chains\footnote{For details, the reader can refer to \citet{emcee}.} to maximize the likelihood of the model, providing as output the probability distributions of the fitted parameters.}
\edit{We derived the disk masses with two different approaches. On one hand, we accounted for the 2D thermal structures obtained by \citet{exoalma_maria} and adopted a stratified model as done in \citet{exoalma_longarini}; on the other hand, we assumed a vertically isothermal scenario where the temperature radial profile matches the brightness temperature measured along the \ce{^12CO} 3 -- 2 layer. The corresponding error bars were obtained by iterating the procedure 100 times, bootstrapping over the parameters that describe the disk thermal structure.}
In this way, we intended to test whether the most important contribution to precise dynamical measurements can be traced back to the knowledge of the 2D thermal structure or rather to the number of molecular tracers for the curves fit and to disentangle which of the two requirements is dominant.

\begin{figure}
   \centering
   \includegraphics[width=\columnwidth]{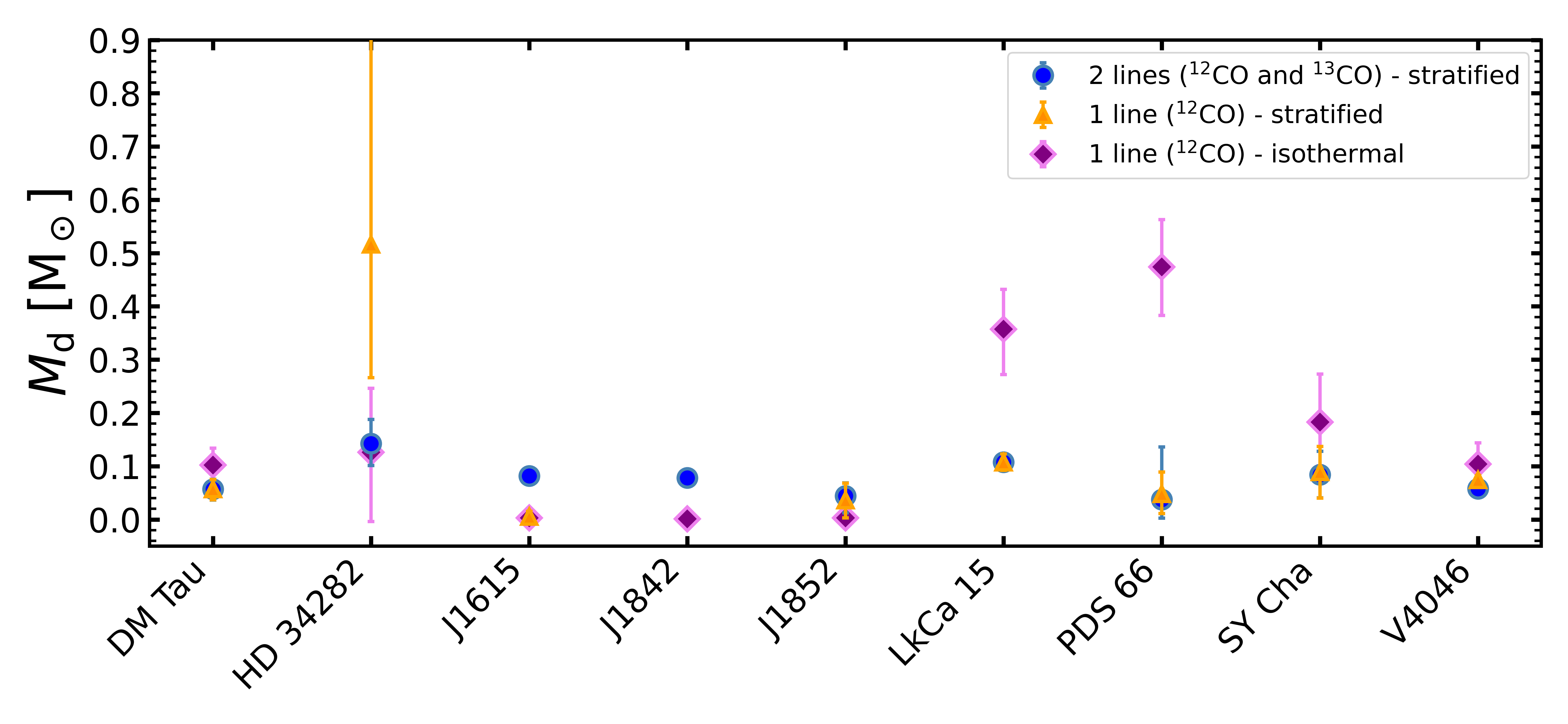}
      \caption{
      Comparison between different dynamical disk mass measurements for the exoALMA disks. Blue dots show best-fit estimates by \citet{exoalma_longarini} obtained by fitting the \ce{^12CO} and \ce{^13CO} 3 -- 2 curves with a stratified model. Orange triangles and purple diamonds show our dynamical estimates obtained by fitting the \ce{^12CO} curve only, with a stratified and an isothermal model, respectively. \edit{The reported uncertainties correspond to $2\sigma$ credible intervals ($\approx95\%$)}. 
      }
      \label{fig:cfr_mdisk}  
\end{figure}

In Fig. \ref{fig:cfr_mdisk} we compare both our estimates with the ones in \citet{exoalma_longarini}. \edit{The reported uncertainties correspond to $2\sigma$ credible intervals ($\approx95\%$), consistently with \citet{exoalma_longarini}.}
We discard AA Tau from the comparison: its high inclination led to anomalous estimates for the emitting surfaces and for the disk mass, marking this disk as an outlier in the exoALMA sample \citep{exoalma_maria,exoalma_longarini}. 
Similarly, we cannot achieve a proper comparison for J1842, as its disk mass could not be recovered by fitting one rotation curve only with the stratified model, as the fit would not converge.
Of the remaining sources, for six out of eight the best-fit estimates from our 1 line-stratified case match the estimates from the 2 lines-stratified case by \citet{exoalma_longarini}, within the corresponding error bars. On the contrary, the 1 line-isothermal scenario cannot recover the reference disk masses within the error bars, leading to large discrepancies in the final estimates: \edit{in the worst scenarios, we get a factor of $>3$ for the LkCa 15 disk and a factor of $\sim 13$ for the most extreme case of PDS 66. For this disk in particular, the determination of well-defined emitting surfaces was complicated and the layers were found to be almost flat, preventing an accurate reconstruction of the 2D temperature structure \citep{exoalma_maria}.}
For the same two disks, our 1 line-stratified estimates agree better with the reference ones, with relative differences of $< 1\%$ for LkCa 15 and $<27\%$ for PDS 66, which is still the same order of magnitude as the usual systematic uncertainties on dynamical mass measurements \citep{veronesi2024,andrews2024}. HD 34282 and J1615 are the only outliers in our sample: in the first case, \edit{the high disk inclination prevents a precise determination of the disk mass in the one line-stratified scenario, which shows higher uncertainties with respect to the other considered sources (in comparison to the rest of the exoALMA sample, large error bars were also obtained by \citet{exoalma_longarini} for the reference value for the mass of this disk)}. In the second case, the fit with one line provides an almost null disk mass in both the stratified and isothermal scenarios, a different output than the fit with two lines; in this case only, the number of fitted lines appears to be particularly significant. 

From these results we infer that in the majority of the cases, once the disk 2D thermal structure is accounted for, fitting only one reliable rotation curve is sufficient to obtain a good estimate of the disk mass, confirming that an accurate thermal structure represents the most relevant ingredient in the path to obtain dynamical disk masses. On the other hand, increasing the number of fitted rotation curves will improve the accuracy of the extracted dynamical estimates, as shown by \citet{pezzotta2025}.

\section{Discminer models best-fit values}\label{app:discminer}

We list in Table \ref{table:discminer_bestfit} the best-fit values for all fitted parameters that we obtained from the \texttt{discminer} models, for all the analyzed tracers. As explained in Sect. \ref{subsec:absorption}, for both sources we fixed the inclination $i$, position angle $\mathrm{PA}$, and systemic velocity $v_\mathrm{sys}$ to the best-fit values of the \ce{^13CO} 3 -- 2 \texttt{discminer} model. The $\mathrm{PA}$ is referred to the redshifted axis of the disk, while the sign of the inclination can be determined by locating the redshifted and blueshifted sides of the disk, and identifying the near side of the disk (see \citealt{exoalma_izquierdo} for details). 

\begin{table*}
\centering
\caption{Best-fit model parameters obtained \rr{with \texttt{discminer}} for HD~97048 and WaOph~6.}
\label{table:discminer_bestfit}
\begin{tabular}{llc|ccc|ccc}
\toprule
 &  &  & \multicolumn{3}{c}{HD 97048} & \multicolumn{3}{c}{WaOph 6} \\
\cmidrule(lr){4-9}
Attribute & Parameter & Unit & \ce{^12CO} 2 -- 1 & \ce{^13CO} 3 -- 2 & \ce{C^18O} 2 -- 1  & \ce{^12CO} 2 -- 1 & \ce{^12CO} 3 -- 2 & \ce{^13CO} 3 -- 2 \\
\midrule
Orientation
& $i$            & [$^\circ$] & $-42.48$ & $-42.48$ & $-42.48$ & $47.56$  & $47.56$  & $47.56$ \\
& $\mathrm{PA}$  & [$^\circ$] & $2.78$   & $2.78$   & $2.78$   & $174.80$ & $174.80$ & $174.80$ \\ 
& $x_\text{c}$          & [au]       & $11.79$  & $9.81$   & $12.53$  & $31.05$  & $25.72$  & $12.99$ \\
& $y_\text{c}$          & [au]       & $2.32$   & $-8.05$  & $5.00$   & $-44.40$ & $-38.64$ & $-34.16$ \\
\midrule
Velocity
& $M_{\star,\mathrm{dm}}$ & [M$_\odot$] & $2.41$ & $2.35$ & $2.37$ & $1.06$ & $1.00$ & $1.03$ \\
& $v_{\rm sys}$ & [km s$^{-1}$] & $4.71$ & $4.71$ & $4.71$ & $4.27$ & $4.27$ & $4.27$ \\
\midrule
Upper surface
& $z_0$ & [au] & 29.00 & 14.98 & 13.70 & 16.06 & 16.41 & 15.85 \\
& $p$   & [--] & 1.26 & 1.75 & 1.18 & 0.86 & 0.96 & 0.77 \\
& $R_\text{t}$ & [au] & 564.64 & 463.39 & 481.85 & 344.58 & 364.34 & 349.65 \\
& $q$   & [--] & 1.78 & 1.33 & 2.18 & 2.54 & 3.02 & 2.24 \\
\midrule
Lower surface
& $z_0$ & [au] & 38.96 & 14.52 & 9.23 & 11.83 & 12.88 & 11.50 \\
& $p$   & [--] & 1.62 & 1.53 & 1.15 & 1.05 & 1.16 & 1.10 \\
& $R_\text{t}$ & [au] & 149.97 & 604.89 & 507.63 & 377.25 & 413.67 & 403.49 \\
& $q$   & [--] & 0.56 & 0.47 & 4.44 & 5.00 & 3.33 & 0.18 \\
\midrule
Peak intensity
& $I_0$ & [Jy px$^{-1}$] & 1.15 & 0.76 & 1.10 & 0.26 & 1.32 & 0.60 \\
& $p$   & [--] & $-2.29$ & $-2.01$ & $-1.72$ & $-2.77$ & $-2.90$ & $-2.52$ \\
& $q$   & [--] & 1.95 & 1.22 & 1.88 & 2.52 & 2.66 & 2.61 \\
\midrule
Line width
& $L_\text{w0}$ & [km s$^{-1}$] & 0.55 & 0.79 & 0.28 & 0.17 & 0.25 & 1.06 \\
& $p$   & [--] & $-0.23$ & $0.54$ & $-0.44$ & 0.13 & $-0.38$ & $-0.74$ \\
& $q$   & [--] & $-0.32$ & $-0.68$ & $-0.36$ & $-0.58$ & $-0.32$ & 0.50 \\
\midrule
Line slope
& $L_\text{s0}$ & [--] & 2.24 & 1.14 & 1.77 & 2.13 & 2.09 & 3.55 \\
& $p$   & [--] & 0.22 & 0.62 & 0.12 & $-0.24$ & $-0.14$ & $-1.59$ \\
\bottomrule
\end{tabular}

\vspace{0.2cm}
\begin{minipage}{0.95\textwidth}
\footnotesize
\textbf{Notes.} For both sources, $i$, $\mathrm{PA}$, and $v_\mathrm{sys}$ were fixed to the corresponding \ce{^13CO} model best-fit.
\end{minipage}
\end{table*}

\section{Maps of the velocity residuals} \label{app:vel_res_maps}
\begin{figure*}
   \centering
   {\LARGE \textbf{HD 97048}\par}
   \vspace{0.3cm}

   \includegraphics[width=0.48\textwidth]{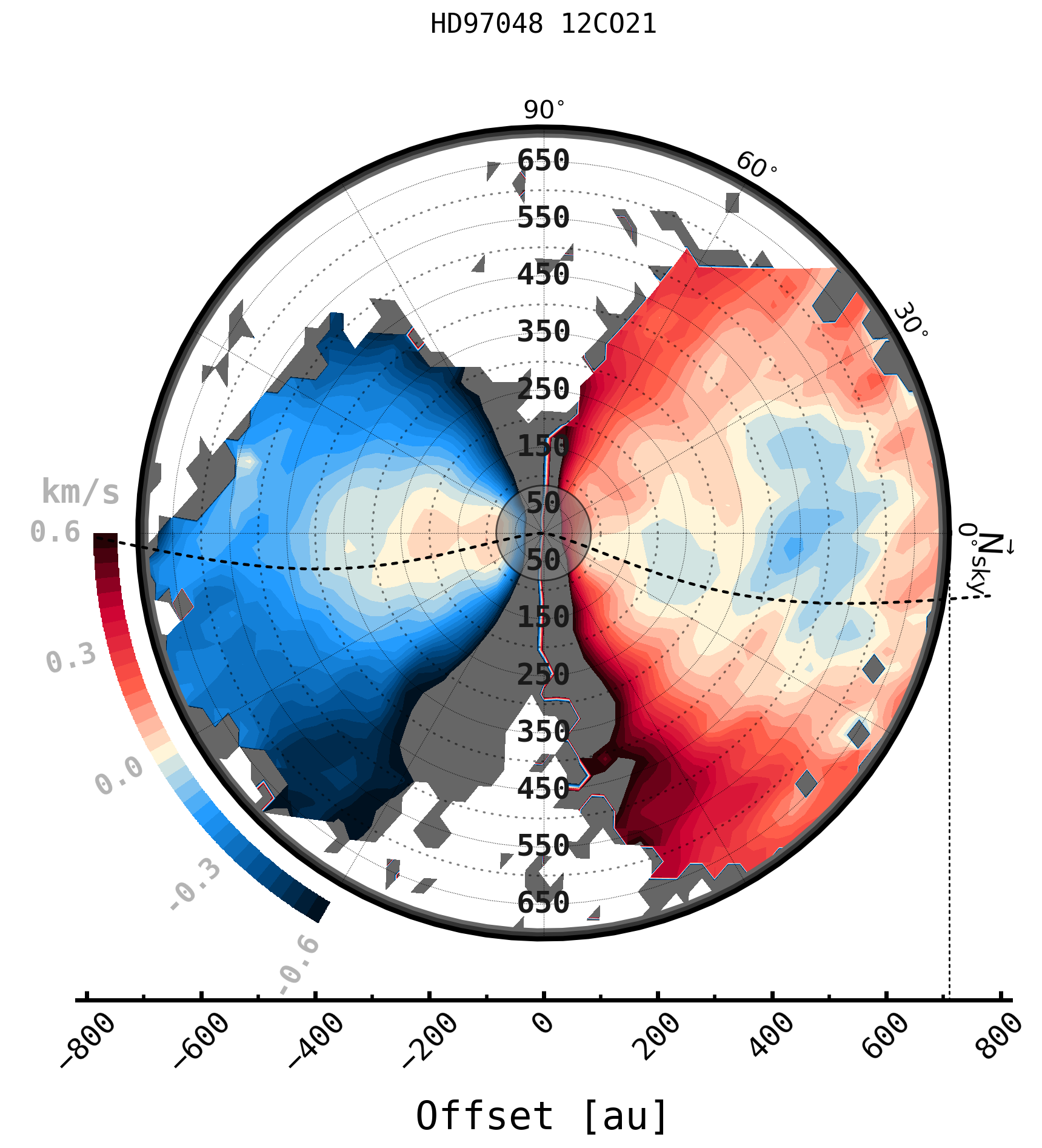}
   \includegraphics[width=0.48\textwidth]{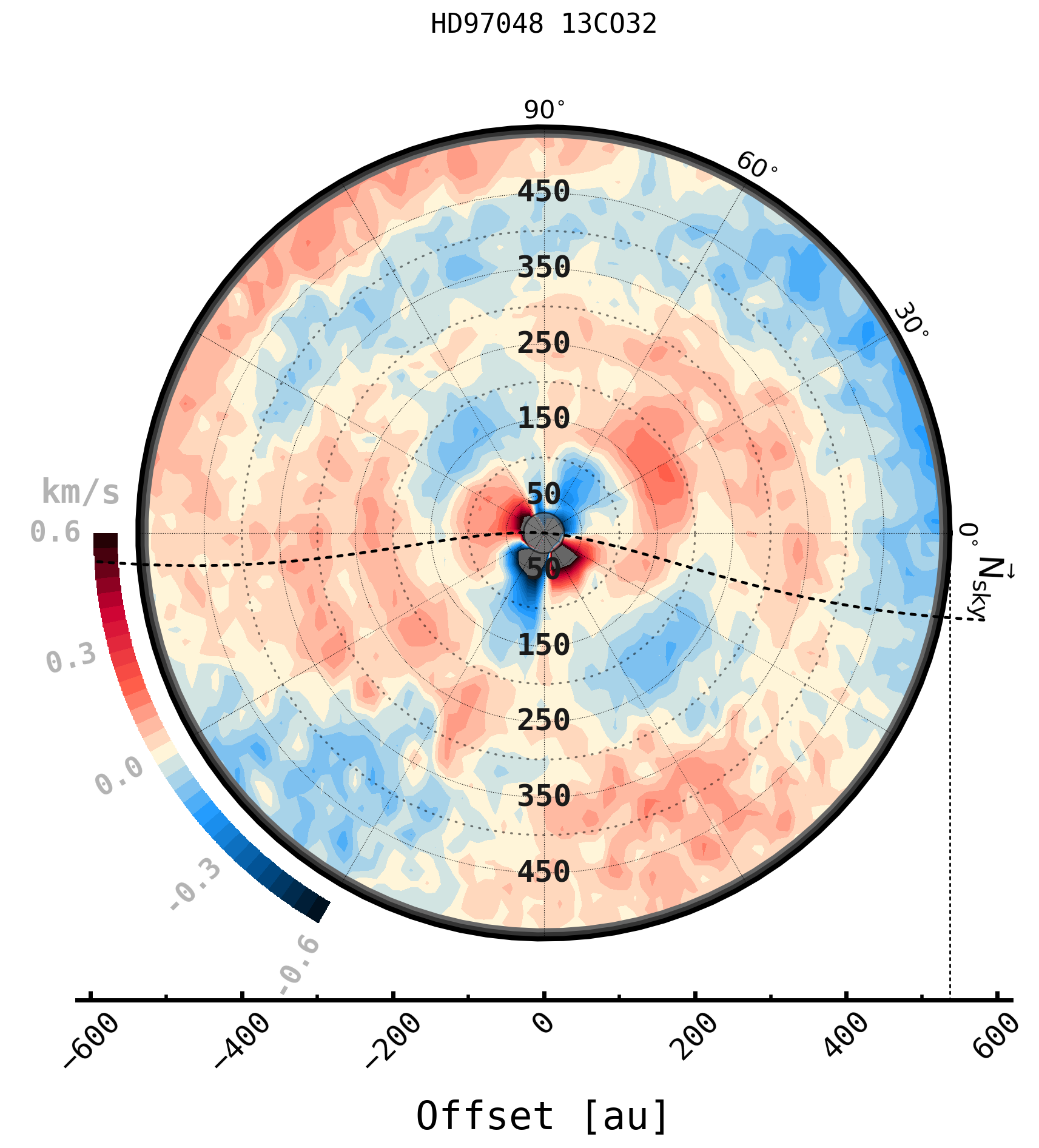}
   \includegraphics[width=0.48\textwidth]{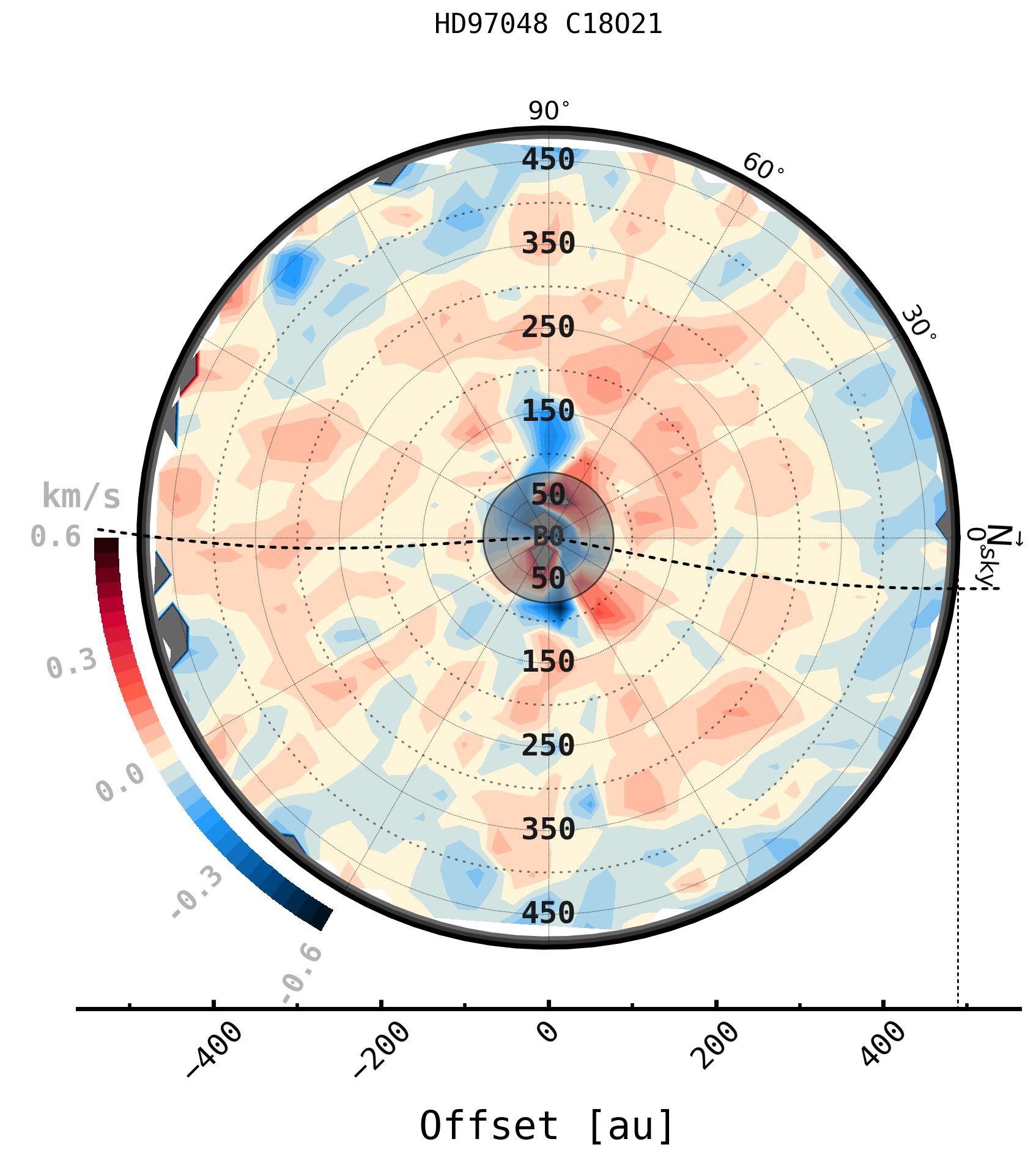}

   \caption{Map of the deprojected velocity residuals for the \ce{^12CO} 2 -- 1, \ce{^13CO} 3 -- 2, and \ce{C^18O} 2 -- 1 lines for HD 97048. The blueshifted side of the disk is located on the left. \edit{The black dashed line shows the $\mathrm{PA}=0$\textdegree\ axis.} Systematically bluer residuals, caused by the presence of absorption, are visible in the blueshifted side of the disk for the \ce{^12CO} 2 -- 1 line, while the other tracers show more uniform residuals.}
   \label{fig:map_vel_res_hd}
\end{figure*}

\begin{figure*}
   \centering
   {\LARGE \textbf{WaOph 6}\par}
   \vspace{0.3cm}

   \includegraphics[width=0.48\textwidth]{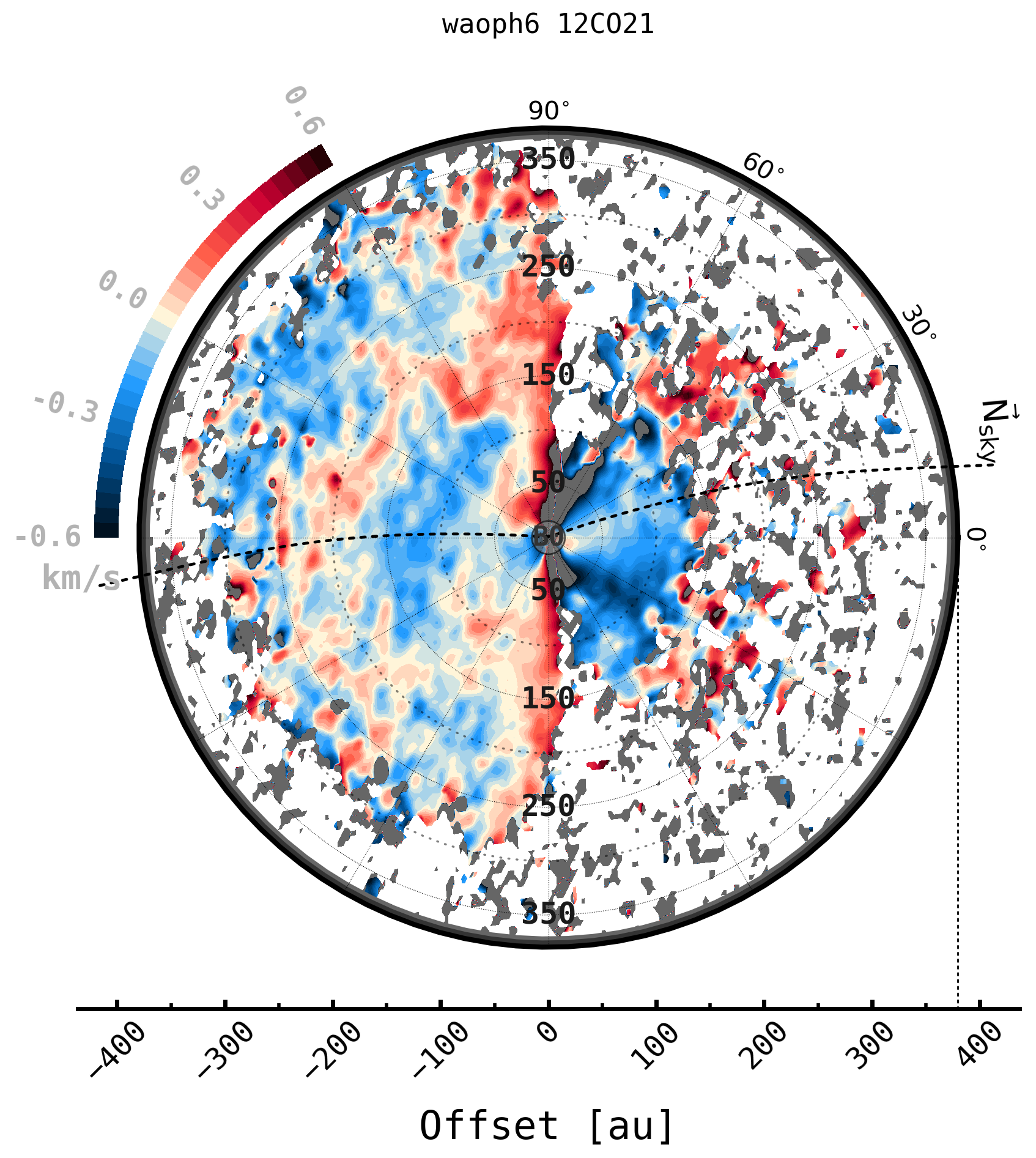}
   \includegraphics[width=0.48\textwidth]{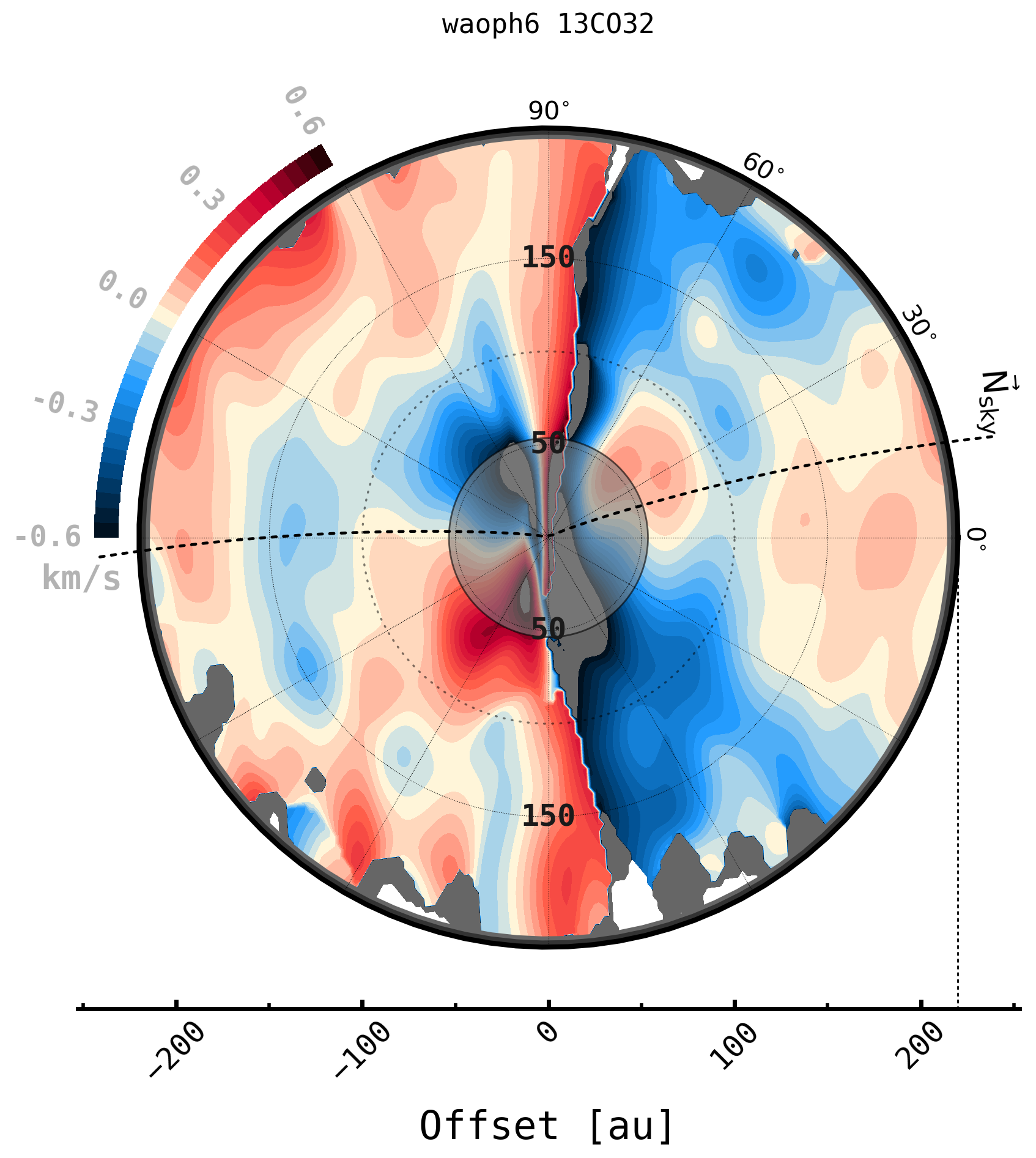}
   \centering
   \includegraphics[width=.9\columnwidth]{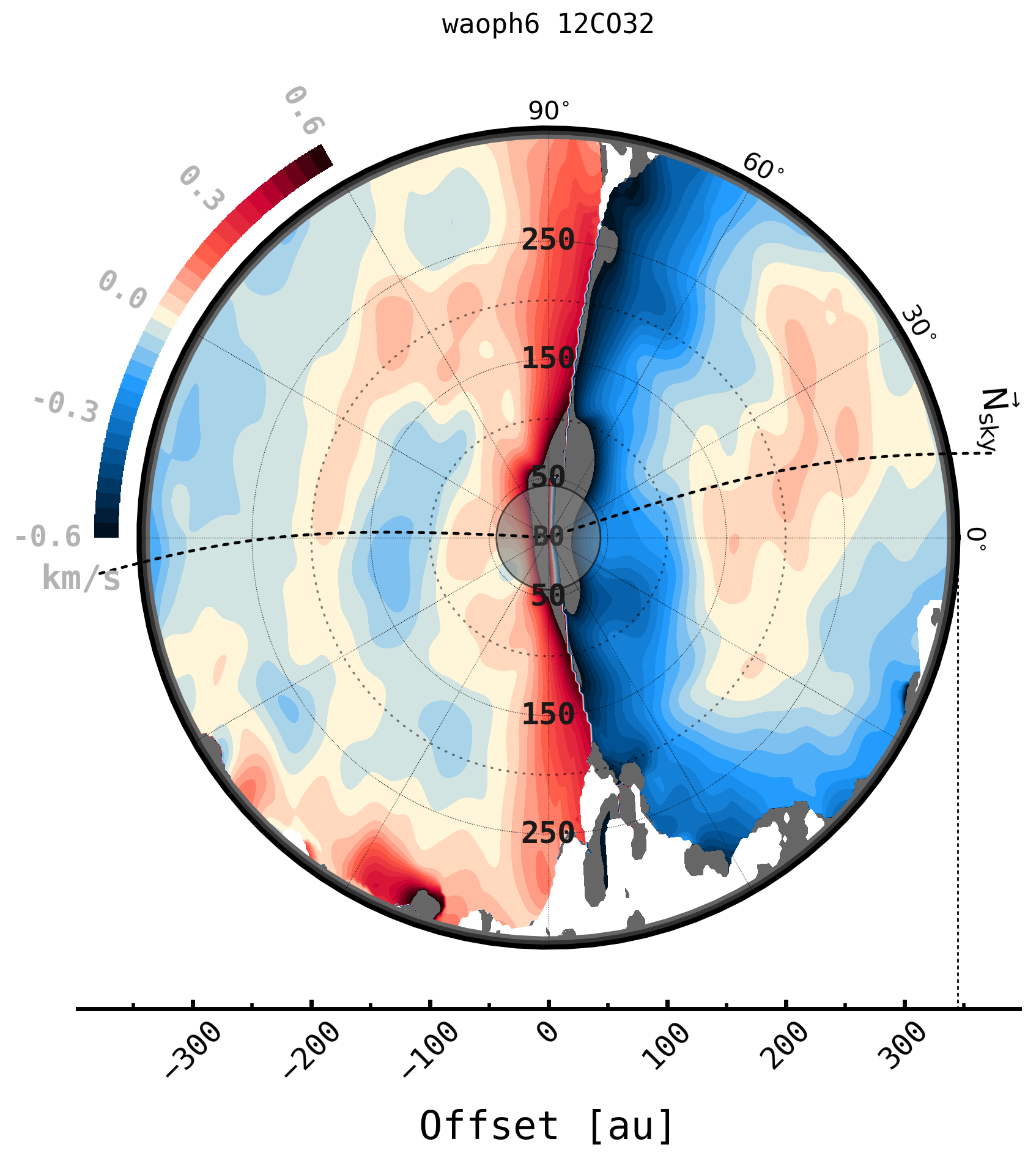}
   \includegraphics[width=.9\columnwidth]{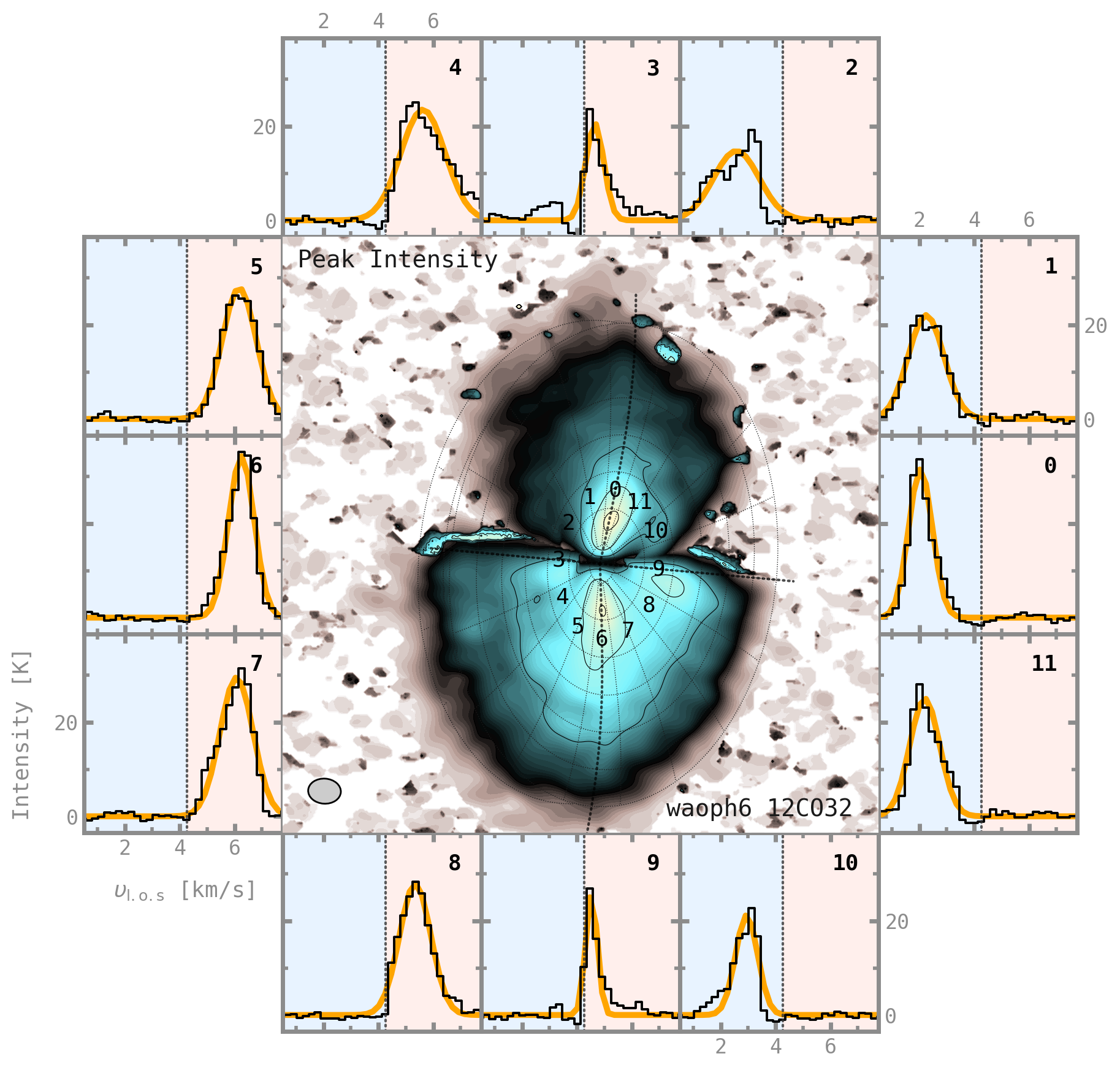}

   \caption{Map of the deprojected velocity residuals for the \ce{^12CO} 2 -- 1, \ce{^13CO} 3 -- 2, and \ce{^12CO} 3 -- 2 lines of WaOph 6. The blueshifted side of the disk is located on the right. \edit{The black dashed line shows the $\mathrm{PA}=0$\textdegree\ axis.} Systematically bluer residuals, caused by the presence of absorption, are visible in the blueshifted side of the disk for all tracers. In the bottom right panel we show the peak intensity map for the \ce{^12CO} 3 -- 2 line of WaOph 6, where the line profiles are extracted along the azimuthal direction in an annulus of fixed radius $R=100$ au. The selected azimuthal locations are specified with numbers on the map, and the corresponding spectra are shown, along with the Gaussian best-fit to the line profiles (orange line). The blueshifted side of the emission is visible in the upper half of the disk.}
   \label{fig:map_vel_res_wa}
\end{figure*}

\edit{We show here the maps of the deprojected velocity residuals for all tracers in HD 97048 (\ref{fig:map_vel_res_hd}) and WaOph 6 (\ref{fig:map_vel_res_wa}). 
As mentioned in Sect. \ref{subsec:absorption}, for the lines with clear absorption in the blueshifted side of the disk (\ce{^12CO} for HD 97048 and all tracers for WaOph 6) we extracted the azimuthal velocities from the redshifted, non-absorbed side of the disk only, to avoid systematic biases in the rotation curves due to cloud absorption.
We here comment on the motivation for the chosen approach, discussing the \ce{^12CO} 3 -- 2 line case for WaOph 6 as an example (shown in the lower panel of Fig. \ref{fig:map_vel_res_wa}), but the same phenomenon is observed for the other tracers with partial absorption in the blueshifted side of the disk. In the left lower panel, the partially absorbed half of the disk shows significantly blueshifted homogeneous residuals, with velocities even $\approx 0.6$ km/s lower than the Keplerian reference; the non-absorbed side, on the contrary, shows much more uniform residuals. For the \ce{^13CO} 3 -- 2 line, absorption causes systematically bluer residuals in the blueshifted side of the disk, while it has a more destructive impact on the retrieved velocities for the \ce{^12CO} 2 -- 1 line. 
As for the \ce{^12CO} 2 -- 1 line of HD 97048, we clearly see that in the blueshifted portion of the disk (on the left in the upper left panel of Fig. \ref{fig:map_vel_res_hd}) the residuals also show homogeneous sub-Keplerian velocities induced by the presence of diffuse absorption. Conversely, for \ce{^13CO} and \ce{C^18O} the residual map shows both blue- and redshifted portions, highlighting the presence of strong kinematic perturbations in the whole disk. }

\edit{We also investigated the shape of the line profiles across the azimuthal direction: we show the \ce{^12CO} 3 -- 2 line for WaOph 6 as an example.} As visible in the lower right panel of Fig. \ref{fig:map_vel_res_wa}, the spectra that are located within the selected azimuthal wedge (the region of [-30\textdegree, 30\textdegree] along the disk major axis) on the blueshifted side show asymmetric line profiles, with a prominent peak systematically shifted toward lower (i.e., bluer) velocities, followed by a sharp decrease in the line slope. This clear asymmetry in the line profiles implies that the presence of absorption in the blueshifted side is causing artifacts in the velocity field.
Whereas excluding the azimuthal regions closer to the disk minor axis helps restricting the problem, the effects of the diffuse absorption in the blueshifted side are still visible within the azimuthal range where average velocities are computed, especially in the inner regions ($R\leq100$ au), preventing the correct identification of the line centroids.
If taken into account for the rotation curves extraction, the retrieved line centroids in the blueshifted side of the disk would systematically bias the rotation velocities, especially at smaller radii, toward lower values.
Moreover, by only considering the non-contaminated side of the disk, we could retrieve precise measurements of the velocities along the whole radial extent of the disk. Conversely, the retrieval of rotational velocities in the disk outer regions was not possible using the disk blueshifted side, as the emission at larger radii is obscured and the signal-to-noise ratio is too low to precisely locate the line centroid from spectra. 
We notice that in all residual velocity maps for WaOph 6 we observe an arc-like feature at $\approx 200$ au in the redshifted side of the disk and, more generally, the overall kinematic pattern seems to be compatible and matching among the different tracers. This would hint at the presence of an actual global perturbation ongoing in the disk, which we are able to recover in all the available tracers.

\FloatBarrier

\section{Beam smearing correction: Rotation curves comparison} \label{app:corr_vs_noncorr}

\begin{figure}
   \centering
   {\LARGE \textbf{HD 97048}\par}
   \vspace{0.2cm}

   \includegraphics[width=0.45\textwidth]{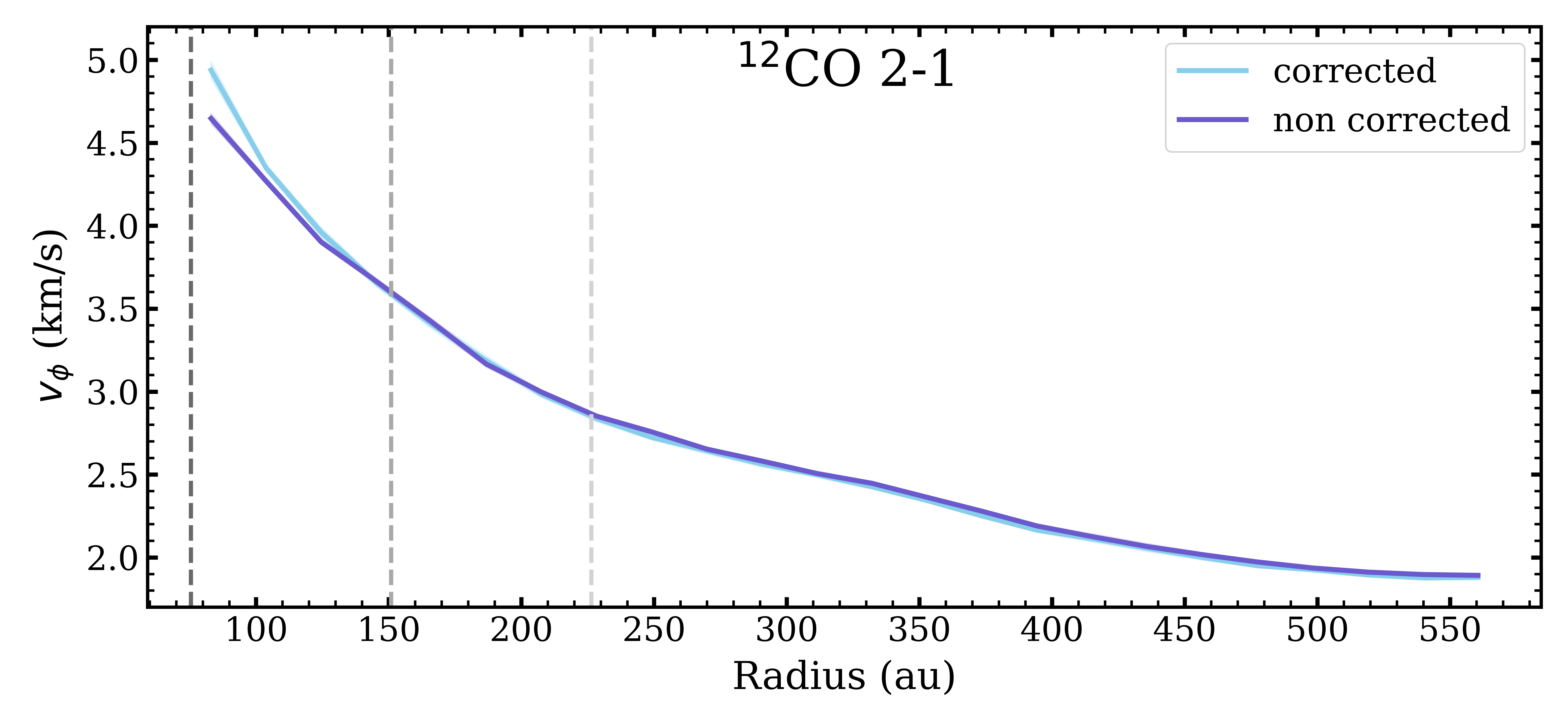}
   \includegraphics[width=0.45\textwidth]{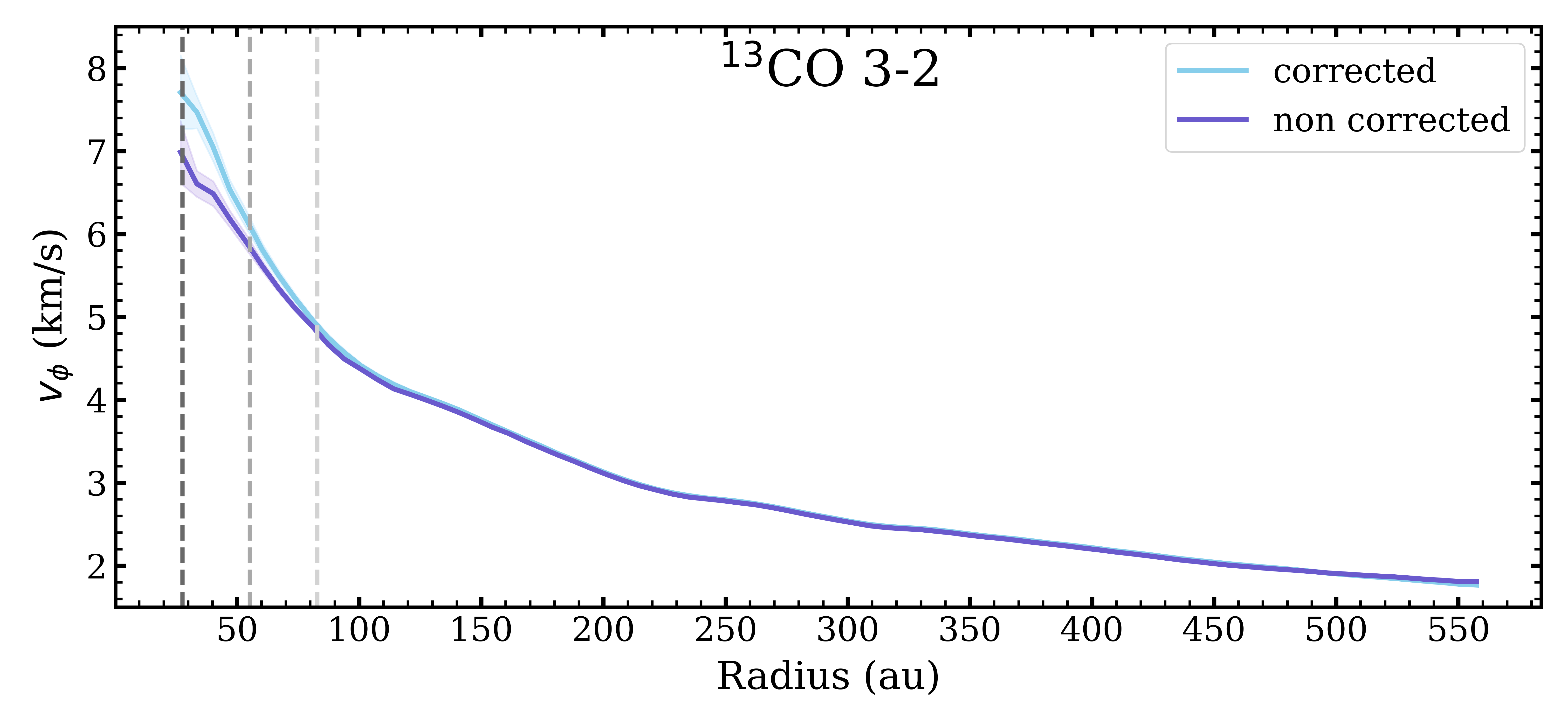}
   \includegraphics[width=0.45\textwidth]{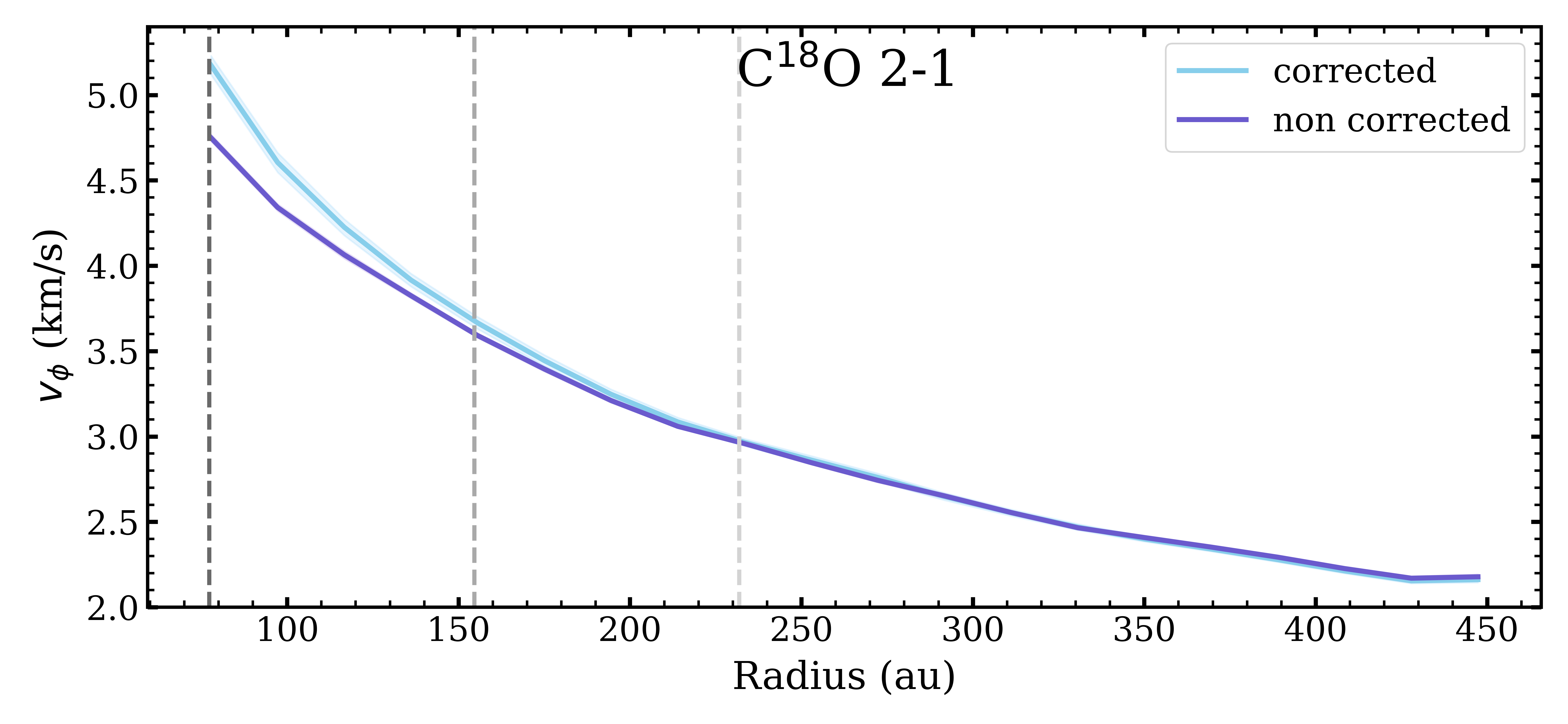}

   \caption{Rotation curves of the \ce{^12CO} 2 -- 1, \ce{^13CO} 3 -- 2, and \ce{C^18O} 2 -- 1 lines
   for HD 97048, before (violet) and after (turquoise) beam smearing correction. \edit{Gray dashed lines represent one, two, and three beams from the central star. Velocity ranges on the y-axis are different for visualization purposes.}}
   \label{fig:corr_vs_noncorr_hd}
\end{figure}

\begin{figure}
   \centering
   {\LARGE \textbf{WaOph 6}\par}
   \vspace{0.2cm}

   \includegraphics[width=0.45\textwidth]{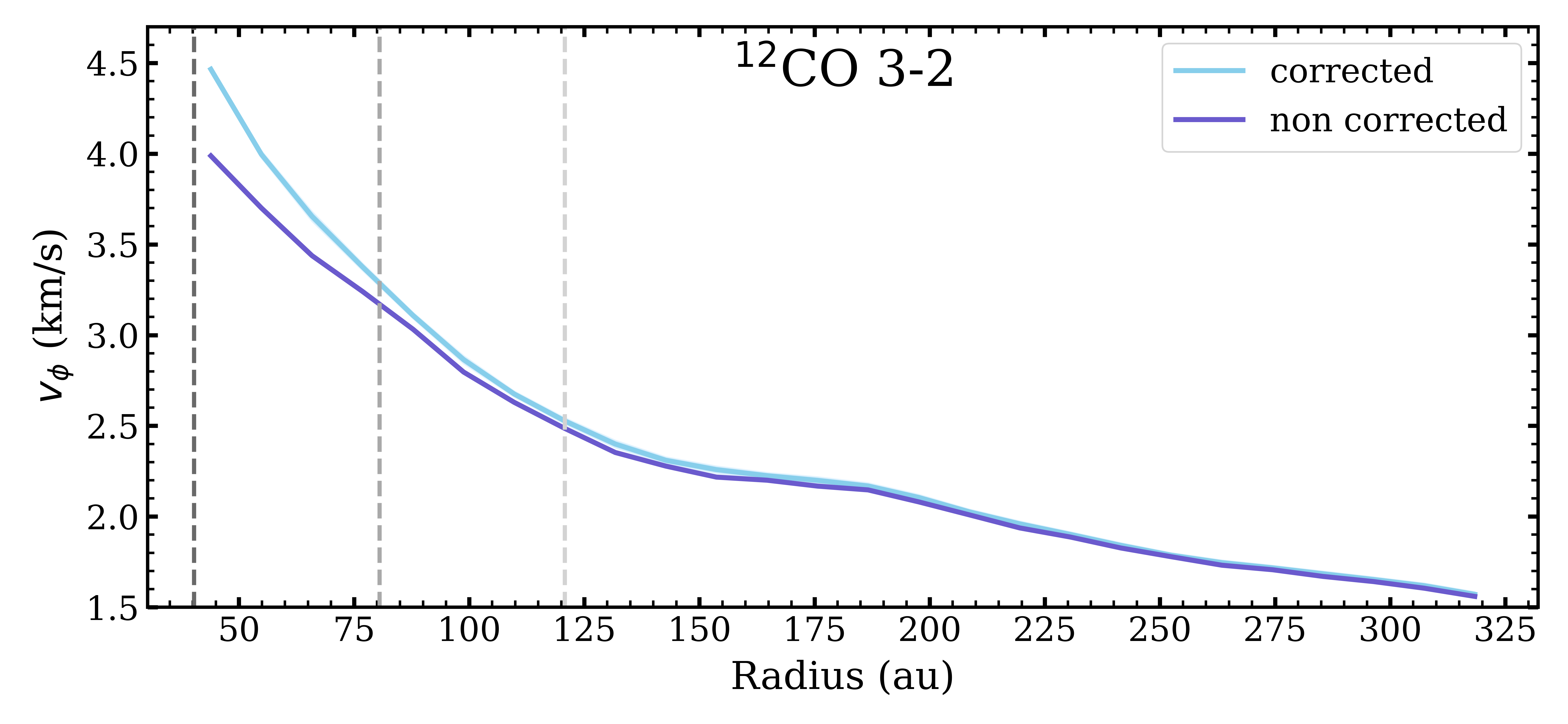}
   \includegraphics[width=0.45\textwidth]{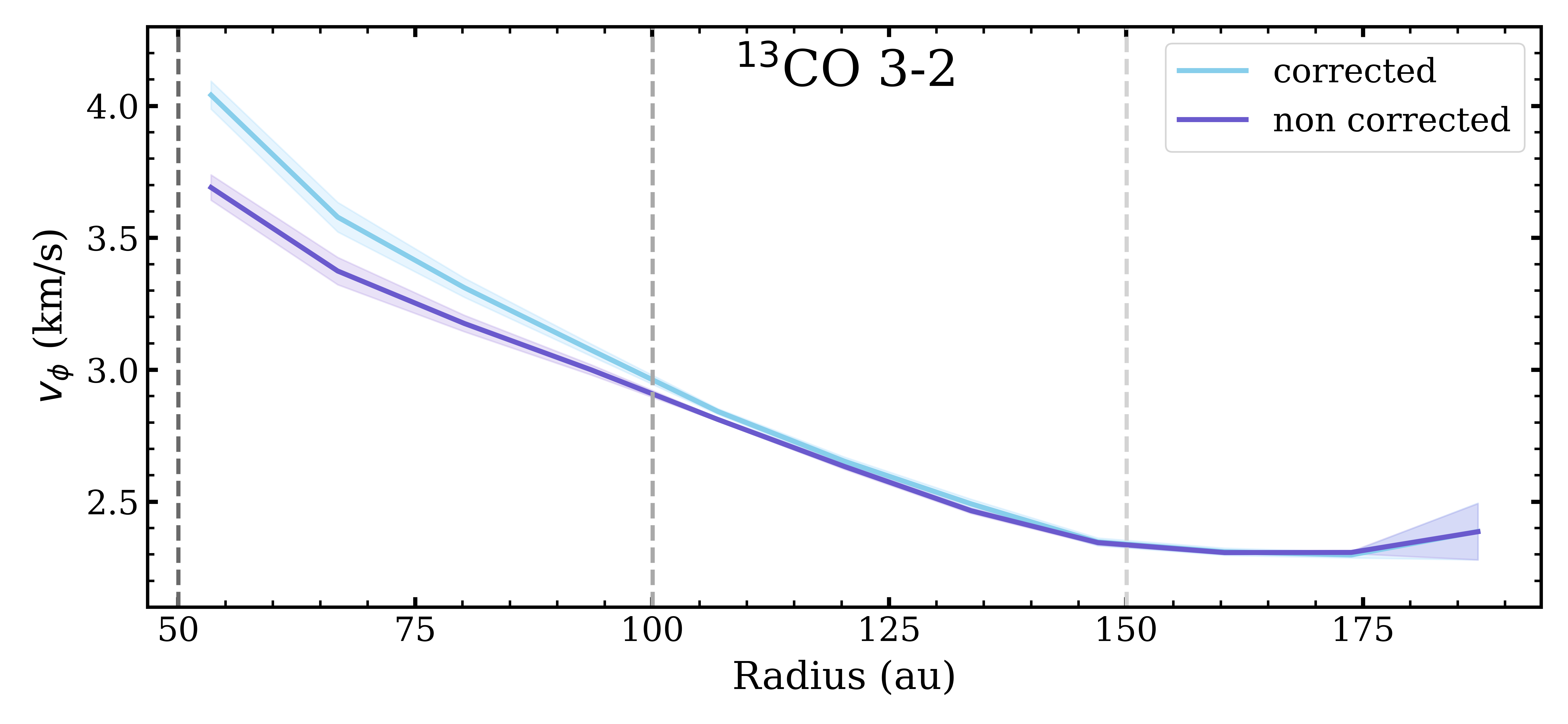}

   \caption{Same as Fig. \ref{fig:corr_vs_noncorr_hd} but for WaOph 6 rotation curves of the \ce{^12CO} and \ce{^13CO} 3 -- 2 lines.
   }
   \label{fig:corr_vs_noncorr_wa}
\end{figure}

In Figs.~\ref{fig:corr_vs_noncorr_hd} and~\ref{fig:corr_vs_noncorr_wa} we compare the corrected and non-corrected rotation curves for all the available tracers for HD 97048 and WaOph 6, respectively.

\section{Emitting layers and brightness temperatures}

Here we show the extracted emitting surfaces for both sources, and the corresponding brightness temperature along them with the color scale. \edit{For WaOph 6, the inner regions above $\approx10$ au appear to be quite cold. However, this area is not probed by any considered emitting surface, and thus its temperature is poorly constrained. This does not affect any of our results, as none of the analyzed tracers emits from this region.}

\begin{figure*}
   \centering
   \includegraphics[width=.8\textwidth]{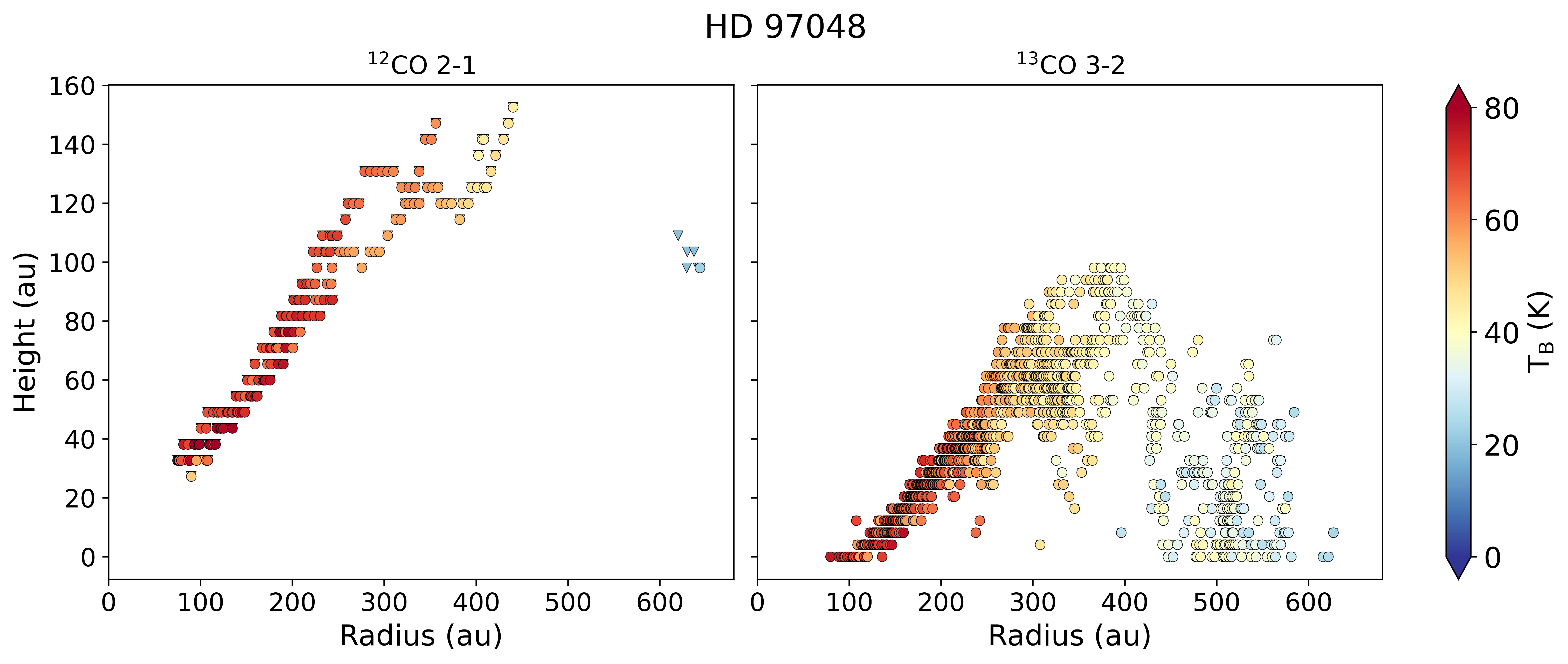}
   \includegraphics[width=.8\textwidth]{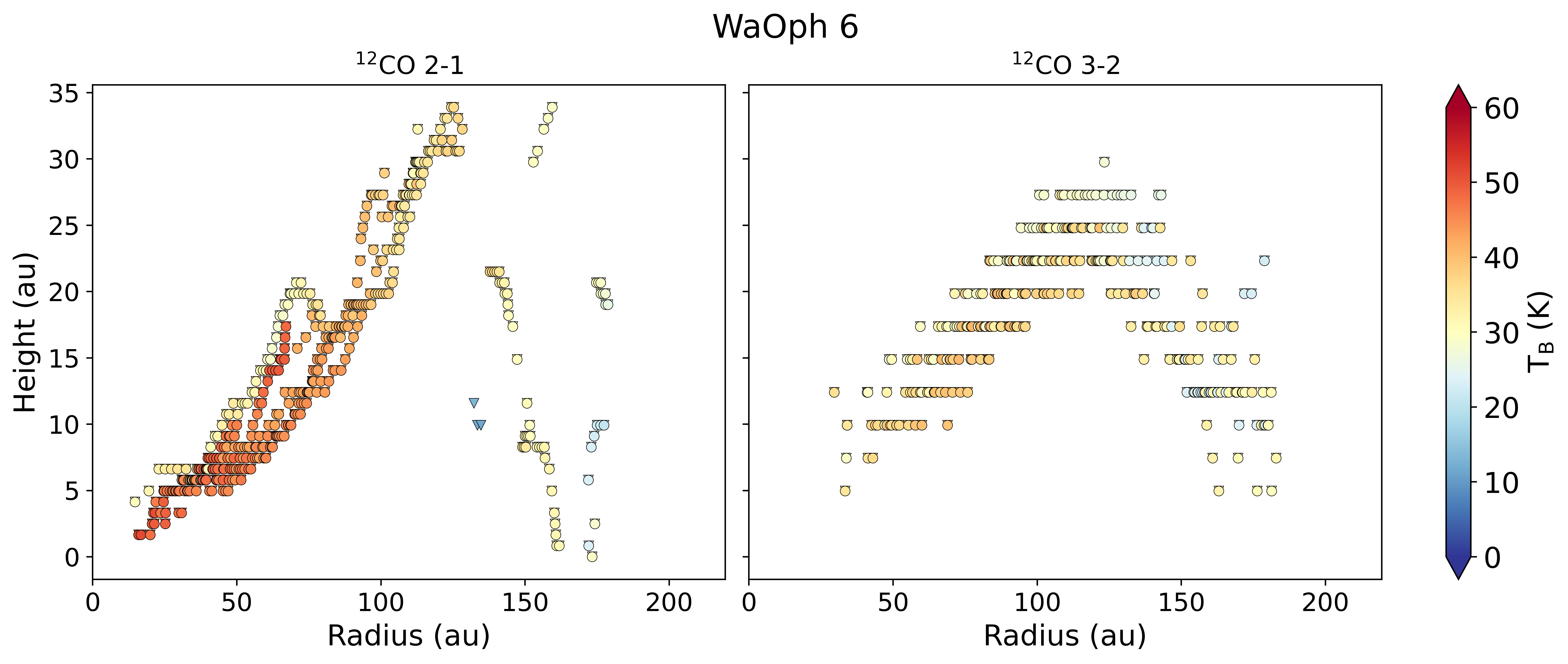}
      \caption{
      Retrieved emitting surfaces for HD 97048 (top panel) and WaOph 6 (bottom panel), for all tracers we considered for the retrieval of the 2D thermal structures. The color bar shows the brightness temperature in K. Points with $T_\mathrm{B}<20$ K are marked by triangles and are not considered in the 2D temperature fit, while all $T_\mathrm{B}\geq20$ points are marked by circles.
      }
         \label{fig:layers}
\end{figure*} 

\section{2D thermal structures}\label{app:2D_T}
Here we show the best-fits of the 2D thermal structures of the two disks, after the bootstrapping procedure.

\begin{figure*}[h!tbp]
    \centering
    \includegraphics[scale=0.5]{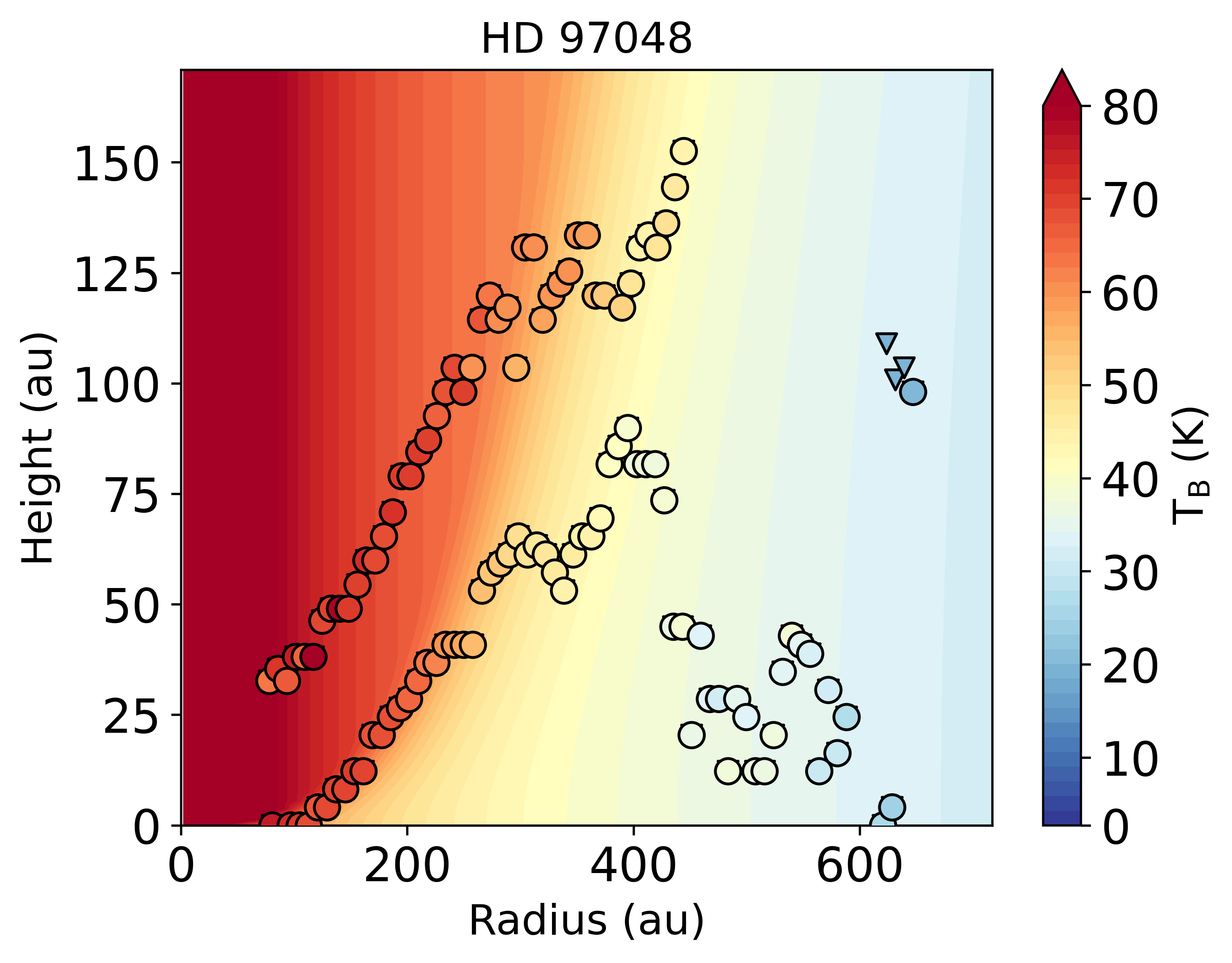}
    \includegraphics[scale=0.5]{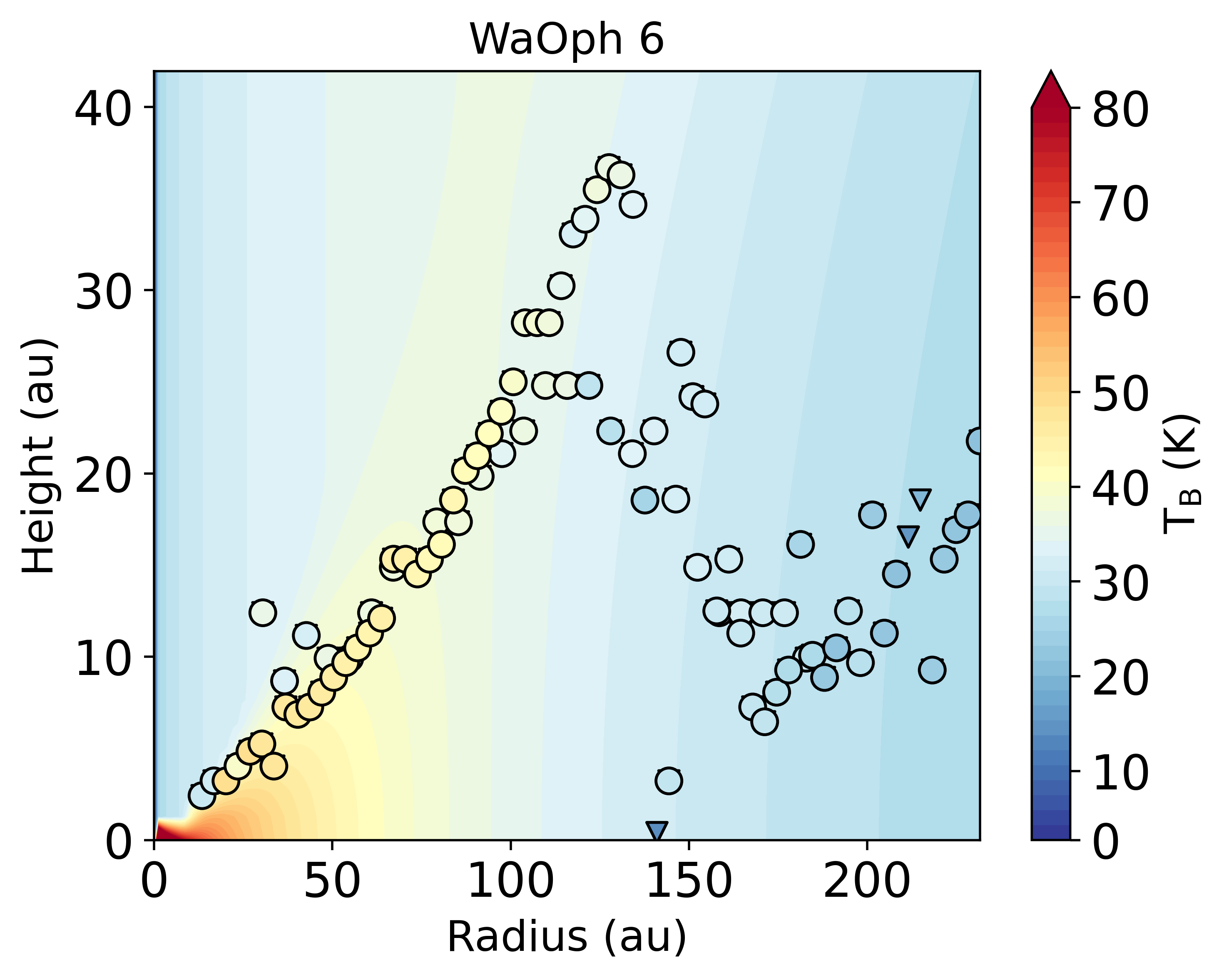} 
    \caption{
    Two-dimensional best-fit thermal structures for HD 97048 (left panel) and WaOph 6 (right panel). The background color represents the parameterized 2D temperature, described by the best-fit parameters after the bootstrapping procedure. The emission surface points with their associated brightness temperatures are binned by a quarter of a beam size and superposed as circles (if $T_\mathrm{B}\geq20$ K) or triangles (if $T_\mathrm{B}<20$ K).}
    \label{fig:2D_T_fits}
\end{figure*} 


\section{Corner plots of the rotation curves fits}\label{app:cornerplots}
We show the corner plots of the rotation curves fit for both sources, after the bootstrapping phase.

\begin{figure*}
   \centering
   \includegraphics[width=0.4\textwidth]{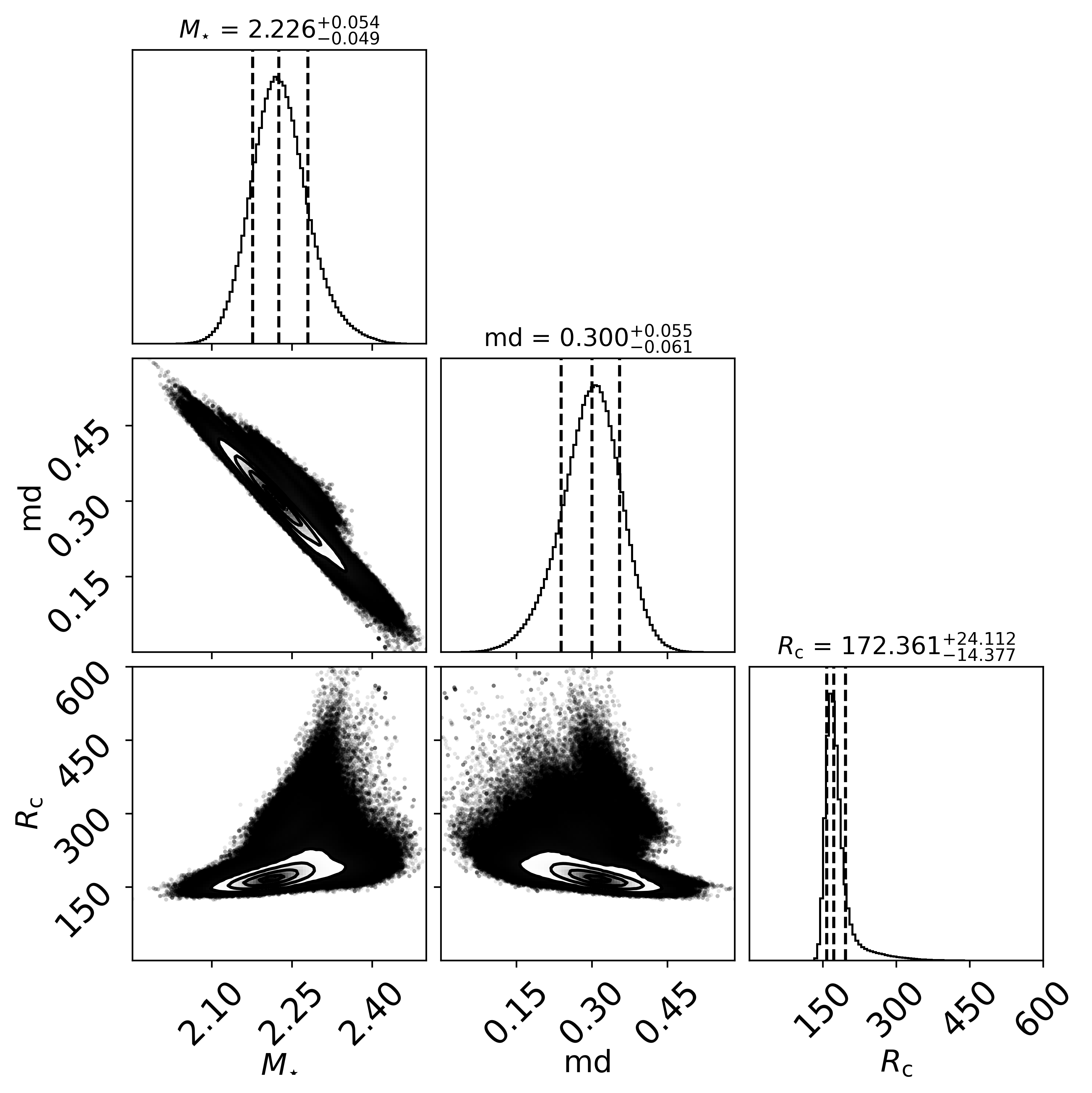} 
   \includegraphics[width=0.4\textwidth]{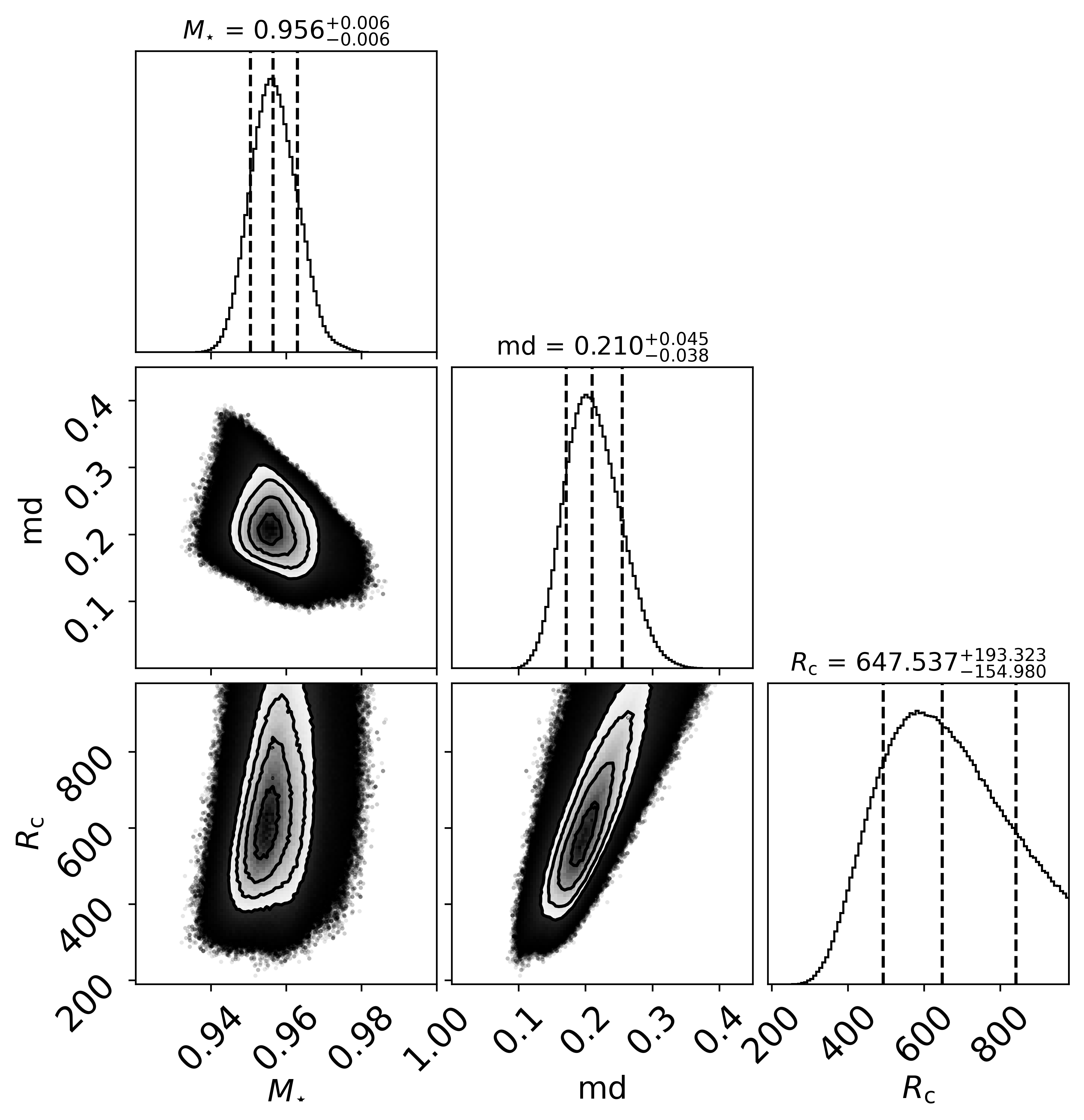} 
      \caption{
      Corner plots of the rotation curves fits for HD 97048 (left panel) and WaOph 6 (right panel), after the bootstrapping procedure. The three dashed lines show respectively the $16^\text{th}$, $50^\text{th}$, and $84^\text{th}$ percentiles of the distributions.}
         \label{fig:total_bootstrapping}
\end{figure*} 

\section{$Q(R)$ radial profiles}\label{app:Q_profiles}

\begin{figure*}[h]
    \centering
    \includegraphics[width=0.8\linewidth]{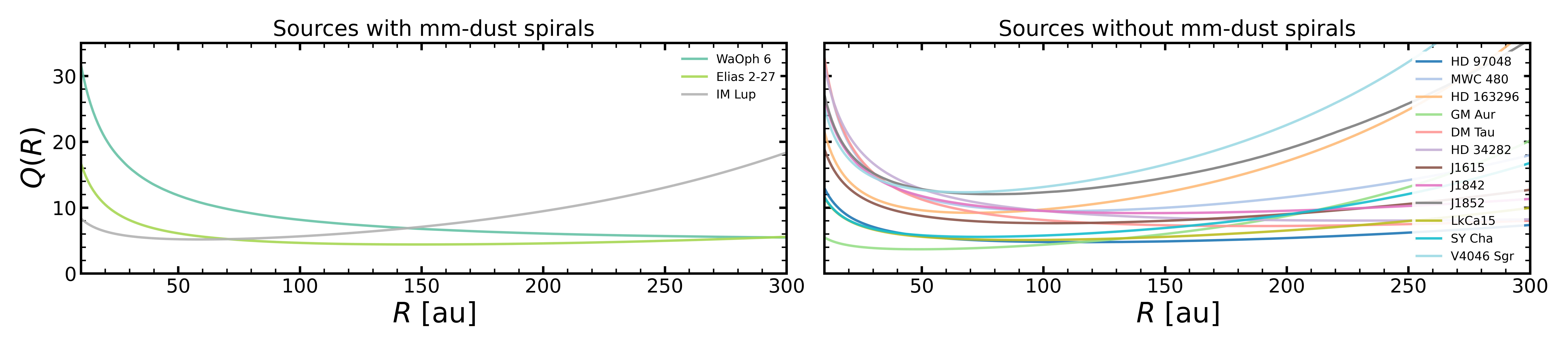}
    \caption{$Q$ profiles for disks with (left panel) and without (right panel) millimeter-dust spirals. The color code indicates the different sources.}
    \label{fig:Q_profiles}
\end{figure*}

\rr{Although a local minimum at any radial location is sufficient to trigger GI, we report in Fig. \ref{fig:Q_profiles} the whole  $Q(R)$ radial profiles for all the sources. This allowed us to show the radial location where the minimum occurs, and whether there is a radially extended region of low $Q$, for each disk. We show sources with (left panel) and without (right panel) millimeter-dust spirals, reporting the corresponding names with the color code. Each line is the median of the 10000 profiles obtained for each disk through the bootstrapping procedure.}

\section{Azimuthal velocity residuals}\label{app:vel_azim_res}

We show here the radial profiles of the azimuthal velocity residuals $\delta v_\phi$ (data-Keplerian) for all the remaining tracers for both sources. The radial derivative of $\delta v_\phi$ is expected to be positive in proximity of a pressure minimum and be negative around the location of pressure maxima, assuming that the observed pressure modulations are due to the presence of millimeter-dust substructures.

We show the velocity residuals linked to pressure substructures using both the corrected (thick black line) and non-corrected (thin gray line) rotation curves, to highlight the differences. The two curves differ mostly inside $R\approx150$ au: including the correction in the curves, we observe pressure modulations that are generally more in agreement with the locations of the millimeter-dust substructures (see also Fig. \ref{fig:deltavphi} for completeness). With our method, we are able to efficiently retrieve the velocity and pressure structures up to a much smaller radius than previously achieved.

\begin{figure*}[t!]
   \centering
   {\LARGE \textbf{HD 97048}\par}
   \includegraphics[width=.4\textwidth]{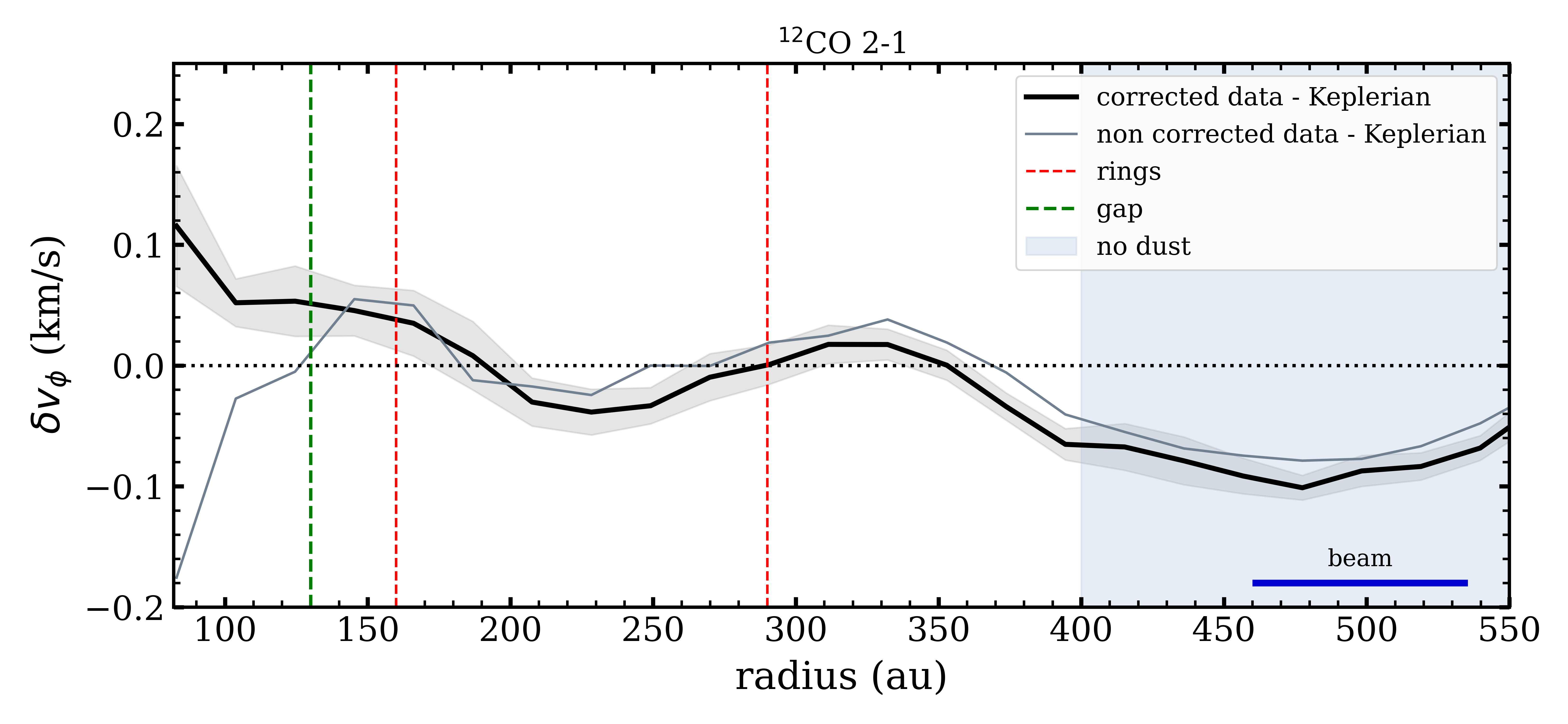}
   \includegraphics[width=.4\textwidth]{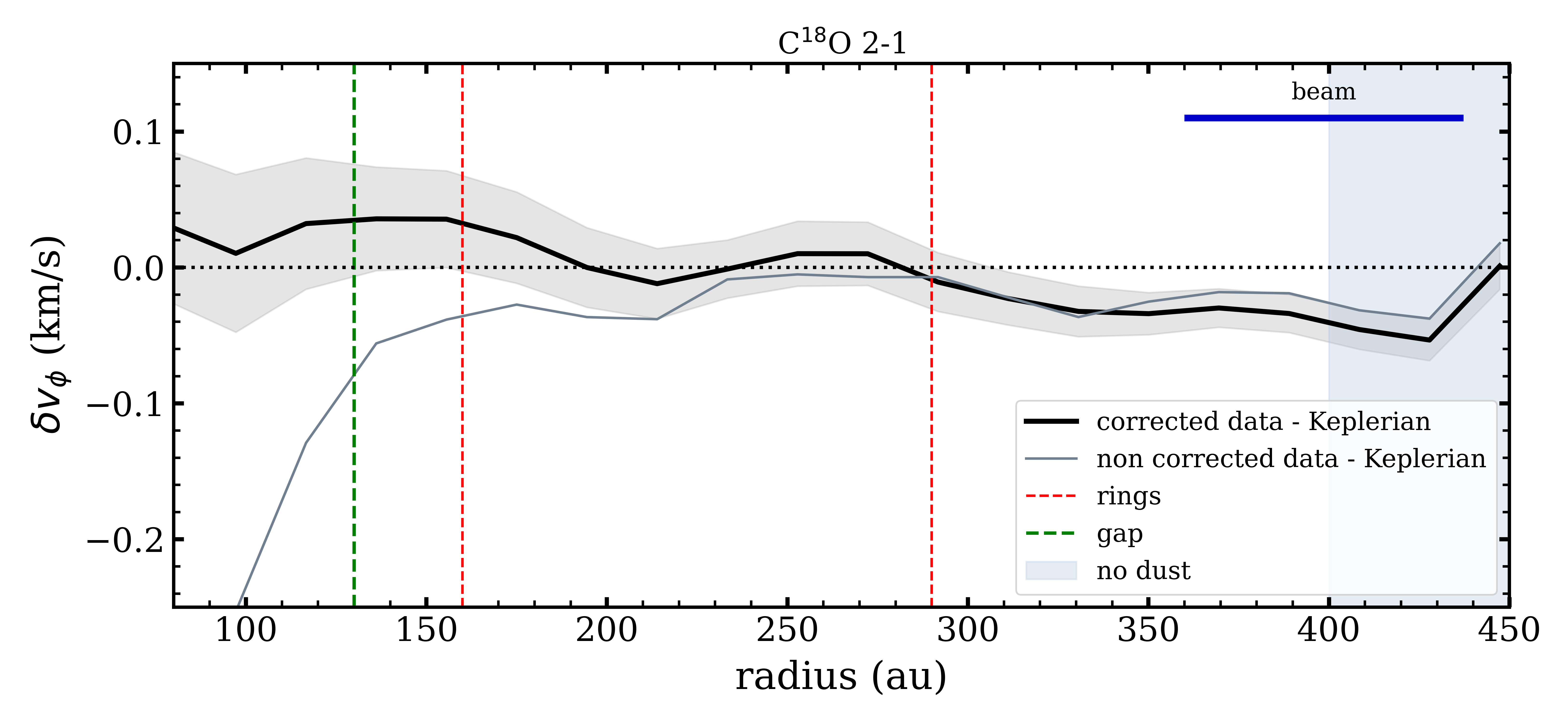}

   {\LARGE \textbf{WaOph 6}\par}
   \includegraphics[width=.4\textwidth]{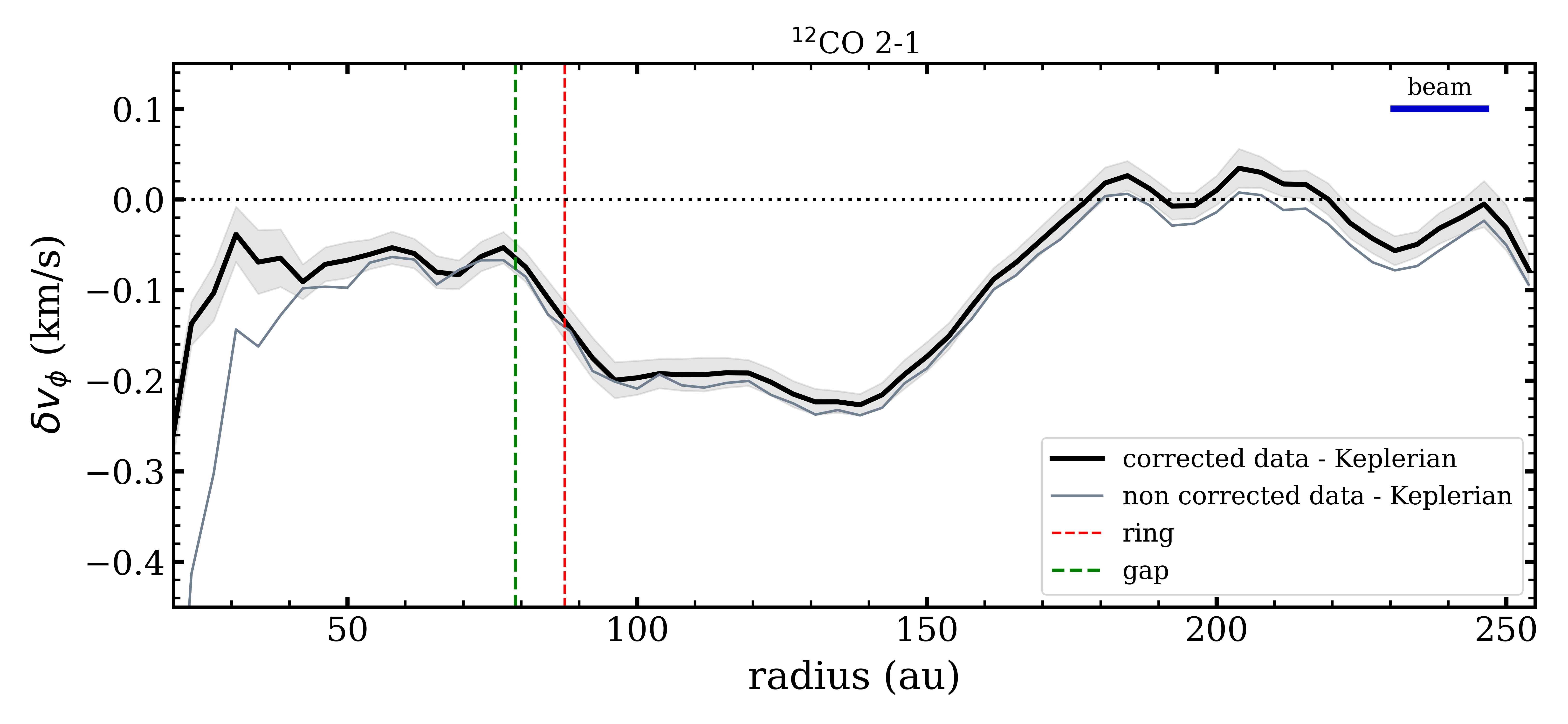}
   \includegraphics[width=.4\textwidth]{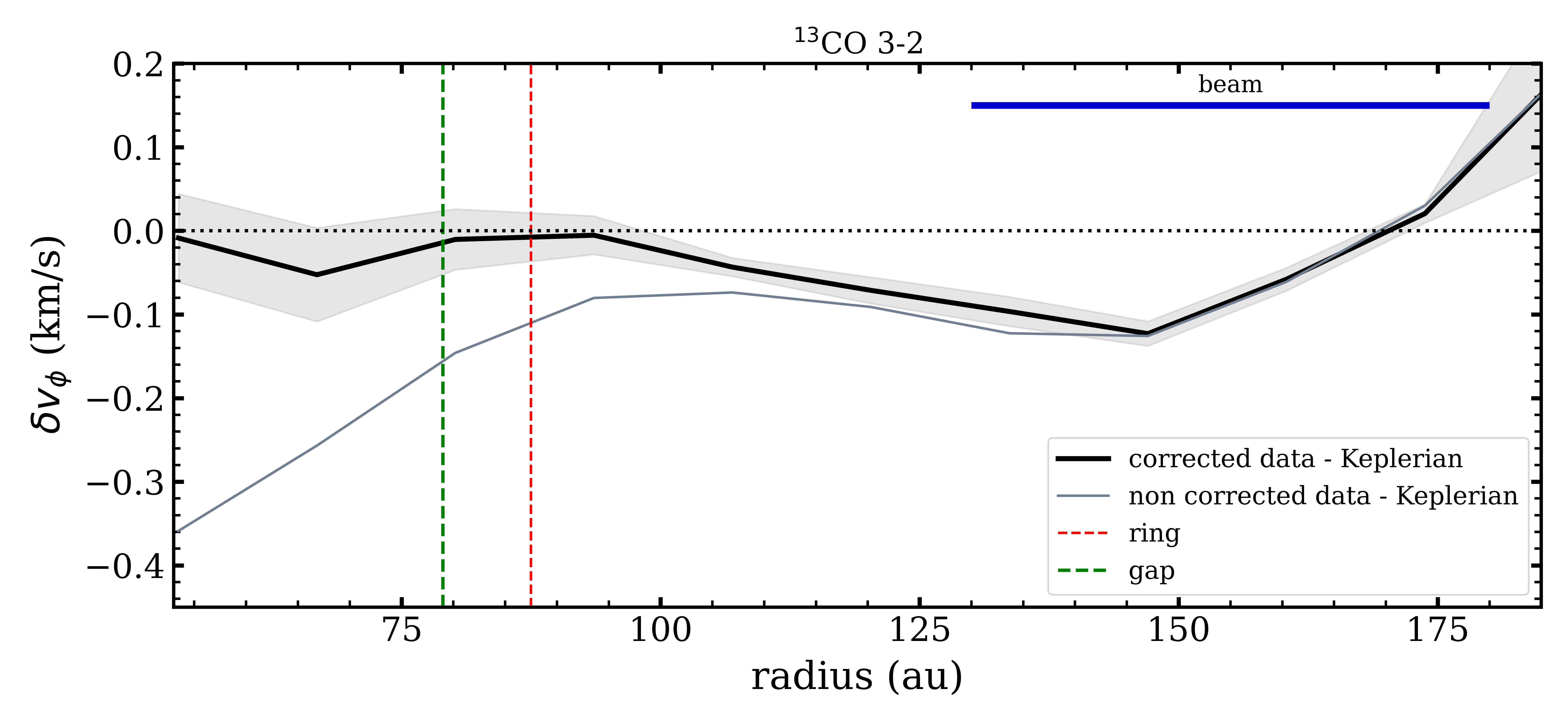}

   \caption{Radial profiles of the azimuthal velocity residuals $\delta v_\phi$ (data-Keplerian) of the \ce{^12CO} and \ce{C^18O} 2 -- 1 lines for HD 97048 (top), and the \ce{^12CO} 2 -- 1 and \ce{^13CO} 3 -- 2 lines for WaOph 6 (bottom), with the corresponding uncertainties. The thick black line is obtained using beam smearing corrected rotation curves, while the thin gray line is obtained with the non-corrected curves. Red lines show the locations of millimeter-dust rings, while green lines mark the millimeter-dust gaps (as reported by \citealt{pinte2019}). \rr{The horizontal blue line in the right corner of each panel shows the beam size of the observations.}}
   \label{fig:deltavphi_others}
\end{figure*}

\end{appendix}

\end{document}